\documentclass[floatfix,prd,epsfig,nofootinbib,superscriptaddress,onecolumn,amssymb]{revtex4}

\usepackage{slashed}
\usepackage{graphicx,color}
\usepackage{epsfig}
\usepackage{dcolumn}
\usepackage{bm}
\usepackage{color}
\usepackage{siunitx}
\usepackage{caption}
\usepackage{subcaption}

\def\lsim{\mathrel{\rlap{\lower4pt\hbox{\hskip1pt$\sim$}}
    \raise1pt\hbox{$<$}}}         
\def\gsim{\mathrel{\rlap{\lower4pt\hbox{\hskip1pt$\sim$}}
    \raise1pt\hbox{$>$}}}         

\newcommand{\beq}{\begin{equation}}  
\newcommand{\eeq}{\end{equation}}  
\newcommand{\beqa}{\begin{eqnarray}}  
\newcommand{\eeqa}{\end{eqnarray}}  
\newcommand{\bea}{\begin{eqnarray}}  
\newcommand{\eea}{\end{eqnarray}}

\def\to{\rightarrow}
\def\Re{{\cal R \mskip-4mu \lower.1ex \hbox{\it e}\,}}
\def\Im{{\cal I \mskip-5mu \lower.1ex \hbox{\it m}\,}}
\def\be{\begin{equation}}
\def\ee{\end{equation}}
\def\bea{\begin{eqnarray}}
\def\eea{\end{eqnarray}}
\def\bit{\begin{itemize}}
\def\eit{\end{itemize}}

\newcommand \eos{\textsc{Eos} }
\newcommand \theia{\textsc{Theia} }

\def\ee{e^+e^-}

\def\msb{{\bar{\ssstyle M \kern -1pt S}}}

\begin{document}

\title{Future Advances in Photon-Based Neutrino Detectors: A SNOWMASS White Paper}

\newcommand{\Editor}{Editor.}
\newcommand{\Contributor}{Primary Contributor.}



\newcommand{\UA}{University at Albany, State University of New York (SUNY), Albany, NY USA}
\newcommand{\ANL}{Argonne National Laboratory, IL, USA}
\newcommand{\APC}{Université de Paris, CNRS, Astroparticule et Cosmologie, F-75013 Paris, France}
\newcommand{\Atlantico}{Universidad del Atl\'antico, Puerto Colombia, Atl\'antico, Colombia}
\newcommand{\UBERN}{Universt\"at Bern, Bern, Switzerland}
\newcommand{\bart}{Bartoszek Engineering, Aurora, IL 60506, USA}
\newcommand{\BEU}{Beykent University, Istanbul, Turkey}
\newcommand{\BU}{Boston University, Boston, MA, USA}
\newcommand{\BNL}{Brookhaven National Laboratory, NY, USA}
\newcommand{\BHU}{Banaras Hindu University, Varanasi, INDIA}
\newcommand{\UoBham}{University of Birmingham, Birmingham B15 2TT, United Kingdom}
\newcommand{\UB}{University of Bristol, Bristol BS8 1TL, United Kingdom}
\newcommand{\UoB}{University of Bucharest, Bucharest, Romania}
\newcommand{\Caltech}{California Institute of Technology, Pasadena, CA, USA}
\newcommand{\UCB}{University of California, Berkeley, Berkeley, CA USA}
\newcommand{\UCD}{University of California, Davis, Davis, CA USA}
\newcommand{\UNC}{University of North Carolina, Chapel Hill, NC USA}
\newcommand{\UCI}{University of California, Irvine, CA, USA}
\newcommand{\JGU}{Johannes Gutenberg University, Mainz, Germany}
\newcommand{\UCSB}{University of California, Santa Barbara, Santa Barbara, CA USA}
\newcommand{\Cambridge}{University of Cambridge, Cambridge CB3 0HE, United Kingdom}
\newcommand{\Catania}{University and INFN Sezione di Catania, Catania, Italy}
\newcommand{\CERN}{CERN, 1 Esplanade des Particules, Geneva, Switzerland}
\newcommand{\CIEMAT}{CIEMAT, Centro de Investigaciones Energéticas, Medioambientales y Tecnólogicas, Madrid, Spain}
\newcommand{\Cincinnati}{University of Cincinnati, Cincinnati, OH, USA}
\newcommand{\CUSB}{Central University of South Bihar, Gaya, INDIA}
\newcommand{\ChU}{Charles University, Faculty of Mathematics and Physics, Prague, Czech Republic}
\newcommand{\Cinvestav}{Centro de Investigación y de Estudios Avanzados del Instituto Politécnico Nacional (Cinvestav), Mexico City, Mexico}
\newcommand{\Columbia}{Columbia University, New York, NY, USA}
\newcommand{\CSU}{Colorado State University, Fort Collins, CO, USA}
\newcommand{\Demokritos}{NCSR Demokritos, Agia Paraskevi, Greece}
\newcommand{\Duke}{Duke University, Durham, NC, USA}
\newcommand{\Drexel}{Drexel University, Philadelphia, PA, USA}
\newcommand{\EDI}{University of Edinburgh, Edinburgh EH8 9YL, United Kingdom}
\newcommand{\EIA}{Universidad EIA, Envigado, Antioquia, Colombia}

\newcommand{\FCUL}{Faculdade de Ci\^{e}ncias da Universidade de Lisboa, Lisboa, Portugal}
\newcommand{\FCTUC}{Faculdade de Ci\^{e}ncias e Tecnologia da Universidade de Coimbra, Coimbra, Portugal}
\newcommand{\FNAL}{Fermi National Accelerator Laboratory, Batavia, IL, USA}
\newcommand{\UF}{University of Florida, Gainesville, FL 32611, USA}
\newcommand{\UGR}{University of Granada \& CAFPE, Granada, Spain}
\newcommand{\Harvard}{Harvard University, Cambridge, MA, USA}
\newcommand{\Hawaii}{University of Hawaii, Honolulu, HI, USA}
\newcommand{\Houston}{University of Houston, Houston, TX, USA}
\newcommand{\ISU}{Idaho State University, Pocatello, ID, USA} 
\newcommand{\IU}{Indiana University, Bloomington, IN, USA} 
\newcommand{\IFIC}{Instituto de F\'isica Corpuscular (IFIC), CSIC \& Universitat de Val\`encia, Paterna, Spain}
\newcommand{\IIT}{Illinois Institute of Technology, Chicago, IL, USA}
\newcommand{\IITH}{Indian Institute of Technology Hyderabad, Kandi, Sangareddy, India}
\newcommand{\CzechAcademyofSciences}{Institute of Physics, Czech Academy of Sciences, 182 00 Prague 8, Czech Republic}
\newcommand{\INFNMi}{Istituto Nazionale di Fisica Nucleare, Sezione di Milano, Milano, Italy}
\newcommand{\INFNMiB}{Istituto Nazionale di Fisica Nucleare, Sezione di Milano Bicocca, Milano, Italy}
\newcommand{\INFNGE}{Istituto Nazionale di Fisica Nucleare, Sezione di Genova, Genova, Italy}
\newcommand{\INFNPd}{Istituto Nazionale di Fisica Nucleare, Sezione di Padova, Padova, Italy}
\newcommand{\INFNNA}{Universit\`a degli Studi di Napoli ``Federico II'' \& Istituto Nazionale di Fisica Nucleare, Sezione di Napoli, Napoli, Italy}
\newcommand{\Insubria}{Universit{\`a} degli Studi dell'Insubria, Dipartimento di Scienza e Alta Tecnologia, Como, Italy}
\newcommand{\UOH}{University of Hyderabad, Hyderabad - 500046, India}
\newcommand{\Iowa}{University of Iowa, Iowa City, IA, USA}
\newcommand{\ISP}{Instituto Superior Técnico, Lisbon, Portugal}
\newcommand{\IowaU}{Iowa State University, Ames, IA, USA}
\newcommand{\Irfu}{IRFU, CEA Saclay, Gif-sur-Yvette, France}
\newcommand{\Iwate}{Iwate University, Morioka, Iwate, Japan}
\newcommand{\JBNU}{Jeonbuk National University, Jeonju, Jeonbuk, Korea}
\newcommand{\KSU}{Kansas State University, Manhattan, KS, USA}
\newcommand{\Kyiv}{Taras Shevchenko National University of Kyiv, 01601 Kyiv, Ukraine}
\newcommand{\NRCKI}{National Research Center Kurchatov Institute, Moscow, Russian Federation}
\newcommand{\JSU}{Jackson State University, Jackson, MS, USA}
\newcommand{\jnu}{Jawaharlal Nehru University, New Delhi 110067,  India}
\newcommand{\JU}{University of Jammu, J$\&$K, India}
\newcommand{\KCL}{King's College London, London WC2R 2LS, United Kingdom }
\newcommand{\Lancaster}{Lancaster University, Lancaster LA1 4YW, United Kingdom}
\newcommand{\LBNL}{Lawrence Berkeley National Laboratory, Berkeley, CA USA}
\newcommand{\LLNL}{Lawrence Livermore National Laboratory, Livermore, CA USA}
\newcommand{\Liverpool}{University of Liverpool, Liverpool L69 7ZE, United Kingdom}
\newcommand{\LANL}{Los Alamos National Laboratory, Los Alamos, NM, USA}
\newcommand{\LIP}{Laboratório de Instrumentação e Física Experimental de Partículas, Lisboa and Coimbra, Portugal}
\newcommand{\LSU}{Louisiana State University, Baton Rouge, LA, USA}
\newcommand{\Lucknow}{University of Lucknow, Lucknow, Uttar Pradesh, India}
\newcommand{\MIT}{Massachusetts Institute of Technology, Cambridge, MA, USA}
\newcommand{\miis}{Middlebury Institute of International Studies at Monterey, Monterey, California 93940, USA}
\newcommand{\MPA}{Max Planck Institute for Astrophysics, Garching, Germany}
\newcommand{\UMN}{University of Minnesota, Minneapolis, MN, USA}
\newcommand{\MSU}{Michigan State University, East Lansing, MI, USA}
\newcommand{\umich}{University of Michigan, Ann Arbor, MI, USA}
\newcommand{\Duluth}{University of Minnesota Duluth, Duluth, MN, USA}
\newcommand{\UMiss}{University of Mississippi, University, MS, USA}
\newcommand{\UCNCl}{Universidad Católica del Norte, Antofagasta, Chile}
\newcommand{\NCCU}{North Carolina Central University, Durham, NC USA}
\newcommand{\UND}{University of North Dakota, Grand Forks, ND, USA}
\newcommand{\Manchester}{University of Manchester, Manchester M13 9PL, United Kingdom}
\newcommand{\ND}{University of Notre Dame, Notre Dame, IN USA}
\newcommand{\UNSPMF}{University of Novi Sad, Faculty of Sciences, Novi Sad, Serbia}
\newcommand{\OakRidge}{Oak Ridge National Laboratory, Oak Ridge, TN, USA}
\newcommand{\OSU}{Ohio State University, Columbus, OH, USA}
\newcommand{\OregonState}{Oregon State University, Corvallis, OR, USA}
\newcommand{\Oxford}{University of Oxford, Oxford, UK}
\newcommand{\UTF}{Universidade Tecnológica Federal do Paraná, Curitiba, PR, Brazil}
\newcommand{\UniPR}{University of Parma, Department of Engineering and Architecture, Parma, Italy}
\newcommand{\Pennsylvania}{Pennsylvania State University, University Park, PA, USA}
\newcommand{\Penn}{University of Pennsylvania, Phildelphia, PA, USA}
\newcommand{\Pinstech}{Pakistan Institute of Nuclear Science and Technology. Islamabad, Pakistan}
\newcommand{\Pisa}{University and INFN Sezione di Pisa, Pisa, Italy}
\newcommand{\Princeton}{Princeton University, Princeton, NJ, USA}
\newcommand{\PNNL}{Pacific Northwest National Laboratory, Richland, WA 99352, USA}
\newcommand{\PAU}{Punjab Agricultural University, Ludhiana, India}
\newcommand{\psu}{Pennsylvania State University, University Park, PA 16802, USA}
\newcommand{\QMUL}{Queen Mary University of London, London E1 4NS, United Kingdom}
\newcommand{\Rochester}{University of Rochester, Rochester, NY, USA}
\newcommand{\Rutgers}{Rutgers University, Piscataway, NJ, USA}
\newcommand{\SJSU}{San Jos\'{e} State University, San Jos\'{e}, CA, USA}
\newcommand{\Sheffield}{University of Sheffield, Department of Physics and Astronomy, Sheffield S3 7RH, United Kingdom}
\newcommand{\SLAC}{SLAC National Accelerator Laboratory, Stanford University, Menlo Park, CA, USA}
\newcommand{\SDSMT}{South Dakota School of Mines and Technology, Rapid City, SD, USA}
\newcommand{\SMU}{Southern Methodist University, Dallas, TX, USA}
\newcommand{\SUSQ}{Department of Physics, Susquehanna University, Selinsgrove, PA, USA}
\newcommand{\SU}{Stockholm University, Stockholm, Sweden}
\newcommand{\SBU}{Stony Brook University, Stony Brook, NY, USA}
\newcommand{\UoS}{University of Sussex, Falmer, East Sussex, UK}
\newcommand{\SYR}{Syracuse University, Syracuse, NY, USA}
\newcommand{\TAU}{Tel-Aviv University, Tel-Aviv, Israel}
\newcommand{\TAMU}{Department of Physics and Astronomy, Texas A\& M University, College Station, TX, USA}

\newcommand{\UGTO}{Universidad de Guanajuato, Leon GTO, MX}
\newcommand{\unam}{Instituto de F{\'i}sica, Universidad Nacional Aut\'onoma de M\'exico, A.P. 20-364, Ciudad de M\'exico 01000, M\'exico}
\newcommand{\UTA}{University of Texas, Arlington, TX, USA}
\newcommand{\UTC}{University of Toronto, Toronto, Canada}
\newcommand{\TUNL}{Triangle Universities Nuclear Laboratory, Durham, NC, USA}
\newcommand{\UCL}{University College London, London WC1E 6BT, United Kingdom}
\newcommand{\UNICAMP}{Universidade Estadual de Campinas (UNICAMP), Campinas, SP, Brazil}
\newcommand{\UNIFAL}{Universidade Federal de Alfenas (Unifal-MG), Po\c{c}os de Caldas, MG, Brazil}
\newcommand{\UFSCAR}{Universidade Federal de São Carlos (UFSCar), Araras, SP, Brazil}
\newcommand{\UNIFESP}{Universidade Federal de S\~ao Paulo (UNIFESP), Diadema, SP, Brazil}
\newcommand{\UFABC}{Universidade Federal do ABC (UFABC), Santo André, SP, Brazil}
\newcommand{\UGA}{University Grenoble Alpes, CNRS, Grenoble INP, LPSC-IN2P3, 38000 Grenoble, France}
\newcommand{\Warwick}{University of Warwick, Coventry, UK}
\newcommand{\Wellesley}{Wellesley College, Physics Department, Wellesley, MA, USA}
\newcommand{\UW}{University of Wisconsin, Madison, WI USA}
\newcommand{\Wisconsin}{University of Wisconsin, Madison, WI USA}
\newcommand{\VT}{Center for Neutrino Physics, Virginia Tech, Blacksburg, Virginia, USA}
\newcommand{\Wright}{Wright Laboratory, Department of Physics, Yale University, New Haven, CT, USA}
\newcommand{\YorkU}{York University, Toronto, ON, Canada}

\newcommand{\ucb}{Physics Department, University of California at Berkeley, Berkeley, CA 94720-7300
}
\newcommand{\lbnl}{
Lawrence Berkeley National Laboratory, 1 Cyclotron Road, Berkeley, CA 94720-8153, USA
}
\newcommand{\jul}{Forschungszentrum J{\"u}lich, Institute for Nuclear Physics, Wilhelm-Johnen-Stra{\ss}e 52425 J{\"u}lich, Germany
}\newcommand{\penn}{Department of Physics and Astronomy, University of Pennsylvania, Philadelphia, PA 19104-6396
}
\newcommand{\fcul}{Universidade de Lisboa, Faculdade de Ci{\^e}ncias (FCUL), Departamento de F{\'i}sica, Campo Grande, Edifacio C8, 1749-016 Lisboa, Portugal
} 
\newcommand{\fctuc}{Universidade de Coimbra, Faculdade de Ci{\^e}ncias e Tecnologia (FCTUC), Departamento de F{\'i}sica, Rua Larga, 3004-516 Coimbra, Portugal 
}\newcommand{\lip}{Laborat{\'o}rio de Instrumenta{}{\c c}{\~a}o e F{\'i}sica Experimental de Part{\'i}culas (LIP), Av. Prof. Gama Pinto, 2, 1649-003, Lisboa, Portugal
}\newcommand{\chic}{
The Enrico Fermi Institute and Department of Physics, The University of Chicago, Chicago, IL 60637, USA
}\newcommand{\bnl}{
Brookhaven National Laboratory, Upton, New York 11973, USA
}\newcommand{\uh}{
University of Hawaiâi at Manoa, Honolulu, Hawaiâi 96822, USA
}\newcommand{\iowa}{
Department of Physics and Astronomy, Iowa State University, Ames, IA 50011, USA
}\newcommand{\jyv}{
Department of Physics, University of Jyv{\"a}skyl{\"a}, Finland
}\newcommand{\ucd}{
University of California, Davis, 1 Shields Avenue, Davis, CA 95616, USA
}\newcommand{\bu}{
Boston University, Department of Physics, Boston, MA 02215, USA
}\newcommand{\mainz}{
Institute of Physics and Excellence Cluster PRISMA, Johannes Gutenberg-Universit{\"a}t Mainz, 55099 Mainz, Germany
}\newcommand{\NIST}{
National Institute of Standards and Technology, Gaithersburg, MD 20899, USA}
\newcommand{\ham}{
Institut f{\"u}r Experimentalphysik, Universit{\"a}t Hamburg, 22761 Hamburg, Germany
}\newcommand{\alb}{
University of Alberta, Department of Physics, 4-181 CCIS, Edmonton, AB T6G 2E1, Canada
}\newcommand{\pnnl}{
Pacific Northwest National Laboratory, Richland, WA 99352, USA
}\newcommand{\laur}{
Laurentian University, Department of Physics, 935 Ramsey Lake Road, Sudbury, ON P3E 2C6, Canada
}\newcommand{\lsu}{
Department of Physics and Astronomy, Louisiana State University, Baton Rouge, LA 70803
}\newcommand{\tub}{
Kepler Center for Astro and Particle Physics, Universit{\"a}t T{\"u}bingen, 72076 T{\"u}bingen, Germany
}\newcommand{\sheff}{
University of Sheffield, Physics \& Astronomy, Western Bank, Sheffield S10 2TN, UK
}\newcommand{\qu}{
Queen's University, Department of Physics, Engineering Physics \& Astronomy, Kingston, ON K7L 3N6, Canada
}\newcommand{\snolab}{
SNOLAB, Creighton Mine 9, 1039 Regional Road 24, Sudbury, ON P3Y 1N2, Canada
}\newcommand{\rut}{
Department of Physics and Astronomy, Rutgers, The State University of New Jersey, 136 Frelinghuysen Road, Piscataway, NJ 08854-8019 USA
}\newcommand{\temp}{
Department of Physics, Temple University, Philadelphia, PA, USA
}\newcommand{\ucla}{
University of California Los Angeles, Department of Physics \& Astronomy, 475 Portola Plaza, Los Angeles, CA 90095-1547, USA
}
\newcommand{\tri}{
SISSA/INFN, Via Bonomea 265, I-34136 Trieste, Italy
}\newcommand{\kav}{
Kavli IPMU (WPI), University of Tokyo, 5-1-5 Kashiwanoha, 277-8583 Kashiwa, Japan
}\newcommand{\kor}{
Center for Underground Physics, Institute for Basic Science, Daejeon 34126, Korea
}\newcommand{\uci}{
University of California, Irvine, Department of Physics, CA 92697, Irvine, USA
}\newcommand{\aach}{
RWTH Aachen University, Physikzentrum, Otto-Blumenthal-Stra{\ss}e, 52074 Aachen, Germany
}\newcommand{\sbu}{
State University of New York at Stony Brook, Department of Physics and Astronomy, Stony Brook, New York, USA
}\newcommand{\tsing}{
Department of Engineering Physics, Tsinghua University, Beijing 100084, China
}\newcommand{\corn}{
Cornell University, Ithaca, NY, USA
}\newcommand{\boul}{
University of Colorado at Boulder, Department of Physics, Boulder, Colorado, USA
}\newcommand{\dres}{
Institut f{\"u}r Kern und Teilchenphysik, TU Dresden, Zellescher Weg 19, 01069, Dresden, Germany
}
\newcommand{\mun}{Physics Department, Technische Universit{\"a}t M{\"u}nchen, 85748 Garching, Germany
}
\newcommand{\mitnew}{
Massachusetts Institute of Technology, Department of Physics and Laboratory for Nuclear Science, 77 Massachusetts Ave Cambridge, MA 02139, USA
}
\newcommand{\kings}{King's College London, Department of Physics, Strand Building, Strand, London WC2R 2LS, UK}
\newcommand{\llnl}{
Lawrence Livermore National Laboratory, Livermore, CA 94550, USA
}
\newcommand{\fnal}{
Fermi National Accelerator Laboratory, Batavia, IL 60510, USA
}
\newcommand{\erc}{Department of Physics, Erciyes University, 38030, Kayseri, Turkey
}
\newcommand{\iow}{Department of Physics and Astronomy, The University of Iowa, Iowa City, Iowa, USA}
\newcommand{\rcns}{Research Center for Neutrino Science, Tohoku University, Sendai, Miyagi, Japan}
\newcommand{\nikhef}{Nikhef and the University of Amsterdam, Science Park, Amsterdam, Netherlands}

\newcommand{\suba}{SUBATECH, Nantes Université, IMT-Atlantique, CNRS/IN2P3, Nantes, France}

\newcommand{\usdm}{Department of Physics, South Dakota School of Mines and Technology, Rapid City, SD 57701, USA}
%
%
%

\affiliation{\UA}
\affiliation{\ANL}
\affiliation{\APC}
\affiliation{\Atlantico}
\affiliation{\UBERN}
\affiliation{\BEU}
\affiliation{\BU}
\affiliation{\BNL}
\affiliation{\BHU}
\affiliation{\UB}
\affiliation{\UoB}
\affiliation{\UoBham}
\affiliation{\Caltech}
\affiliation{\UCD}
\affiliation{\UCB}
\affiliation{\UNC}
\affiliation{\UCI}
\affiliation{\JGU}
\affiliation{\UCSB}
\affiliation{\Cambridge}
\affiliation{\Catania}
\affiliation{\CERN}
\affiliation{\ChU}
\affiliation{\CIEMAT}
\affiliation{\Cincinnati}
\affiliation{\CUSB}
\affiliation{\Cinvestav}
\affiliation{\Columbia}
\affiliation{\CSU}
\affiliation{\Drexel}
\affiliation{\Duke}
\affiliation{\EDI}
\affiliation{\EIA}
\affiliation{\FCUL}
\affiliation{\FNAL}
\affiliation{\UF}
\affiliation{\UGR}
\affiliation{\Harvard}
\affiliation{\Hawaii}
\affiliation{\Houston}
\affiliation{\ISU}
\affiliation{\IU}
\affiliation{\IFIC}
\affiliation{\IIT}
\affiliation{\IITH}
\affiliation{\CzechAcademyofSciences}
\affiliation{\INFNMiB}
\affiliation{\INFNNA}
\affiliation{\INFNGE}
\affiliation{\INFNPd}
\affiliation{\Insubria}
\affiliation{\UOH}
\affiliation{\Iowa}
\affiliation{\ISP}
\affiliation{\IowaU}
\affiliation{\Irfu}
\affiliation{\Iwate}
\affiliation{\jnu}
\affiliation{\JBNU}
\affiliation{\JSU}
\affiliation{\JU}
\affiliation{\KSU}
\affiliation{\Kyiv}
\affiliation{\KCL}
\affiliation{\NRCKI}
\affiliation{\Lancaster}
\affiliation{\LBNL}
\affiliation{\LLNL}
\affiliation{\Liverpool}
\affiliation{\LANL}
\affiliation{\LIP}
\affiliation{\LSU}
\affiliation{\Lucknow}
\affiliation{\Manchester}
\affiliation{\MIT}
\affiliation{\MPA}
\affiliation{\nikhef}
\affiliation{\UMN}
\affiliation{\Duluth}
\affiliation{\UMiss}
\affiliation{\NIST}
\affiliation{\UCNCl}
\affiliation{\NCCU}
\affiliation{\UND}
\affiliation{\ND}
\affiliation{\UNSPMF}
\affiliation{\OakRidge}
\affiliation{\OSU}
\affiliation{\OregonState}
\affiliation{\Oxford}
\affiliation{\PNNL}
\affiliation{\UniPR}
\affiliation{\Pennsylvania}
\affiliation{\Penn}
\affiliation{\Pinstech}
\affiliation{\UTF}
\affiliation{\PAU}
\affiliation{\Pisa}
\affiliation{\Princeton}
\affiliation{\QMUL}
\affiliation{\Rochester}
\affiliation{\Rutgers}
\affiliation{\SJSU}
\affiliation{\Sheffield}
\affiliation{\SLAC}
\affiliation{\SDSMT}
\affiliation{\SMU}
\affiliation{\SU}
\affiliation{\SUSQ}
\affiliation{\SBU}
\affiliation{\UoS}
\affiliation{\SYR}
\affiliation{\TAU}
\affiliation{\TAMU}
\affiliation{\UGTO}
\affiliation{\UTA}
\affiliation{\UTC}
\affiliation{\TUNL}
\affiliation{\UNICAMP}
\affiliation{\UNIFAL}
\affiliation{\UFSCAR}
\affiliation{\UNIFESP}
\affiliation{\UFABC}
\affiliation{\UGA}
\affiliation{\UNIFAL}
\affiliation{\Warwick}
\affiliation{\Wellesley}
\affiliation{\UW}
\affiliation{\VT}
\affiliation{\Wright}
\affiliation{\YorkU}
\affiliation{\suba}
\affiliation{\unam}




\author{Joshua~R.~Klein}
\altaffiliation{\Editor}
\affiliation{\Penn}



\author{Tomi Akindele}
\altaffiliation{\Contributor}
\affiliation{\LLNL}

\author{Adam Bernstein}
\altaffiliation{\Contributor}
\affiliation{\LLNL}

\author{Steven Biller}
\altaffiliation{\Contributor}
\affiliation{\Oxford}

\author{Nathaniel Bowden}
\altaffiliation{\Contributor}
\affiliation{\LLNL}

\author{Jason Brodsky}
\altaffiliation{\Contributor}
\affiliation{\LLNL}

\author{D.F.~Cowen}
\altaffiliation{\Contributor}
\affiliation{\Pennsylvania}

\author{Michael Ford}
\altaffiliation{\Contributor}
\affiliation{\LLNL}

\author{Julieta Gruszko}
\altaffiliation{\Contributor}
\affiliation{\UNC}
\affiliation{\TUNL}

\author{Logan Lebanowski}
\altaffiliation{\Contributor}
\affiliation{\Penn}

\author{Aobo Li}
\altaffiliation{\Contributor}
\affiliation{\UNC}
\affiliation{\TUNL}

\author{Viacheslav A. Li}
\altaffiliation{\Contributor}
\affiliation{\LLNL}

\author{Wei Mu}
\altaffiliation{\Contributor}
\affiliation{\FNAL}

\author{J. Pedro Ochoa-Ricoux}
\altaffiliation{\Contributor}
\affiliation{\UCI}

\author{Gabriel~D.~Orebi~Gann}
\altaffiliation{\Contributor}
\affiliation{\UCB}
\affiliation{\LBNL}

\author{Mayly Sanchez}
\altaffiliation{\Contributor}
\affiliation{\Iowa}

\author{Robert Svoboda}
\altaffiliation{\Contributor}
\affiliation{\UCD}

\author{Matthew Wetstein}
\altaffiliation{\Contributor}
\affiliation{\Iowa}

\author{Michael Wurm}
\altaffiliation{\Contributor}
\affiliation{\JGU}

\author{Minfang Yeh}
\altaffiliation{\Contributor}
\affiliation{\BNL}

\collaboration{Editors and Contributors}
\noaffiliation



\author{M.~Askins}\affiliation{\ucb}\affiliation{\lbnl}
\author{Z.~Bagdasarian}\affiliation{\ucb}\affiliation{\lbnl}
\author{N.~Barros}\affiliation{\penn}
\affiliation{\fctuc}
\affiliation{ \lip}
\author{E.W.~Beier}\affiliation{\penn}
\author{A.~Bernstein}\affiliation{\llnl}
\author{M.~B\"ohles}\affiliation{\mainz}
\author{E.~Blucher}\affiliation{\chic}
\author{R.~Bonventre}\affiliation{\lbnl}
\author{E.~Bourret}\affiliation{\lbnl}
\author{E.~J.~Callaghan}\affiliation{\ucb}\affiliation{\lbnl}
\author{J.~Caravaca}\affiliation{\ucb}\affiliation{\lbnl}
\author{T.~Classen}\affiliation{\llnl}
\author{M.~Diwan}\affiliation{\bnl}
\author{S.T.~Dye}\affiliation{\uh}
\author{J.~Eisch}\affiliation{\fnal}
\author{A.~Elagin}\affiliation{\chic}
\author{T.~Enqvist}\affiliation{\jyv}
\author{U.~Fahrendholz}\affiliation{\mun}
\author{V.~Fischer}\affiliation{\ucd}
\author{K.~Frankiewicz}\affiliation{\bu}
\author{C.~Grant}\affiliation{\bu}
\author{D.~Guffanti}\affiliation{\mainz}
\author{C.~Hagner}\affiliation{\ham}
\author{A.~Hallin}\affiliation{\alb}
\author{C.~M.~Jackson}\affiliation{\pnnl}
\author{R.~Jiang}\affiliation{\chic}
\author{T.~Kaptanoglu}\affiliation{\ucb}\affiliation{\lbnl}
\author{J.R.~Klein}\affiliation{\penn}
\author{Yu.~G.~Kolomensky}\affiliation{\ucb}\affiliation{\lbnl}
\author{C.~Kraus}\affiliation{\snolab}\affiliation{\laur}
\author{F.~Krennrich}\affiliation{\iowa}
\author{T.~Kutter}\affiliation{\lsu}
\author{T.~Lachenmaier}\affiliation{\tub}
\author{B.~Land}\affiliation{\ucb}\affiliation{\lbnl}\affiliation{\penn}
\author{K.~Lande}\affiliation{\penn}
\author{L.~Lebanowski}\affiliation{\penn}
\author{J.G.~Learned}\affiliation{\uh}
\author{V.A.~Li}\affiliation{\llnl}
\author{V.~Lozza}\affiliation{\fcul}
\affiliation{ \lip}
\author{L.~Ludhova}\affiliation{\jul}\affiliation{\aach}
\author{M.~Malek}\affiliation{\sheff}
\author{S.~Manecki}\affiliation{\laur}\affiliation{\qu}\affiliation{\snolab}
\author{J.~Maneira}\affiliation{\fcul}
\affiliation{ \lip}
\author{J.~Maricic}\affiliation{\uh}
\author{J.~Martyn}\affiliation{\mainz}
\author{A.~Mastbaum}\affiliation{\rut}
\author{C.~Mauger}\affiliation{\penn}
\author{M.~Mayer}\affiliation{\mun}
\author{J.~Migenda}\affiliation{\kings}
\author{F.~Moretti}\affiliation{\lbnl}
\author{J.~Napolitano}\affiliation{\temp}
\author{B.~Naranjo}\affiliation{\ucla}
\author{F.M.~Newcomer}\affiliation{\penn}
\author{M.~Nieslony}\affiliation{\mainz}
\author{L.~Oberauer}\affiliation{\mun}
\author{G.~D.~Orebi~Gann}\affiliation{\ucb}\affiliation{\lbnl}
\author{J.~Ouellet}\affiliation{\mitnew}
\author{T.~Pershing}\affiliation{\ucd}
\author{S.T.~Petcov}\affiliation{\tri,\kav}
\author{L.~Pickard}\affiliation{\ucd}
\author{R.~Rosero}\affiliation{\bnl}
\author{M.~C.~Sanchez}\affiliation{\iowa}
\author{J.~Sawatzki}\affiliation{\mun}
\author{S.~Schoppmann}\affiliation{\ucb}\affiliation{\lbnl}
\author{S.H.~Seo}\affiliation{\kor}
\author{M.~Smiley}\affiliation{\ucb}\affiliation{\lbnl}
\author{M.~Smy}\affiliation{\uci}
\author{A.~Stahl}\affiliation{\aach}
\author{H.~Steiger}\affiliation{\mainz}\affiliation{\mun}
\author{M.~R.~Stock}\affiliation{\mun}
\author{H.~Sunej}\affiliation{\bnl}
\author{R.~Svoboda}\affiliation{\ucd}
\author{E.~Tiras}\affiliation{\erc}\affiliation{\iow}
\author{W.~H.~Trzaska}\affiliation{\jyv}
\author{M.~Tzanov}\affiliation{\lsu}
\author{M.~Vagins}\affiliation{\uci}
\author{C.~Vilela}\affiliation{\sbu}
\author{Z.~Wang}\affiliation{\tsing}
\author{J.~Wang}\affiliation{\usdm}
\author{M.~Ward}\affiliation{\qu}
\author{M.~Wetstein}\affiliation{\iowa}
\author{M.J.~Wilking}\affiliation{\sbu}
\author{L.~Winslow}\affiliation{\mitnew}
\author{P.~Wittich}\affiliation{\corn}
\author{B.~Wonsak}\affiliation{\ham}
\author{E.~Worcester}\affiliation{\bnl}\affiliation{\sbu}
\author{M.~Wurm}\affiliation{\mainz}
\author{G.~Yang}\affiliation{\sbu}
\author{M.~Yeh}\affiliation{\bnl}
\author{E.D.~Zimmerman}\affiliation{\boul}
\author{S.~Zsoldos}\affiliation{\ucb}\affiliation{\lbnl}
\author{J.R.~Wilson}\affiliation{\KCL}
\author{K.~Zuber}\affiliation{\dres}

\noaffiliation
\collaboration{Co-Signers from Theia}

\author{T. Akindele	}\affiliation{\llnl}
\author{T. Anderson	}\affiliation{\psu}
\author{M.~Askins}\affiliation{\lbnl}
\author{A. Baldoni	}\affiliation{\psu}
\author{A. Barna	}\affiliation{\uh}
\author{T. Benson	}\affiliation{\Wisconsin}
\author{M. Bergevin	}\affiliation{\llnl}
\author{A. Bernstein	}\affiliation{\llnl}
\author{B. Birrittella	}\affiliation{\Wisconsin}
\author{J. Boissevain	}\affiliation{\penn}
\author{J. Borusinki	}\affiliation{\uh}
\author{D. Cowen	}\affiliation{\psu}
\author{B. Crow	}\affiliation{\uh}
\author{F. Dalnoki-Veress	}\affiliation{\miis}
\author{D. Danielson	}{\affiliation{\chic}
\author{S. Dazeley	}\affiliation{\llnl}
\author{M. Diwan	}\affiliation{\bnl}
\author{A. Druetzler	}\affiliation{\uh}
\author{S. Dye	}\affiliation{\uh}
\author{A. Fienberg	}\affiliation{\psu}
\author{V. Fischer	}\affiliation{\ucd}
\author{K. (Kat) Frankiewicz}\affiliation{\bu}
\author{D. Gooding	}\affiliation{\bu}
\author{C. Graham	}\affiliation{\umich}
\author{C. Grant	}\affiliation{\bu}
\author{J. Griskevich	}\affiliation{\uci}
\author{P. Haugen	}\affiliation{\llnl}
\author{J. He	}\affiliation{\ucd}
\author{J. Hecla	}\affiliation{\ucb}\affiliation{\llnl}
\author{I. Jovanovic	}\affiliation{\umich}
\author{M. Keenan	}\affiliation{\psu}
\author{P. Keener	}\affiliation{\penn}
\author{P. Kunkle	}\affiliation{\bu}
\author{J. Learned	}\affiliation{\uh}
\author{V. Li	}\affiliation{\llnl}
\author{J. Maricic	}\affiliation{\uh}
\author{P. Marr-Laundrie	}\affiliation{\Wisconsin}
\author{J. Moore	}\affiliation{\psu}
\author{A. Mullen	}\affiliation{\ucb}\affiliation{\llnl}
\author{E. Neights	}\affiliation{\psu}
\author{K. Nishimura	}\affiliation{\uh}
\author{B. O'Meara	}\affiliation{\psu}
\author{K. Ogren	}\affiliation{\umich}
\author{S. Olivier	}\affiliation{\llnl}
\author{G. Orebi Gann	}\affiliation{\lbnl}\affiliation{\ucb}
\author{L. Oxborough	}\affiliation{\Wisconsin}
\author{A. Papatyi	}\affiliation{\pnnl}
\author{B. Paulos	}\affiliation{\Wisconsin}
\author{T. Pershing	}\affiliation{\llnl}
\author{L. Pickard	}\affiliation{\ucd}
\author{L. Sabarots	}\affiliation{\Wisconsin}
\author{V. Shebalin	}\affiliation{\uh}
\author{M. Smy	}\affiliation{\uci}
\author{H. Song	}\affiliation{\bu}
\author{F. Sutanto	}\affiliation{\umich}
\author{R. Svoboda	}\affiliation{\ucd}
\author{M. Vagins	}\affiliation{\uci}
\author{R. (Rick) Van Berg}\affiliation{\penn}
\author{G. Varner	}\affiliation{\uh}
\author{V. Veeraraghavan}\affiliation{\iowa}
\author{S. Ventura	}\affiliation{\uh}
\author{B. Walsh	}\affiliation{\bnl}
\author{G. Wendel	}\affiliation{\psu}
\author{D. Westphal	}\affiliation{\llnl}
\author{M. Wetstein	}\affiliation{\iowa}
\author{A. Wilhelm	}\affiliation{\umich}
\author{S. Wolcott	}\affiliation{\Wisconsin}
\author{M. Yeh	}\affiliation{\bnl}

\collaboration{Co-Signers from WATCHMAN} 

\author{T.~Anderson}\affiliation{\psu}
\author{E.~Anderssen}\affiliation{\lbnl}
\author{M.~Askins}\affiliation{\lbnl}\affiliation{\ucb}
\author{A.~J.~Bacon}\affiliation{\penn}
\author{Z.~Bagdasarian}\affiliation{\lbnl}\affiliation{\ucb}
\author{A.~Baldoni}\affiliation{\psu}
\author{N.~Barros}\affiliation{\fcul}\affiliation{ \lip}
\author{L.~Bartoszek}\affiliation{\bart}
\author{A.~Bat}\affiliation{\erc}
\author{M.~Bergevin}\affiliation{\llnl}
\author{A.~Bernstein}\affiliation{\llnl}
\author{E.~Blucher}\affiliation{\chic}
\author{J.~Boissevain}\affiliation{\bart}
\author{R.~Bonventre}\affiliation{\lbnl}
\author{D.~Brown}\affiliation{\lbnl}
\author{E.~J.~Callaghan}\affiliation{\lbnl}\affiliation{\ucb}
\author{D.~F.~Cowen}\affiliation{\psu}
\author{S.~Dazeley}\affiliation{\llnl}
\author{M.~Diwan}\affiliation{\bnl}
\author{K.~Frankiewicz}\affiliation{\bu}
\author{C.~Grant}\affiliation{\bu}
\author{T.~Kaptanoglu}\affiliation{\lbnl}\affiliation{\ucb}
\author{J.R.~Klein}\affiliation{\penn}
\author{C.~Kraus}\affiliation{\snolab}\affiliation{\laur}
\author{T.~Kroupova}\affiliation{\penn}
\author{B.~Land}\affiliation{\lbnl}\affiliation{\ucb}\affiliation{\penn}
\author{L.~Lebanowski}\affiliation{\penn}
\author{V.~Lozza}\affiliation{\fcul}\affiliation{ \lip}
\author{A.~Marino}\affiliation{\boul}
\author{A.~Mastbaum}\affiliation{\rut}
\author{C.~Mauger}\affiliation{\penn}
\author{M.~Newcomer}\affiliation{\penn}
\author{A.~Nikolica}\affiliation{\penn}
\author{G.~D.~Orebi~Gann}\affiliation{\lbnl}\affiliation{\ucb}
\author{L.~Pickard}\affiliation{\ucd}
\author{J.~Saba}\affiliation{\lbnl}
\author{S.~Schoppmann}\affiliation{\lbnl}\affiliation{\ucb}
\author{J.~Sensenig}\affiliation{\penn}
\author{M.~Smiley}\affiliation{\lbnl}\affiliation{\ucb}
\author{H.~Steiger}\affiliation{\mainz}\affiliation{\mun}
\author{R.~Svoboda}\affiliation{\ucd}
\author{E.~Tiras}\affiliation{\erc}\affiliation{\iow}
\author{W.~H.~Trzaska}\affiliation{\jyv}
\author{R.~van~Berg}\affiliation{\penn}\affiliation{\bart}
\author{G.~Wendel}\affiliation{\psu}
\author{M.~Wetstein}\affiliation{\iowa}
\author{M.~Wurm}\affiliation{\mainz}
\author{G.~Yang}\affiliation{\lbnl}\affiliation{\ucb}
\author{M.~Yeh}\affiliation{\bnl}
\author{E.D.~Zimmerman}\affiliation{\boul}

\collaboration{Co-Signers from Eos}



\author{M.~Chen}\affiliation{\qu}
\author{A.~Bialek}\affiliation{\snolab}\affiliation{\laur}
\author{E.~Caden}\affiliation{\snolab}\affiliation{\laur}
\author{E.~J.~Callaghan}\affiliation{\ucb}\affiliation{\lbnl}
\author{D.~Cookman}\affiliation{\Oxford}
\author{E.~Falk}\affiliation{\UoS}
\author{D.M.~Gooding}\affiliation{\bu}
\author{A.~Hallin}\affiliation{\alb}
\author{R.~Hunt-Stokes}\affiliation{\Oxford}
\author{A.S.~In\'{a}cio}\affiliation{\fcul}\affiliation{ \lip}
\author{T.~Kroupova}\affiliation{\P\penn}
\author{Y.H.~Lin}\affiliation{\snolab}\affiliation{\qu}
\author{M.~Luo}\affiliation{\penn}
\author{S.~Manecki}\affiliation{\snolab}\affiliation{\laur}\affiliation{\qu}
\author{A.B.~McDonald}\affiliation{\qu}
\author{S.~Naugle}\affiliation{\penn}
\author{L.J.~Nolan}\affiliation{\QMUL}
\author{J.~Paton}\affiliation{\Oxford}
\author{M.~Rigan}\affiliation{\UoS}
\author{J.~Rumleskie}\affiliation{\laur}
\author{B.~Tam}\affiliation{\qu}
\author{S.~Valder}\affiliation{\UoS}
\author{E.~V\'azquez-J\'auregui}\affiliation{\unam}
\author{A.~Wright}\affiliation{\qu}
\author{S.~Yang}\affiliation{\alb}
\author{Y.~Zhang}\affiliation{\alb}
\author{A.~Zummo}\affiliation{\penn}
\author{C.~Mills}\affiliation{\UoS}
\author{S.~Andringa}\affiliation{\lip}

\collaboration{Co-Signers from SNO+}


\author{N. S. Bowden}\affiliation{\LLNL}
\author{B. R. Littlejohn}\affiliation{\IIT}
\author{H. P. Mumm}\affiliation{\NIST}
\author{R.~Neilson}\affiliation{\Drexel}
\author{C. Roca}\affiliation{\LLNL}
\author{F.~Sutanto}\affiliation{\LLNL}
\author{A. B. Hansell}\affiliation{\SUSQ}
\author{P.~Weatherly}\affiliation{\Drexel}
\author{K. M. Heeger}\affiliation{\Wright}
\author{P.~T.~Surukuchi}\affiliation{\Wright}
\author{P. Kunkle}\affiliation{\bu}
\collaboration{Co-Signers from PROSPECT}


\collaboration{Co-Signers from Borexino}


\author{C.~Grant}\affiliation{\BU}
\author{T.~O'Donnell}\affiliation{\VT}
\author{H.~Karwowski}\affiliation{\UNC}
\author{K.~Inoue}\affiliation{\rcns}
\author{Y.~Kamei}\affiliation{\rcns}
\author{Y.~Kishimoto}\affiliation{\rcns}
\author{A.~Gando}\affiliation{\rcns}
\author{Y.~Gando}\affiliation{\rcns}
\author{D.M.~Markoff}\affiliation{\NCCU,\TUNL}
\collaboration{Co-Signers from KamLAND-Zen}


\author{T.~Heinz}\affiliation{\tub}
\author{A.~Stahl}\affiliation{\aach}
\author{H.~Steiger}\affiliation{\mainz}\affiliation{\mun}
\author{T.~Sterr}\affiliation{\tub}
\author{M.C.~Vollbrecht}\affiliation{\jul}\affiliation{\aach}
\author{V.~Vorobel}\affiliation{\ChU}

\collaboration{Co-Signers from JUNO}


\author{T.J.C.~Bezerra}\affiliation{\UoS}
\author{S.~Dusini}\affiliation{\INFNPd}
\author{W.C.~Griffith}\affiliation{\UoS}
\author{J.~Hartnell}\affiliation{\UoS}
\author{J.C.C.~Porter}\affiliation{\UoS}

\author{M. Bongrand}\affiliation{\suba}
\author{B. Viaud}\affiliation{\suba}
\author{F. Yermia}\affiliation{\suba}

\author{B.~Roskovec}\affiliation{\ChU}

\author{A.~Weber}\affiliation{\JGU}
\author{L.~Koch}\affiliation{\JGU}

\author{E. Calvo}\affiliation{\CIEMAT}
\author{C.~Palomares}\affiliation{\CIEMAT}
\author{A.~Verdugo}\affiliation{\CIEMAT}
\author{M.~Chen}\affiliation{\qu}

\collaboration{Co-Signers from LiquidO}

\author{M.~Ascensio-Sosa}\affiliation{\ISU}
\author{A.~Bat}\affiliation{\erc}
\author{J.~Beacom}\affiliation{\OSU}
\author{M.~Bergevin}\affiliation{\LLNL}
\author{P.~Boerstoel}\affiliation{\Rutgers}
\author{M.~Breisch}\affiliation{\tub}
\author{S.~Dazeley}\affiliation{\LLNL}
\author{E.~Drakopoulou}\affiliation{\Demokritos}
\author{S.~Edayath}\affiliation{\ISU}
\author{R.~Edwards}\affiliation{\Warwick}
\author{J.~Eisch}\affiliation{\FNAL}
\author{S.~Gardiner}\affiliation{\FNAL}
\author{V.~Fischer}\affiliation{\ISU}
\author{H.~Frisch}\affiliation{\chic}
\author{P.~Hackspacher}\affiliation{\UCD}
\author{C.~Hagner}\affiliation{\ham}
\author{J.~He}\affiliation{\UCD}
\author{S.~Husnugil}\affiliation{\erc}
\author{B.~Kaiser}\affiliation{\tub}
\author{M.~Kamislioglu}\affiliation{\erc}
\author{F.~Krennich}\affiliation{\ISU}
\author{T.~Lachenmaier}\affiliation{\tub}
\author{F.~Lemmons}\affiliation{\SDSMT}
\author{K.~Lin}\affiliation{\Rutgers}
\author{D.~Maksimovic}\affiliation{\mainz}
\author{M.~Malek}\affiliation{\Sheffield}
\author{A.~Mastbaum}\affiliation{\Rutgers}
\author{M.~Needham}\affiliation{\EDI}
\author{C.~McGivern}\affiliation{\FNAL}
\author{J.~Minock}\affiliation{\Rutgers}
\author{M.~Nieslony}\affiliation{\mainz}
\author{M.~O'Flaherty}\affiliation{\Warwick}
\author{L.~Pickard}\affiliation{\UCD}
\author{N.~Poonthottathil}\affiliation{\ISU}
\author{E.~Pottenbaum}\affiliation{\ISU}
\author{B.~Richards}\affiliation{\Warwick}
\author{M.~Smy}\affiliation{\UCI}
\author{M.~Stender}\affiliation{\ham}
\author{E.~Tiras}\affiliation{\erc}
\author{V.~Veeraraghavan}\affiliation{\ISU}
\author{G.~Vera}\affiliation{\UCD}
\author{J.~Wang}\affiliation{\SDSMT}
\author{A.~Weinstein}\affiliation{\ISU}
\author{B.~Wonsak}\affiliation{\ham}
\author{T.~Zhang}\affiliation{\UCD}
\collaboration{Co-Signers from ANNIE}





\vspace{3cm}

\begin{abstract}

\end{abstract}

\maketitle

\newpage
\textsf{\tableofcontents}

\clearpage

\section{Executive Summary}

 Between the last Snowmass and now, there has been an explosion of new ideas and new, enabling technologies that will significantly expand the capabilities of next-generation photon-based neutrino detectors.
Detectors that use photons as the primary carrier of interaction information have a long, rich history in neutrino physics, going as far back as the discovery of the neutrino itself.  Large-scale, monolithic detectors that use either Cherenkov or scintillation light have played major roles in nearly every discovery of neutrino oscillation phenomena~\cite{superk,sno,dayabay,kamland,t2k} or observation of astrophysical neutrinos~\cite{kii,sno,imb,borexino,borexinopep,icecube}.  New detectors at even larger scales are being built right now, including JUNO~\cite{juno}, Hyper-Kamiokande~\cite{hyperk}, and DUNE~\cite{dunearapucas}.
    Photon-based detectors have been so successful because they are inexpensive, remarkably versatile, and have dynamic ranges that reach all the way from tens of keV~\cite{snoplus} to PeV~\cite{icecube}.  
    
        The new technologies described in this white paper will lead to neutrino physics and astrophysics programs of great breadth: from high-precision accelerator neutrino oscillation measurements, to detection of reactor and solar neutrinos, and even to neutrinoless double beta decay measurements that will probe the normal hiearachy regime.   They will also be valuable for neutrino applications, such as non-proliferation via reactor monitoring.
        
        Of particular community interest is the development of {\it hybrid} Cherenkov/scintillation detectors, which can simultaneously exploit the advantages of Cherenkov light---reconstruction of direction and related high-energy PID---and the advantages of scintillation light---high light-yield, low-threshold detection with low-energy PID. Hybrid Cherenkov/scintillation detectors could have an exceptionally broad dynamic range in a single experiment, allowing them to have both high-energy, accelerator-based sensitivity while also achieving a broad low-energy neutrino physics and astrophysics program.  Recently the Borexino Collaboration~\cite{borexinocherscint} has published results showing that even in a detector with standard scintillator and no special photon sensing or collecting, Cherenkov and scintillation light can be discriminated well enough on a statistical basis that a sub-MeV solar neutrino direction peak can be seen.  Thus the era of hybrid detectors has begun, and many of the enabling technologies described here will make full event-by-event direction reconstruction in such detectors possible.
        
        We summarize the new technologies discussed in this white paper below:
        \begin{itemize}
        \item {\bf Water-based liquid scintillator (WbLS)}: WbLS allows tunable high-light yield scintillation with long absorption lengths seen in water Cherenkov detectors.  The reduced scintillation light compared to standard liquid scintillator allows Cherenkov light to be seen more easily, thus creating a hybrid Cherenkov/scintillation detector. WbLS also allows a wide variety of isotopes to be loaded (see below).
        \item {\bf Slow Fluors}: Slowing the scintillation light down can allow standard photon sensor timing to be used to discriminate Cherenkov and scintillation light, thus allowing hybrid Cherenkov/scintillation detectors with no expensive sensors or collectors and with Cherenkov sensitivity down into the short wavelength regime, preserving good Cherenkov light yield.  
        \item {\bf Isotopic Loading}: New approaches have allowed a wide variety of isotopes to be ``loaded'' into either liquid or plastic scintillator, including Gd or $^6$Li for neutron capture and PSD; Te or Xe for neutrinoless double beta decay; indium or $^7$Li for solar neutrino detection, and others. These approaches significantly extend the physics programs of photon-based detectors at very low cost.
        \item {\bf New Photon Sensors}: New advances in the efficiency of photomultiplier tubes, including long-wavelength sensitivity, and significant improvements in timing even with devices as large as 8'', make hybrid Cherenkov/scintillation detectors even better, with high light yields for both Cherenkov and scintillation light with good separation between the two types of light (``chertons'' and ``scintons''). Large Area Picosecond Photon Detectors ({\bf LAPPDs}) have pushed photon timing into the picosecond regime, allowing Cherenkov/scintillation separation to be done even with standard scintillation time profiles. The fast timing of LAPPDs also makes reconstruction of event details fine enough to track particles with the produced photons.
        
        \item {\bf New Photon Collectors}:  {\bf Dichroicons}, which are Winston-style light concentrators made from dichroic mirrors, allow photons to be sorted by wavelength thus directing the long-wavelength end of broad-band Cherenkov light to photon sensors that have good sensitivity to those wavelengths, while directing narrow-band short-wavelength scintillation light to other sensors.  Dichroicons are particularly useful in high-coverage hybrid Cherenkov/scintillation detectors.  
        {\bf ARAPUCA}s use wavelength shifters and dichroic mirrors to trap photons and thus improve the overall collection efficiency, even for small-area silicon photomultipliers. 
        Distributed imaging uses lenses to exploit the directionality of photons, even in a non-imaging photon-based detector. 
        \item {\bf New Simulation Approaches}: Simulating each photon in a model of a scintillation detector can be slow, particularly when geometric details are needed in a precision model.  New GPU-accelerated photon ray-tracers like {\bf  Chroma} can speed these simulations up by factors of at least 100, and machine learning techniques can also be used to quickly predict the photon counts from particular emission points in a large detector.
        \item {\bf New Analysis Approaches}: Machine learning techniques have only begun to be used in photon-based detectors, but they are already showing great value in rejecting complex backgrounds that have been difficult to deal with in past experiments, such as the use of {\bf KamNET} in the KamLAND-Zen experiment.
        \item {\bf New DAQ and Readout Techniques}: Photon detectors typically see a small number of photons (single photoelectrons to perhaps hundreds) on each sensor.  Each pulse from the photon detection is a known pulse shape, with little variation. Thus, feature extraction either by firmware in inexpensive FPGAs using low-cost sampling ADCs, or even with analog techniques, can lead to very low-cost per channel for very large detectors. 
        \end{itemize}
        
        These technologies are already becoming part of conceptual designs for new-large scale detectors:
        \begin{itemize} 
        \item {\bf \theia}: is a planned large-scale hybrid Cherenkov/scintillation detector that would use one or more of the many technologies described above---WbLS, dichroicons, fast PMTs, LAPPDs, slow fluors, or a combination---to accomplish a broad physics program, including accelerator-based neutrino oscillations, solar and supernova neutrinos, and in a future phase, neutrinoless double beta decay through isotopic loading of a $0\nu\beta\beta$ isotope. One option would be a \theia that fits into one of the caverns being excavated for DUNE, another is a 100~ktonne scale detector located at SURF in a new cavern.
        \item {\bf SLIPS}: The SLIPS idea eliminates the need for ``containment'' of inner sensitive volume by floating scintillator on top of another liquid, and placing photon sensors at the bottom of the detector with reflectors directing light down to the bottom phototubes.  The design would be both low-cost and reduce the radioactivity that comes with an inner vessel.
        \item {\bf LiquidO}: LiquidO goes in an entirely new direction for photon-based detectors by deliberately choosing a nearly-opaque scintillator that effectively self-segments a large-volume detector. The light is extracted with optical fibers, which see light only created nearby, thus creating a photon-based, segmented tacking detector.
        \item {\bf Slow-Fluor Detectors}: As decribed above, slow-fluors allow the creation of a hybrid Cherenkov/scintillation detector without any special photon detection, and large-scale detector built out of them would yield a broad range of physics, including neutrinoless double beta decay with rejection of solar neutrino backgrounds using direction.
        \item {\bf ArTEMIS}: The development of Xe doping, ARAPUCAs, and new cryogenic PMTs, make the possibility of a DUNE-scale hybrid Cherenkov/scintillation detector in LAr possible.  Such a detector would observe the UV scintillation light via wavelength shifter, while identifying Cherenkov light with sensors able only to see optical photons.
        \end{itemize}
        
            With many of the technologies being well-developed, a broad program of mid-scale demonstrators has begun. These demonstrators are major steps on the road to a full-scale neutrino experiment:
            \begin{itemize}
            \item {\bf ANNIE Phase 3}: The third phase of the Accelerator  Neutrino Neutron Interaction Experiment will explore the capabilities for neutron detection in neutrino interactions using gadolinium-loaded water-based scintillator.  ANNIE Phase 3 will be a critical step towards high-energy neutrino oscillation physics in a large-scale hybrid/Cherenkov experiment like \theia.
            \item {\bf NuDot}: The existing NuDot experiment builds on the successful benchtop-scale ``FlatDot'' experiment, and is aimed at a demonstration of hybrid Cherenkov/scintillation detection capabilities, and eventually a physics measurement of $2\nu\beta\beta$ spectrum.
            \item {\bf Eos}: The recently-approved Eos demonstrator will initially use WbLS with new fast-timing PMTs and a small array of dichroicons to demonstrate Cherenkov/scintillation separation and associated position and direction reconstruction in a detector at the few-tonne scale.
           \end{itemize} 
        
            Taken together, these new, enabling technologies, new detector ideas and the program of prototype demonstrators paint a picture of an exciting new path for a broad neutrino physics program that will be done by photon-based detectors. 
            
            \clearpage


\section{Introduction}

Detectors that use photons as the primary carrier of interaction information have a long, rich history in neutrino physics, going as far back as the discovery of the neutrino itself.  Large-scale, monolithic detectors that use either Cherenkov or scintillation light have played major roles in nearly every discovery of neutrino oscillation phenomena~\cite{superk,sno,dayabay,kamland,t2k} or observation of astrophysical neutrinos~\cite{kii,sno,imb,borexino,borexinopep,icecube}.  New detectors at even larger scales are being built right now, including JUNO~\cite{juno}, Hyper-Kamiokande~\cite{hyperk}, and DUNE~\cite{dunearapucas}. 

Despite the maturity of the technologies being used, there have been recent advances across a broad range of areas that are creating new, enabling technologies that will allow much broader or more sensitive physics programs with such detectors.  Of particular interest have been ``hybrid'' detectors, that observe both Cherenkov and scintillation light while providing some form of discrimination between them.  Such detectors would have very broad physics programs which leverage the differences between these types of light: good high-energy particle ID and direction reconstruction from Cherenkov light, with the excellent energy resolution and low-energy particle ID possible with scintillation light.
Many different approaches to developing hybrid detectors have been developed in recent years, from tunable light-yield scintillators like water-based liquid scintillator (WbLS~\cite{wbls}), to slow fluors~\cite{biller}, to fast timing~\cite{chesslappd,chicago}, to spectral sorting of photons~\cite{dichroicons}. A significant program of prototypes of these detectors is underway now, including work at the CHESS~\cite{chess} array, NuDOT, ANNIE~\cite{annieLOI}, and the newly approved Eos demonstrator.  

In addition to these new technologies are new approaches to simulation and reconstruction, including machine learning techniques and GPU-acclerated ray-tracing, new scintillating liquids, new DAQ paradigms, and new ways to load isotopes into liquids to change the accessible physics programs.  

We discuss here these enabling technologies and their context for future physics experiment.  We believe that with these advances, there is an exciting future ahead for neutrino detectors that are able to exploit them.


\section{Hybrid Cherenkov/Scintillation Detectors}

 The physics accessible in large Water Cherenkov (WC) detectors such as Super-Kamiokande (SK) is limited in many areas of interest by the inability to detect particles with energy below the Cherenkov threshold.  For example, this limits sensitivity to the Diffuse Supernova Neutrino Background (DSNB)~\cite{SK_DSNB:2021} due to enhanced backgrounds from low-energy atmospheric neutrino interactions, and reduced signal from the inability to detect positron annihilation, which enhances the prompt signal from the leading reaction $\overline{\nu}_e + p\rightarrow e^{+} + n$. In the area of proton decay, the kaon from $p\rightarrow \overline{\nu}K^{+}$ is below the Cherenkov threshold, and in the area of solar neutrinos the $^{7}$Be and CNO neutrinos are practically undetectable as much of the energy from the neutrino electron scattering reaction is invisible.

Organic liquid scintillators (LS) have been used to enhance sensitivity for particle below Cherenkov threshold, and to provide high light yield and thus narrow energy resolution needed to see monoenergetic signals like the $0\nu$ peak in neutrinoless double beta decay. LS is currently being used in the KamLAND, Borexino, and SNO+ detectors, and the JUNO detector now under construction. While this is very effective at increasing sensitivity at low energies, it comes at the loss of the directional sensitivity and multi-track resolution that is a hallmark of WC detectors. 

Hybrid neutrino detectors, which leverage both the unique topology of Cherenkov light and the high light yield of scintillation, have the potential to revolutionize the field of low- and high-energy neutrino detection, offering unprecedented event imaging capabilities and resulting background rejection.  These performance characteristics would substantially increase sensitivity to reactor signals, as well as a broad program of fundamental physics.

Discrimination between ``chertons'' and ``scintons'' can be achieved in several ways.
The use of a cocktail like water-based liquid scintillator (WbLS) provides a
favorable ratio of Cherenkov/scintillation light ~\cite{chess}.
Combining angular
and timing information allows discrimination between Cherenkov and
scintillation light for high-energy events even in a standard scintillator like
LAB-PPO~\cite{chess}.  Slowing scintillator emission time down by using slow
secondary fluors can also provide excellent separation~\cite{biller}, while using very fast timing~\cite{LAPPDtiming} can pick off the prompt Cherenkov photons from scintillators with faster time profiles.
Recent R\&D with dichroic filters to sort photons by wavelength has shown
separation of long-wavelength chertons from the typically shorter-wavelength
scintons even in LAB-PPO, with only small reductions in the total scintillation
light~\cite{dichroicons}.  In principle, all of these techniques could be
deployed together if needed to achieve a particular full physics program.  New
reconstruction techniques, to leverage the multi-component light detection, are
being developed and with the fast timing of newly available PMTs and the
ultrafast timing of LAPPDs (Large Area Picosecond Photon Detectors), allow effective tracking for high-energy
events and excellent background rejection at low energies.

\section{New Scintillating Materials and Fluors}
\subsection{Slow Fluors}
The properties of four slow fluors have been studied in the context of LAB-based liquid scintillator mixtures to provide a means to effectively separate Cherenkov light in time from the scintillation signal with high efficiency \cite{Biller:2020uoi}. This allows for directional and particle ID information while also maintaining good energy resolution. Such an approach is highly economical (i.e. small compared to other experimental costs) and can be readily applied to existing and planned large-scale liquid scintillator instruments without the need of additional hardware development and installation. Using this technique, we have explicitly demonstrated Cherenkov separation on a bench-top scale, showing clear directionality, for electron energies extending below 1 MeV. 

Timing measurements were made using a 90Sr source directed through the sample vial and either towards or away from a measurement PMT. Electrons from supported beta decay of 90Y (Q=2.28 MeV) first pass through a 2mm diameter scintillating fibre coupled to a trigger PMT. Typical electron energies that make it through the fibre and vial wall are less than 1 MeV.

Figure \ref{fig:SlowFluor1} shows the measured time spectrum for acenaphthene for the case where electrons are directed through the scintillator sample and towards the measurement PMT.  It is used here as a primary fluor, with a concentration of 4g/L in LAB. Absorption and emission spectra were separately measured and are shown in the inset of the figure. 

\begin{figure}[htp]
\centering 
\includegraphics[width=0.7\textwidth]{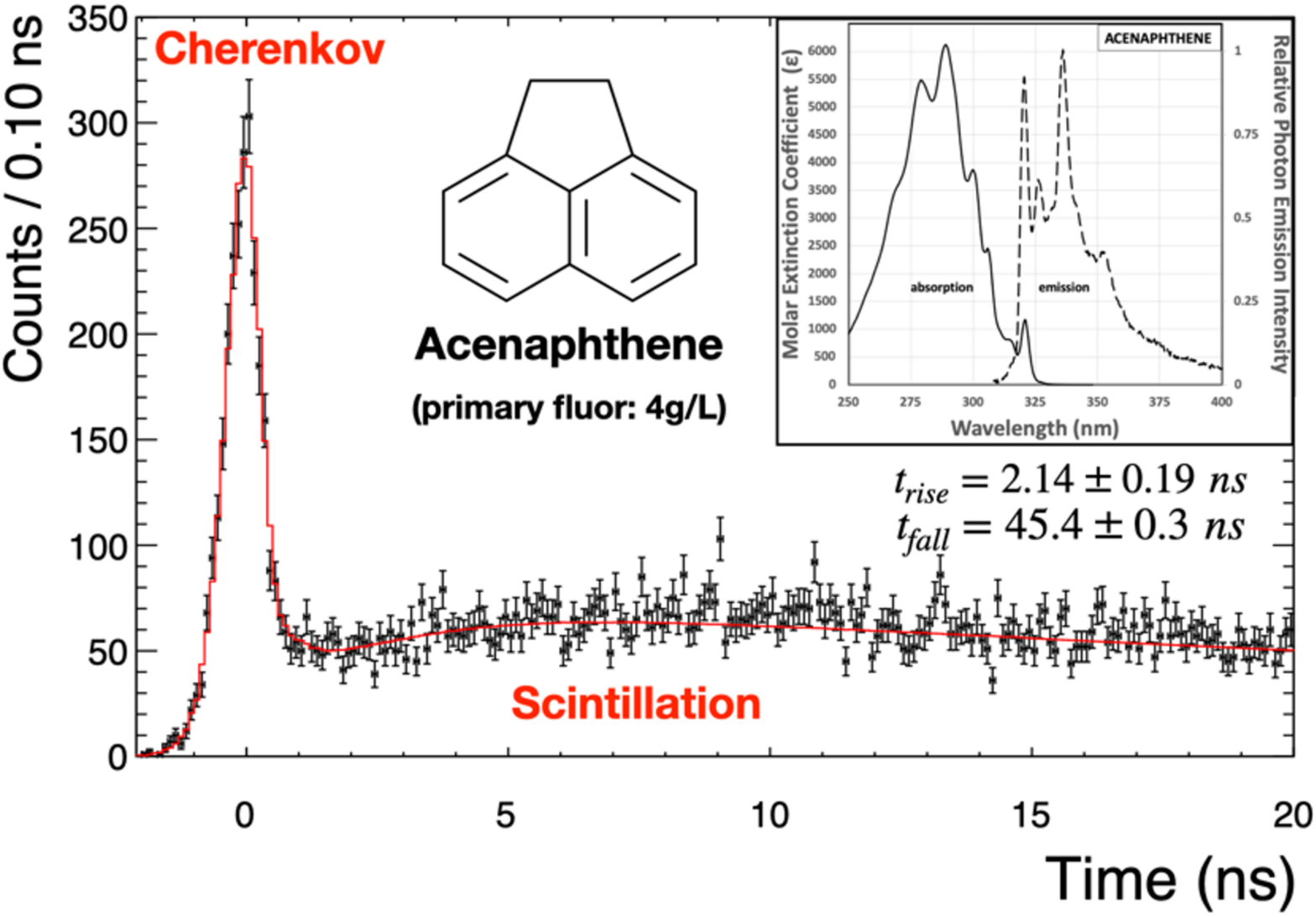} 
\caption{Zoomed in time spectrum for 4 g/L acenapththene in LAB with clear Cherenkov peak.} 
\label{fig:SlowFluor1}
\end{figure}

Figure \ref{fig:SlowFluor2} shows the full time spectrum for acenaphthene in LAB, comparing configurations when the source is directed towards and away from the measurement PMT. A well-separated, directional Cherenkov signal is clearly observed for the forward configuration. The Cherenkov signal and scintillator time constants are extracted using a combined fit of data from both configurations. The primary scintillator time constant values are given in each figure.

\begin{figure}[htp]
\centering 
\includegraphics[width=0.7\textwidth]{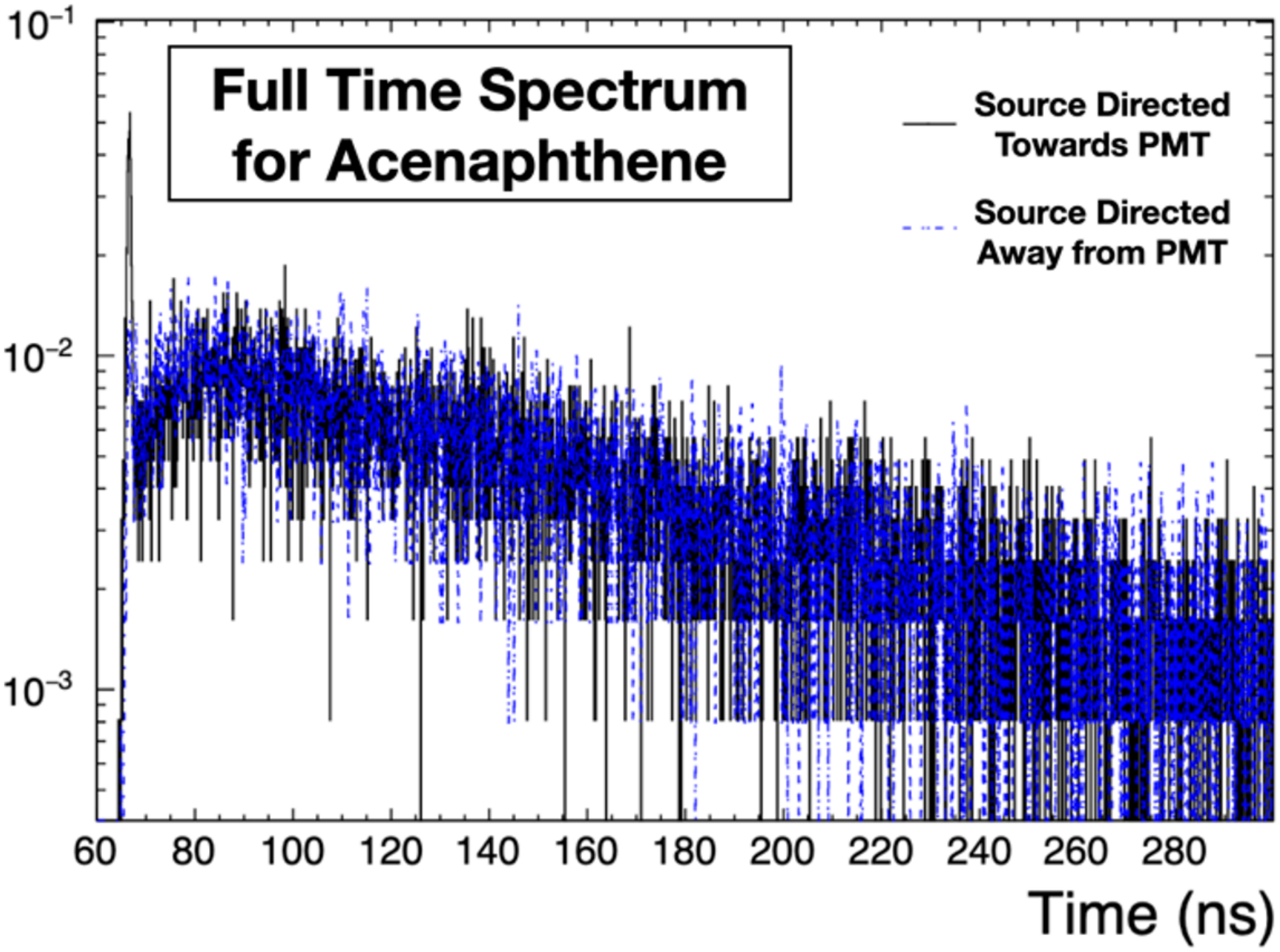} 
\caption{Full time spectrum for 4 g/L acenapththene in LAB for source directed towards (solid) and away (dashed) from the PMT.} 
\label{fig:SlowFluor2}
\end{figure}

\begin{figure}[htb!]
\centering 
\includegraphics[width=0.9\textwidth]{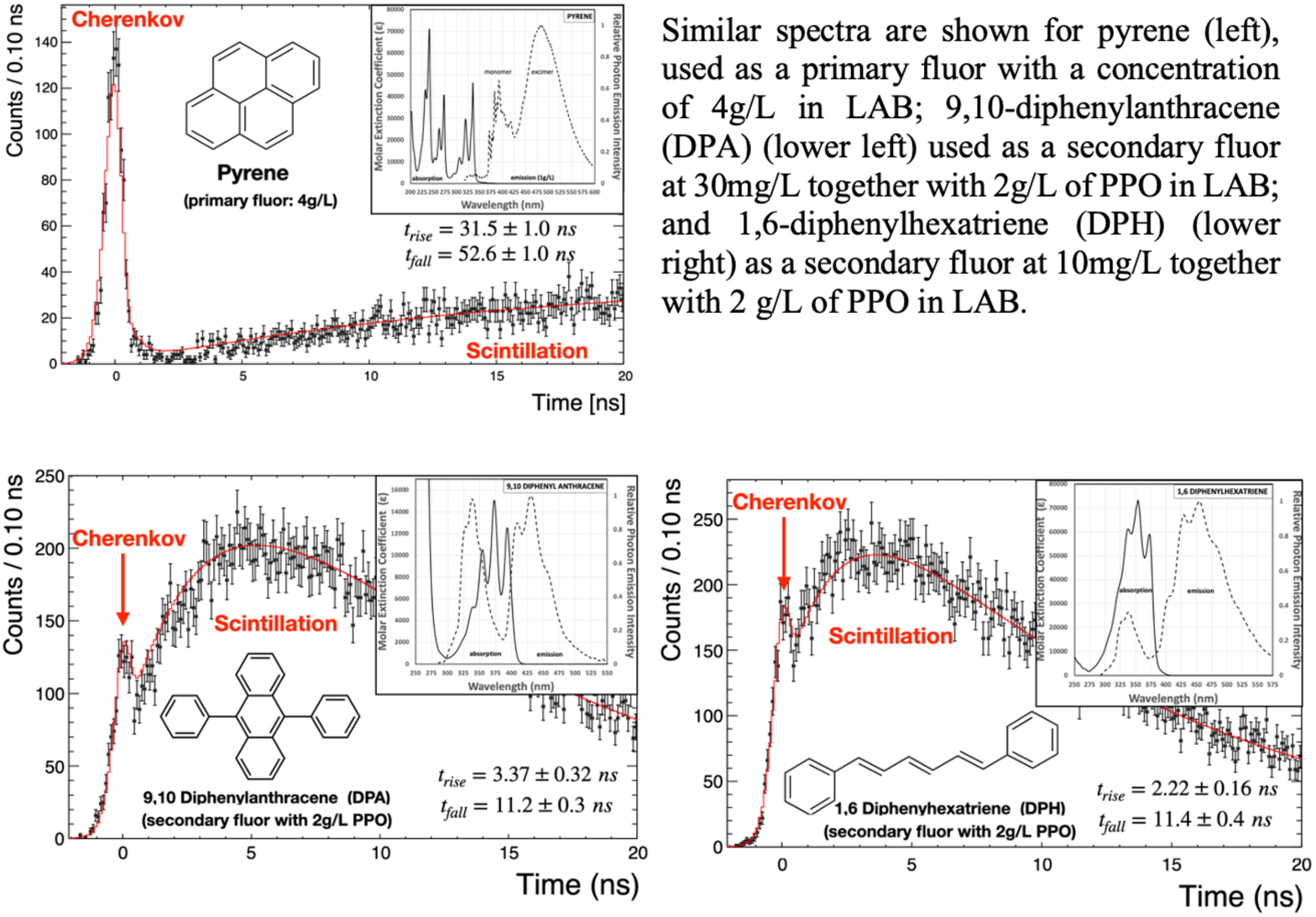} 
\label{fig:SlowFluor3}
\end{figure}


The table below lists some of the fluor combinations studied along with the measured intrinsic light yield, which were deconvolved from the PMT response and normalised relative to the previously determined yield of PPO in LAB from studies by SNO+.

\begin{figure}[htb!]
\centering 
\includegraphics[width=0.6\textwidth]{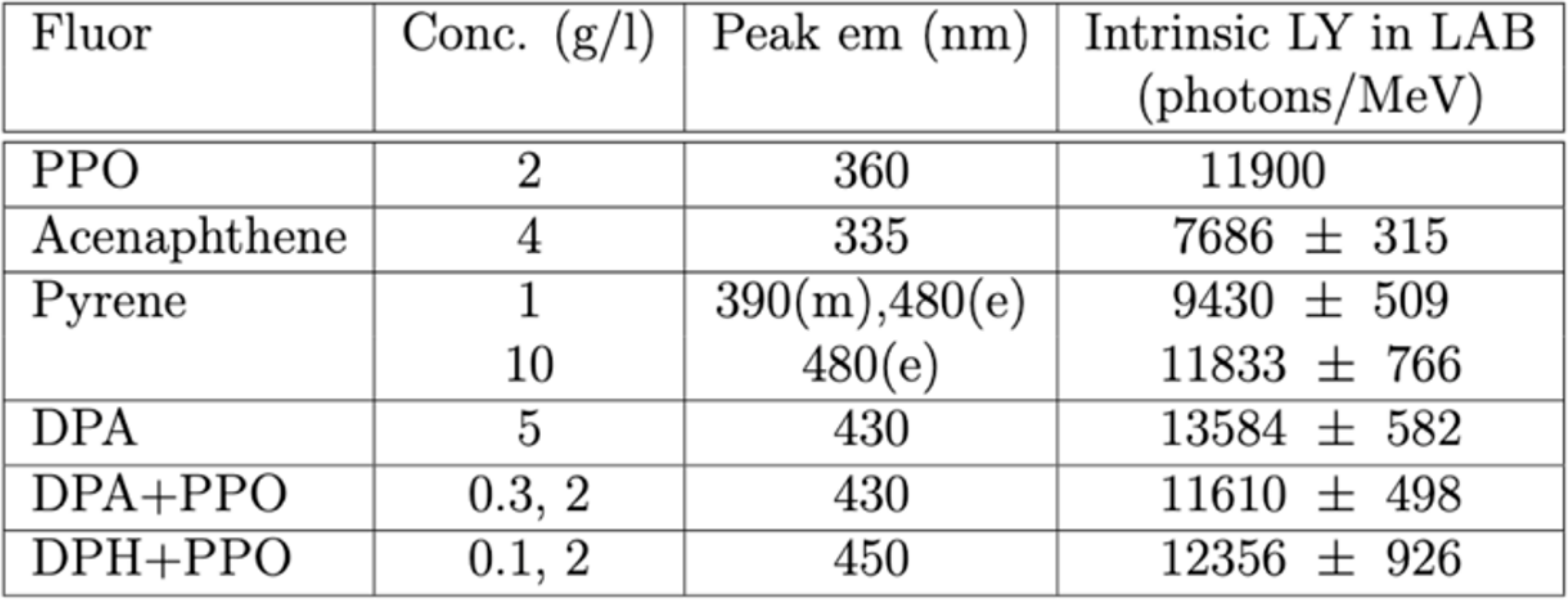} 
\label{fig:SlowFluor4}
\end{figure}

Different fluor combinations may prove suitable for different applications. This study has potentially important consequences for a variety of future instruments, including measurements of low energy solar neutrinos and searches for neutrinoless double beta decay in loaded scintillator detectors. This also opens the possibility of obtaining good directional information for elastic scattering events from supernovae neutrinos and reactor anti-neutrinos in large scale liquid scintillation detectors. While the use of slow fluors means that the vertex resolution may be worse than typical large-scale liquid scintillator detectors (but better than typical large-scale Cherenkov detectors), the balance between position resolution, Cherenkov separation purity and energy resolution can be tuned for a particular physics objective by modifying the fluor mixture. This balance is also affected by the presence of fluorescence quenchers, which may be naturally present in the case of loaded scintillator mixtures or could be purposely introduced to change the balance.

\subsection{Water-based Liquid Scintillator}
\label{sec:WbLS}

The development of Water-based Liquid Scintillator (WbLS)~\cite{wbls} has the potential to signifcantly impact and enhance  hybrid particle detection capabilities.
WbLS is essentially liquid scintillator encapsulated in surfactant micelles that are thermodynamically stable in water (see Fig.~\ref{fig:WbLS}).
By introducing varying amounts (typically 1\%-10\%) of liquid scintillator into water, the liquid yield can be adjusted to allow detection of particles below Cherenkov threshold  while not sacrificing directional capability, cost, or environmental friendliness. First developed at Brookhaven National Lab (BNL), WbLS is a leading candidate for the main target medium for the proposed \theia detector, which would enhance the scientific program at the LBNF significantly, as described in the \theia White Paper ~\cite{theiawp}.

\begin{figure}[htb!]
\centering
\includegraphics[width=0.4\textwidth]{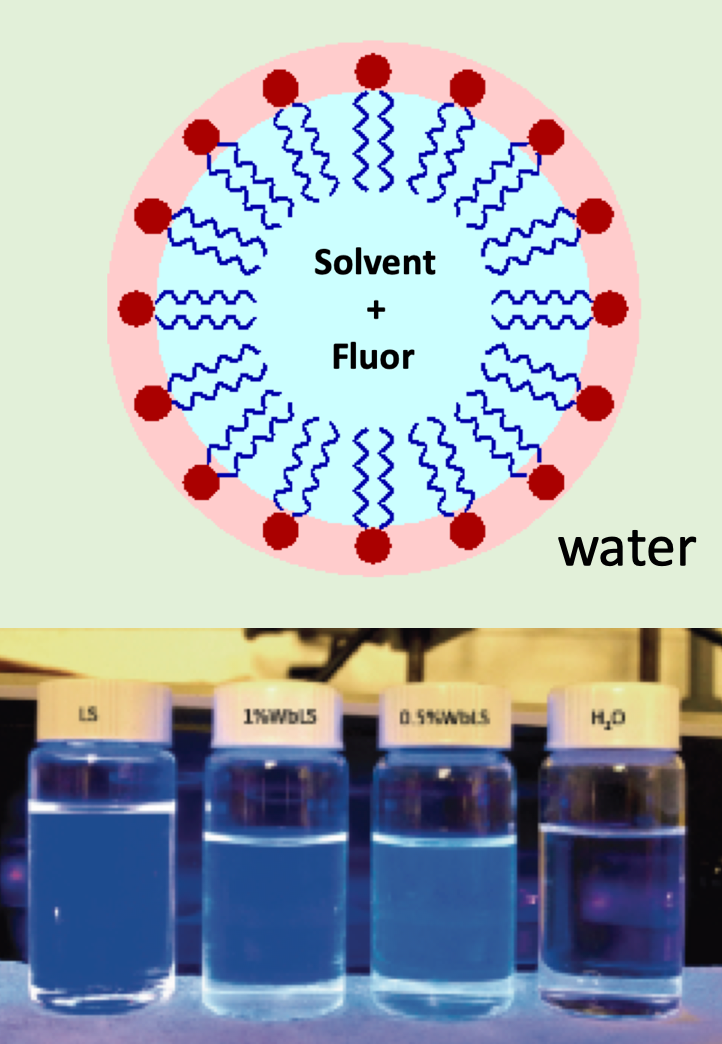}
\caption{Top: Liquid scintillator encapsulated in a surfactant micelles, which are typically around 10 nm in size. Bottom: From right to left: water, 0.5\% WbLS, 1\% WbLS, pure liquid scintillator.}
\label{fig:WbLS}
\end{figure}

There is an active R\&D effort to realize the novel WbLS liquid target being considered for \theia and other detectors. These are mainly focused on gadolinium-loaded WbLS (GdWbLS) in order to facilitate neutron detection. Used in ANNIE and SK-Gd, gadolinium doping to 0.1\% by weight will result in 90\% of neutron captures being on gadolinium due to the large cross-section. Gadolinium capture results in an 8 MeV gamma cascade versus the 2.2 MeV from capture on hydrogen. These efforts include a precision measurement of attenuation at long distances, demonstration of material compatibility with detector components, and characterization to tune the simulations. Several demonstrators are now funded and being built at several US and European labs, in addition to programs at several universities:\\
\begin{itemize}
\item BNL 30-ton deployment demonstrator. Described in more detail below, this 3-year program will construct a fully-operational prototype at BNL for as the next step towards the kilotin scale.
\item LBNL The \eos reconstruction and characterization demonstrator (see Sec.~\ref{sec:eos}) is a 3-year program at LBNL will allow testing of the performance parameters of WbLS using a several ton-scale detector.
\item The SANDI 0.4 ton inner vessel GdWbLS upgrade for ANNIE will take place in summer 2022. Described in more detail in Sec.~\ref{sec:ANNIE}, this will allow neutrino event reconstruction in WbLS at the GeV-scale.
\item The STFC-supported BOELYN 25-ton WbLS demonstrator at Boulby Underground Lab will be constructed in 2022-23 and will enable characterization of radioactive backgrounds for WbLS in an underground environment.
\end{itemize}


\subsubsection{Brookhaven 30-ton Demonstrator and UC Davis Recirculation R\& D}

While WbLS and GdWbLS have been produced and deployed in 1 ton scale quantities, it is important to demonstrate the practical deployment on a scale that would allow addressing the issues of long-term stability and {\it in situ} purification. It is well-know that optical impurities may leach from detector material such as stainless steel and some plastics, and if not removed these will degrade the transparency over time. With pure water, this is done using fairly standard off-the-shelf techniques. For gadolinium-doped water (e.g. ANNIE and SK-Gd), specialized systems had to be developed that would remove contaminants, but not gadolinium. For GdWbLS/WbLS the situation is even more complicated, as typical water systems cannot handle organic concentrations at these levels - the organics much be separated from the liquid before optical contaminants can be removed using standard techniques.\\

To address this, the Liquid Scintillator Development Lab (LSDL) at UC Davis has been developing a organics separator system designed to remove the GdWbLS organics to the level of 1ppm (from the starting point of 10,000 ppm). This is done by pushing the liquid through a series of nanofilters that strain out the micelles, but leave the other contaminants intact. The gradual scaling up of this effort is shown in Fig.~\ref{fig:UCD_WbLS}. To date, this system has been able to separate 99.2\% of the organics out (as measured via fluorescence and XRF) while passing more than 97\% of the gadolinium. Thus, this first stage is ready for deployment at the existing 1-ton BNL prototype, starting with the first NF201 device before summer 2022. \\

A 30-ton demonstrator is currently under construction at BNL. This demonstrator will have the ability to test large-scale deployment of different WbLS formulas without or with loading of metallic ions (i.e. Gd). Two vital subsystems: Gd-water purification and organics separation systems are included in the demonstrator configuration. This demonstrator will determine the circulation requirement and if crucial, demonstrate the purification efficacy to maintain optical stability of WbLS. The main goals are to retire risks that could be derived from deployment and operation of a kiloton-scale WbLS detector by examining (1) feasibility of in-situ sequential mixing scheme, (2) compatibility, in terms of flow rate and temperature, between Gd-H2O and WbLS systems, (3) separation and recombination efficacy of nanofiltration system and (4) detector performance in light collection and optical stability. This program also aims to test detector cleanliness and material compatibility (varied PMTs, cable, calibration, etc.) under different detector configurations, and to provide liquid-handling and ESH training for the scientific community. The facility to house 30-ton demonstrator is expected to receive occupancy by end of 2022. The installation of subsystems is planned to be completed in spring 2023. The full operation of this 30-ton demonstrator is scheduled to start in summer 2023. \\

\begin{figure}[htb!]
\centering
\includegraphics[width=0.80\textwidth]{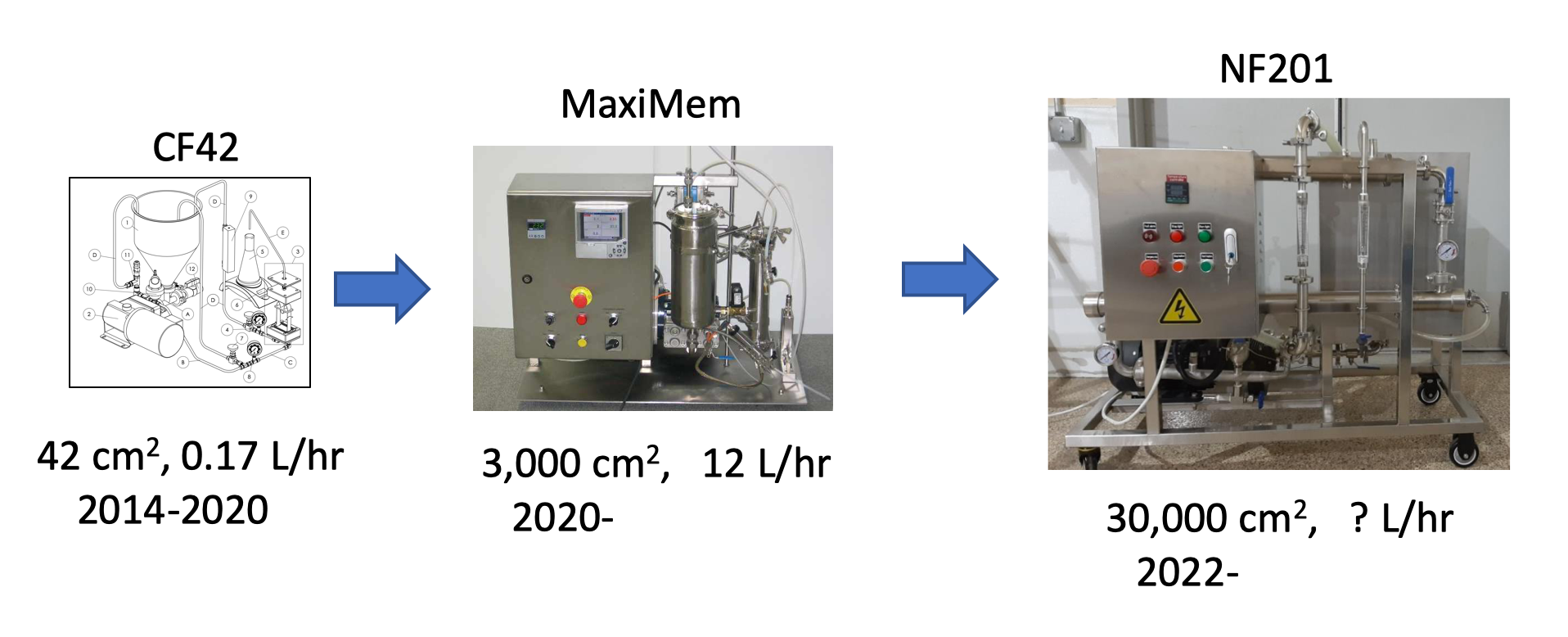}
\caption{Gradual scale up of the LSDL nanofiltration development testbed. The NF201 device is now on order for the BNL 1-ton system, with anticipated installation in 2022. This is a key development towards the full 30-ton scale system.}
\label{fig:UCD_WbLS}
\end{figure}

\begin{figure}[htb!]
\centering
\includegraphics[width=\textwidth]{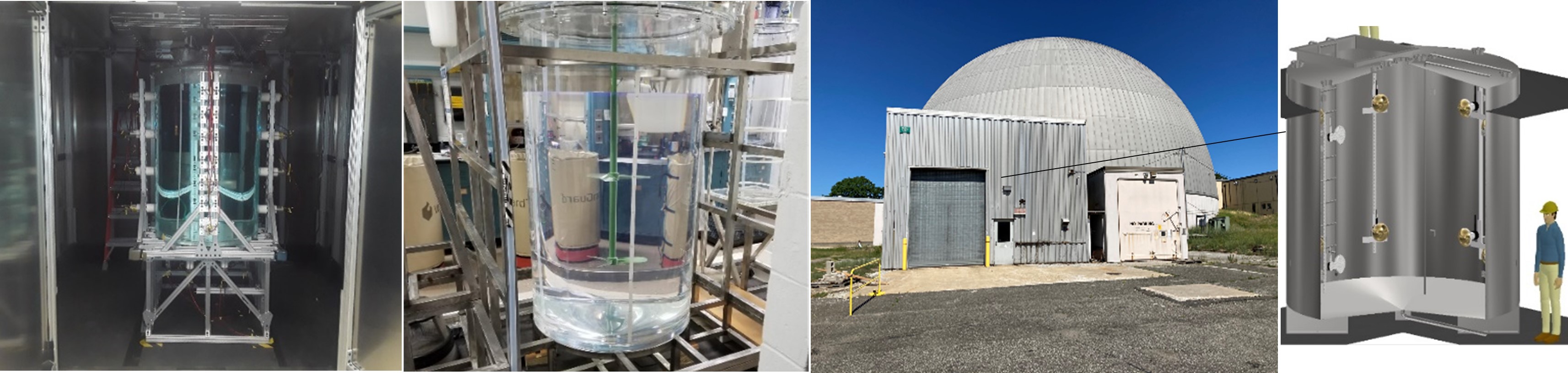}
\caption{The BNL 1-ton WbLS prototype and production facility (left), and the future home of the 30-ton demonstrator (right).}
\label{}
\end{figure}

While this is going on at BNL, the LSDL is now turning attention to the second stage development, as it is necessary to reduce organics by another order of magnitude. This effort, which might take 1-2 years would be the last step in finalizing this system for use in the design of a kiloton scale device, such as \theia, which will be further validated at the 30-ton demonstrator.\\


\subsubsection{Scintillation Pulse-Shape Discrimination in WbLS}
\label{subsec:WbLS_PSD}
Researchers at LLNL have developed and demonstrated the ability to distinguish gamma-rays and neutrons in WbLS using the pulse shape of the scintillation light~\cite{Fordpsd}. Pulse-shape discrimination has been used in liquid scintillators for decades to identify neutrons from gamma-rays. Here, the fast and slow decay times of scintillation light caused from the excitation and decay of singlet and triplet states respectively. The contributions of the slow decay light is due to the density of triplet states along the particle track. As a result the delayed light is directly dependent on the dE/dx of the particle. While other proponents of WbLS have addressed the potential for particle identification through the use of Cherenkov and Scintillation light separation and/or ratios~\cite{dsnb_psd}. The use of the scintillation pulse-shape in combination with other techniques for Ch/S separation would allow for better particle identification for a larger range of particles. 
\medskip\\
The material was synthesized utilizing an approach referred to as the hydrophilic-lipophilic difference in which the the salinity, temperature, oil fraction, and characteristics of the surfactant is balanced to create a stable emulsion~\cite{hydro-lipo}. This approach was used to generate WbLS with both LAB and xylene. Given that the light-yield and pulse-shape discrimination figure of merit are characteristics of the scintillation component; the scintillation fraction is directly proportional to these properties. Future simulation studies can inform requirements for this material by balancing the light yield, optical properties, and scintillation times for a specific application. 
\begin{figure}[ht]
     \centering
     \begin{subfigure}{0.45\textwidth}
         \centering
         \includegraphics[width=\textwidth]{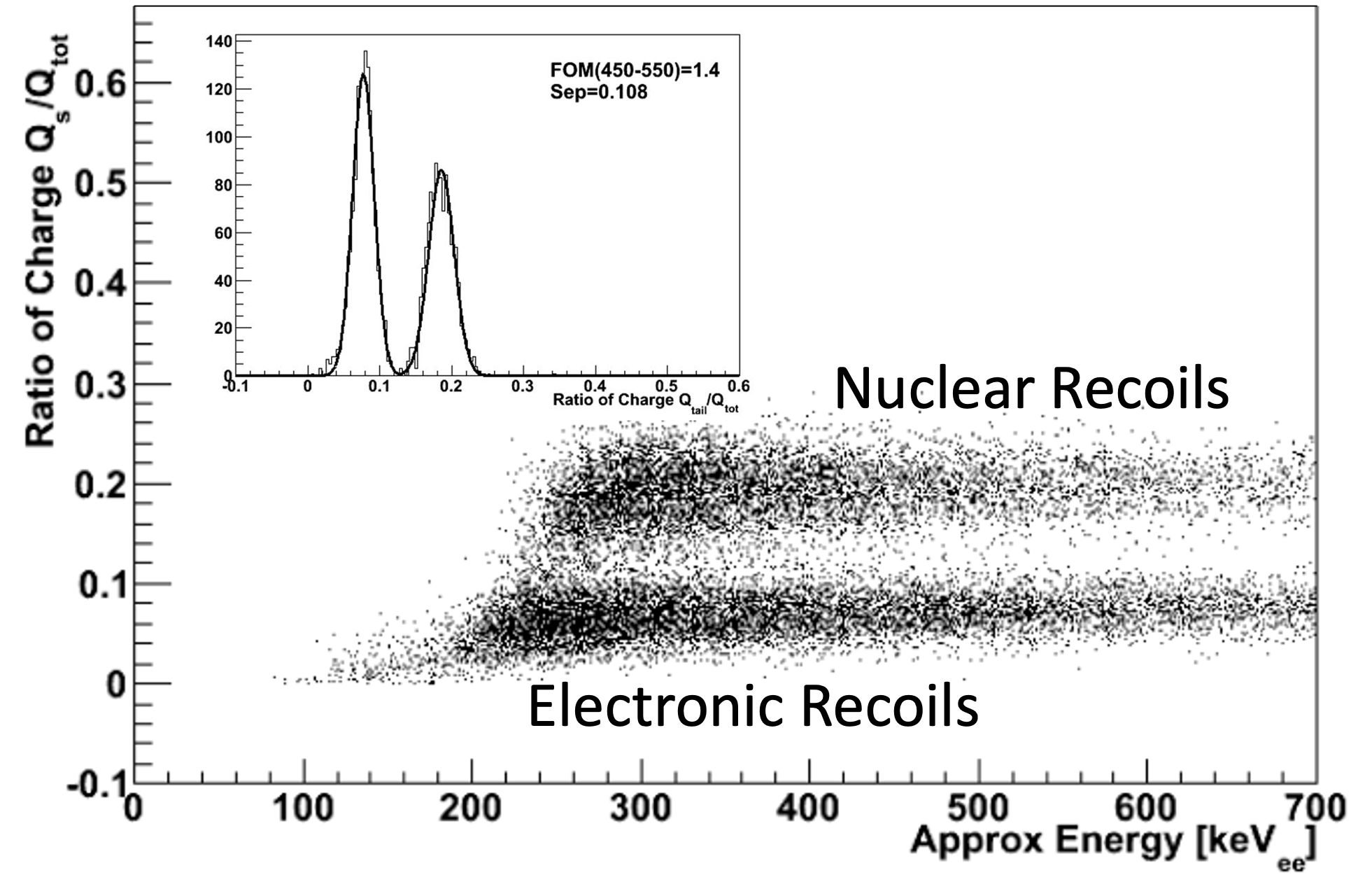}
         \caption{The neutron/gamma PSD performance of a sample of WbLS with 33 wt$\%$ oil phase as a function of energy.This sample was exposed to $^{252}$Cf source. Here clear separation of nuclear and electronic recoils can be observed.}
         \label{fig:WbLS-PSD}
     \end{subfigure}
     \hfill
     \begin{subfigure}{0.45\textwidth}
         \centering
         \includegraphics[width=\textwidth]{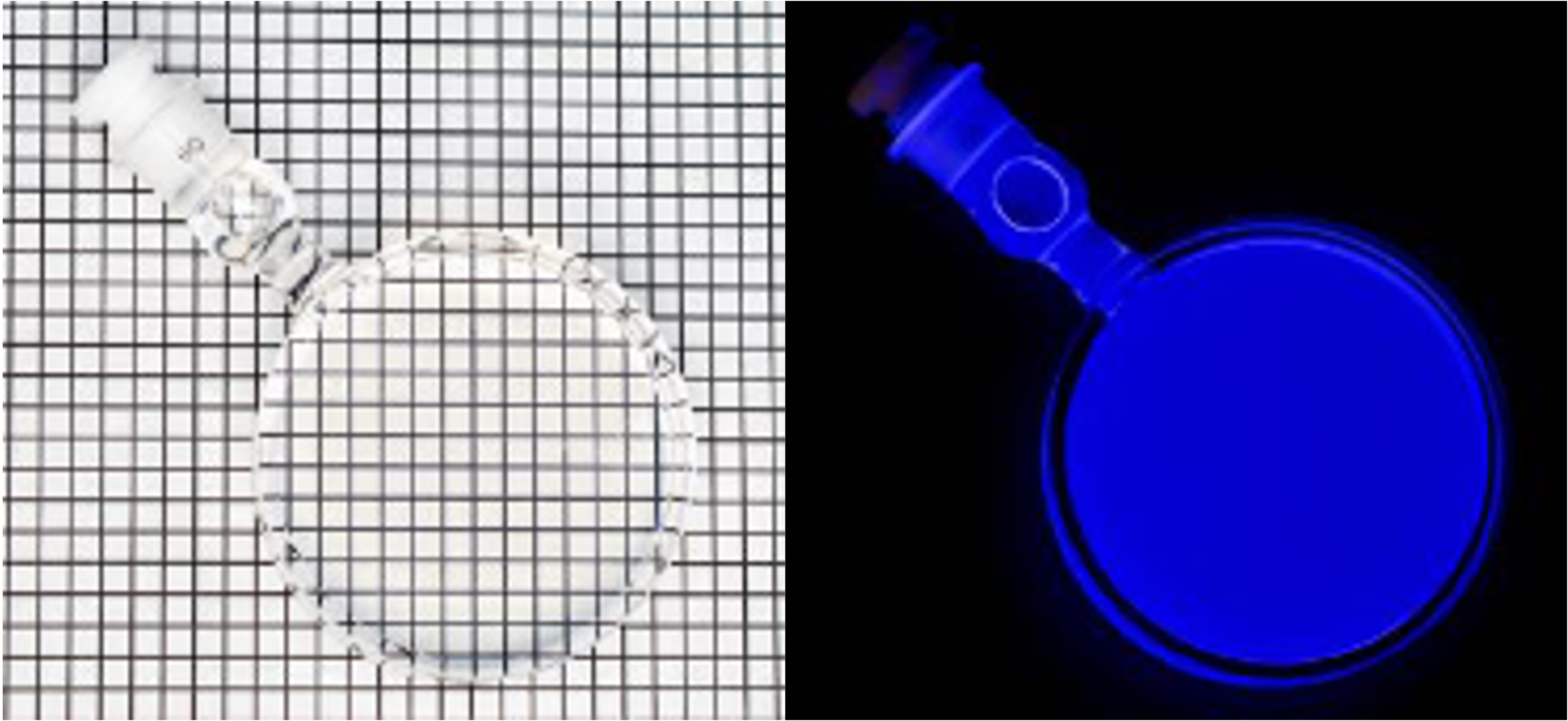}
         \caption{Photographs of a 50 mm diameter cuvette containing a transparent formulation of WbLS with PSD capability for (n,g). {\it (Left)} A photograph of a transparent formulation and {\it (Right)} the same sample under a UV lamp.}
         \label{fig:WbLS-PSD-Photo}
     \end{subfigure}
     \hfill
        \caption{Examples of Pulse Shape Discrimination of Scintillation Light in WbLS}
        \label{fig:WbLS-PSD-Figure}
\end{figure}

%

\subsection{$^6$Li-doped Pulse-Shape-Discriminating Plastic Scintillators}

Plastic scintillators are comprised of dopants entrained in a polymer matrix. In some cases, the scintillation signal from these dopants can be used to discriminate between particle types, e.g. electromagnetic depositions vs those from heavy ions, via ``Pulse Shape Discrimination'' (PSD)~\cite{ZAITSEVA201288}. Further particle discrimination can be enabled by other dopants that have specific neutron capture reactions, which can also be identified via a combination of energy gating and PSD. For example, when loaded with lithium-6 and high amounts a primary dye along with a secondary dye that shifts the wavelength of scintillation plastic scintillators can use PSD to distinguish between electrons or positrons, neutrons, and thermalized neutron captures~\cite{ ZAITSEVA2013747, MABE201680}. This new class of materials, primarily developed at the Lawrence Livermore National Laboratory,  can provide a combination of important capabilities for future neutrino detectors, particularly those that use Inverse Beta Decay where efficient and distinct detection of the low energy final state neutron is critical. 
At the current stage of development these plastics have broadly similar performance in terms of light yield, optical attenuation, and PSD separation to liquids (e.g. Fig.~\ref{fig:Li6-PSD}). Elements have been produced with greater than 50-cm length and 7.5-cm width, and a commercial vendor is developing a production process (Fig.~\ref{fig:Li6-PS}). 

A key challenge to the production of these $^6$Li-doped PSD plastics is the balance between scintillation properties and solubility of the dopants in the polymer matrix and the polymer precursors (i.e., monomers). 
Typical primary dyes like m-terphenyl (mTP) and 2,5-diphenyloxazole (PPO) have high solubility in monomers like styrene. However, organic salts that contain lithium-6 as the cation do not typically have high solubility in monomers like styrene. 
One option to improve the solubility of these lithium-6 salts is to incorporate a co-monomer with styrene that can dissolve the lithium-6 salts. 
Methacrylic acid is one possible co-monomer that has polar substituents that can solubilize the lithium-6 salts~\cite{osti_1490925,Frangville}. 
This co-monomer has been used primarily with organic salts that contain aliphatic organic groups. When used with dyes that contain heteroatoms like oxygen and nitrogen (e.g., PPO), this co-monomer may detrimentally affect the scintillation performance. 
Another option for soluble lithium-6 salts is to use organic salts with aromatic organic groups that have good solubility in primary dyes like PPO. 
A co-monomer like methyl methacrylate or a small amount of solvent (e.g., tetrahydrofuran) may be necessary to improve solubility of these aromatic lithium-6 salts without drastically reducing scintillation performance~\cite{osti_1490925}.
However, the aromatic groups may absorb energy and affect scintillation performance. 
Additionally, the aromatic acids that are used to synthesize these aromatic salts are solid at room temperature, which makes purification of the lithium-6 salts difficult. 
Overall, the balance of scintillation properties and solubility of dopants that enable multiple modes of PSD remains an interesting challenge to investigate and solve.

Plastic scintillators with PSD and $^6$Li-doping provide the same capabilities as the PSD $^6$Li-doped liquid scintillator used for  PROSPECT that is described in Sec.~\ref{sec:mdwbls}. 
As described in a recent LOI~\cite{Li6Organic:2021loi}, the new solid form makes self-supporting fine-grained segmentation on a large scale possible, enables light guiding approaches based on air gaps, and eliminates the need for containment vessels and supporting infrastructure. Several detector implementations are proposed and under development using $^6$Li-doped PSD plastics. The SANDD concept (Fig.~\ref{fig:Li6-PS}) has implemented mm-scale segmentation to study reactor antineutrino directionality and cosmogenic background rejection~\cite{Li:2019sof,Sutanto:2021xpo}. Segmentation at the 5--10-cm scale in 2D bar geometries is being developed for efficient, aboveground reactor antineutrino detection (e.g. ROADSTR~\cite{ROADSTR:2021loi}) with the goal of a mobile system appropriate for monitoring applications and measurements at  multiple reactors with correlated systematics.

\begin{figure}[ht]
     \centering
     \begin{subfigure}{0.45\textwidth}
         \centering
         \includegraphics[width=\textwidth]{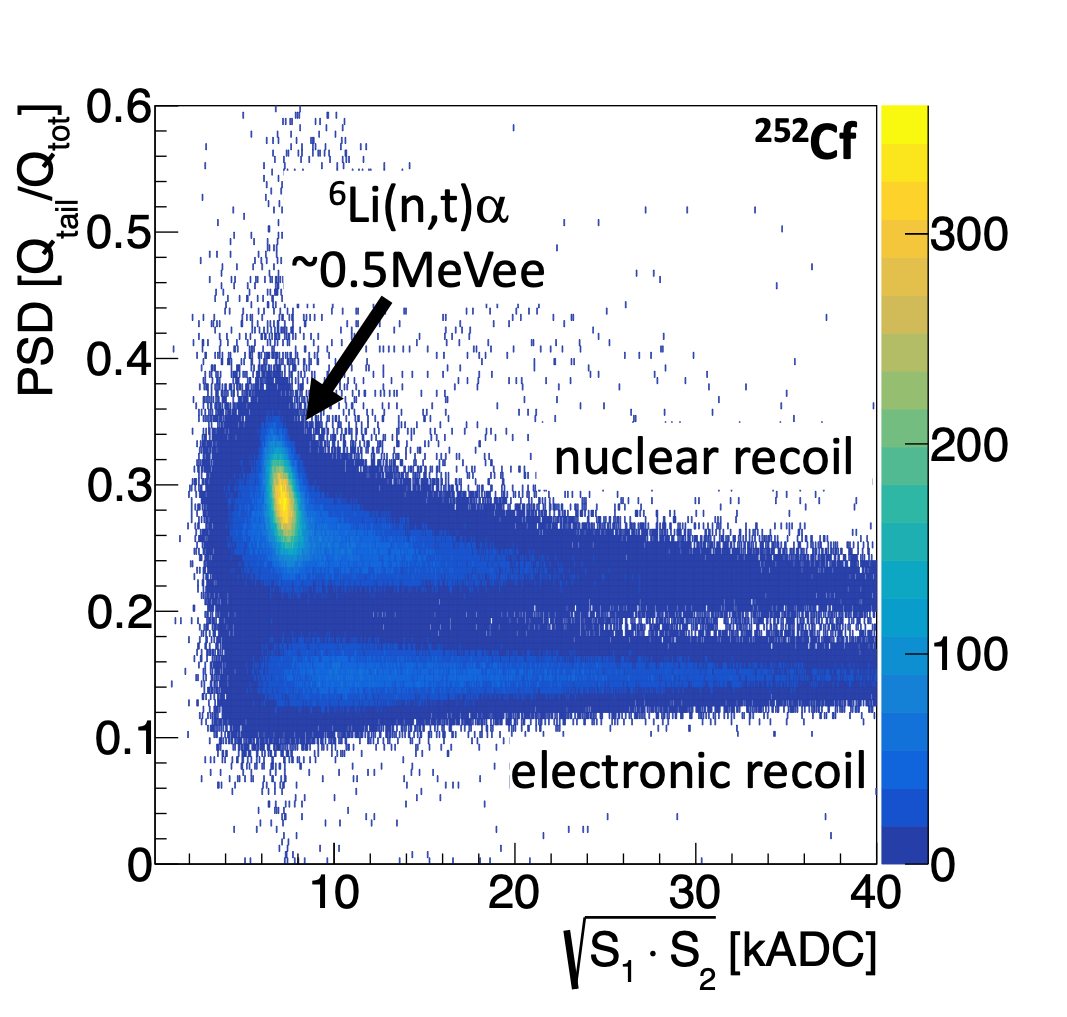}
         \caption{Pulse Shape Discrimination performance of the 5.5~cm $\times$ 5.5~cm $\times$ 50~cm $^6$Li-doped PSD plastic scintillator shown in panel (b) under $^{252}$Cf flood-field illumination , in terms of uncalibrated ADC units. A clear neutron capture feature is observed at about 0.5~MeV$_{ee}$, along with electronic and nuclear recoil separation.}
         \label{fig:Li6-PSD}
     \end{subfigure}
     \hfill
     \begin{subfigure}{0.5\textwidth}
         \centering
         \includegraphics[width=\textwidth]{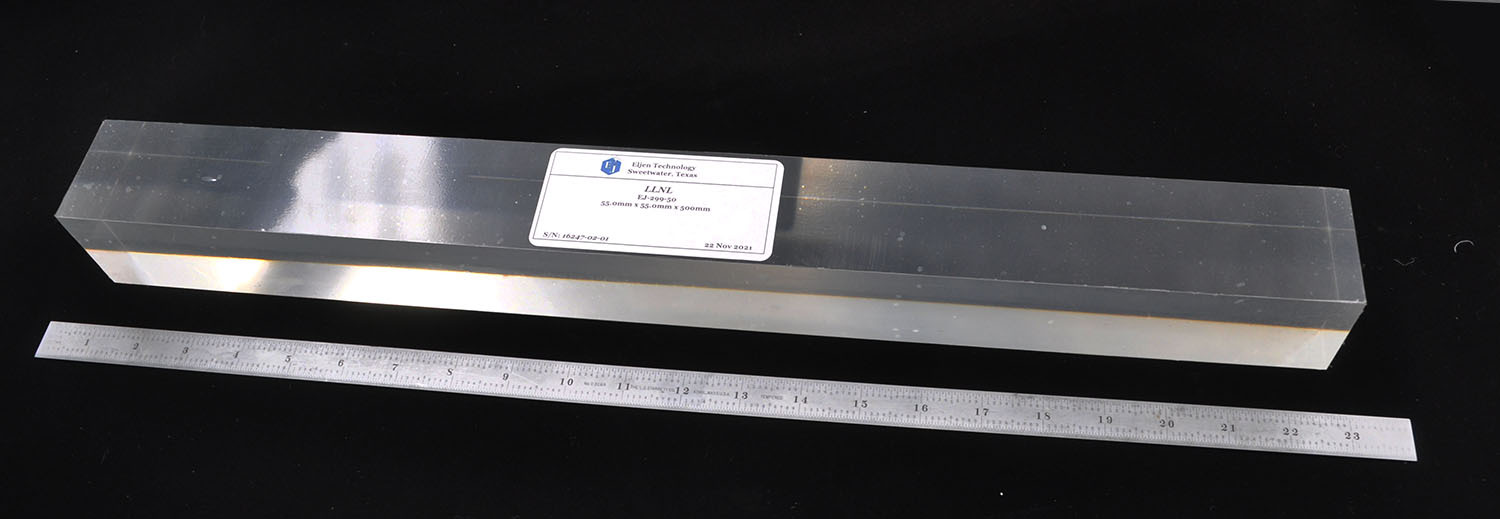}
          \includegraphics[width=0.6\textwidth]{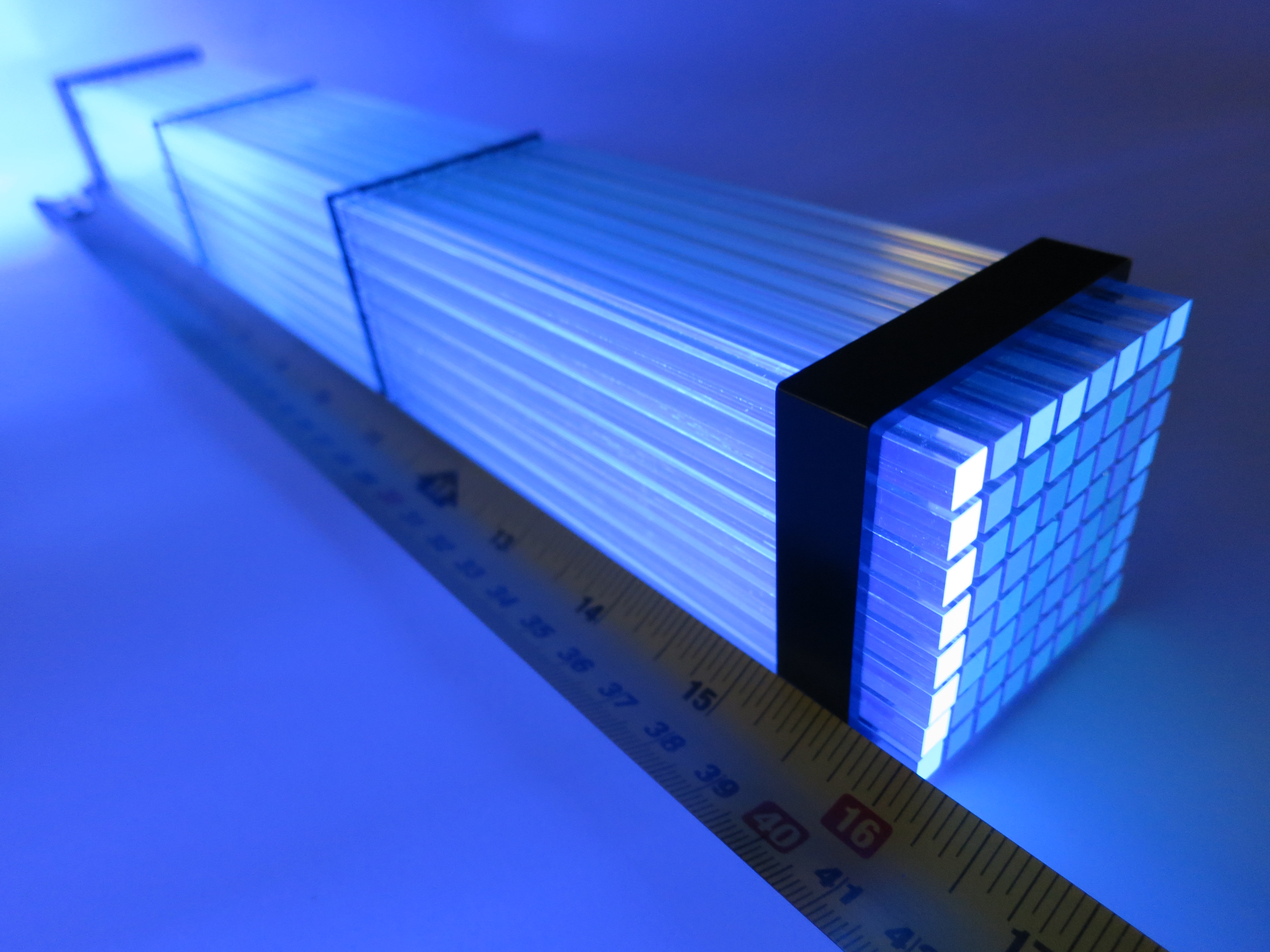}
         \caption{Photographs of $^6$Li-doped PSD plastic scintillator examples. {\it Top:} A 5.5~cm $\times$ 5.5~cm $\times$ 50~cm bar produced by Eljen Technology. {\it Bottom:} 64 rods of 5.4~mm $\times$ 5.4~mm$ \times$ 40~cm dimensions, as used in the SANDD central module~\cite{Sutanto:2021xpo}.}
         \label{fig:Li6-PS}
     \end{subfigure}
     \hfill
        \caption{Examples of $^6$Li-doped PSD plastic scintillator.}
        \label{fig:Li6_Scintillators}
\end{figure}

 
\subsection{Additively Manufactured Scintillators}
 
Efforts are underway to produce scintillators that can be additively manufactured (AM, i.e. 3D printed) into structures beneficial for neutrino detection. One application is as a replacement for existing methods of construction for pixelated (i.e. segmented) detectors, where AM scintillators would enable low-cost automated manufacture, finer pixelization, and precise control over the reflector/air-gap between segments. Another application is the production of heterogeneous scintillators with structured features between 10--\SI{1000}{\micro\meter} designed to detect the location and direction of ionizing recoil tracks at that small scale, a focus of work at LLNL \cite{zhang_architected_2021}. These fine structures are impossible to manufacture with conventional scintillators, but they offer position resolution, directionality, and particle discrimination abilities that can enhance neutrino detection.
 
Several forms of AM scintillators have been demonstrated in recent literature (Figure \ref{fig:AM_Scintillators}). Solid scintillator can be converted into filament and extruded from a hot nozzle (fused filament fabrication) \cite{berns_additive_2022}. Alternatively, uncured scintillator can be extruded as a thixotropic ink that cures in place (direct ink write), or precisely photocured using directed light (stereolithography or digital light processing) \cite{lee_characterization_2019,kim_fabrication_2019}. Recent demonstrations show great promise, but in general show reduced light output compared to conventional cast scintillators. Ongoing improvements to the printing process and material formulations may overcome this hurdle in the near term. The benefits of AM scintillators may also be worthwhile to neutrino detectors even with some attendant compromise in light output.

\begin{figure}[ht]
     \centering
     \begin{subfigure}{0.55\textwidth}
         \centering
         \includegraphics[width=\textwidth]{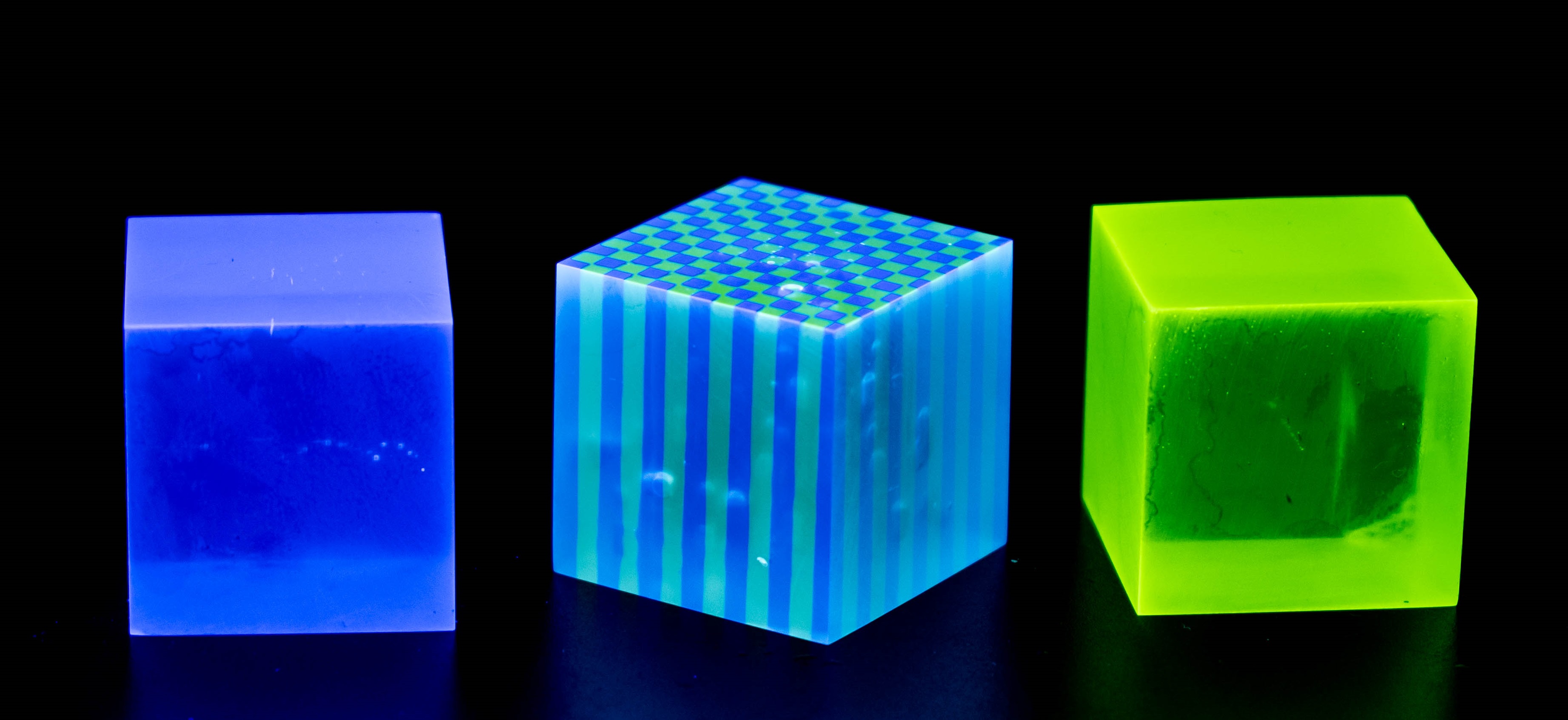}
         \caption{A heterogenous scintillator printed at LLNL with a structure of alternating blue- and green-dyed zones, as well as conventionally-cast reference samples of each color. The scintillator is a \SI{26}{\mm} cube.}
         \label{fig:AMSS_sample}
     \end{subfigure}
     \hfill
     \begin{subfigure}{0.35\textwidth}
         \centering
         \includegraphics[width=\textwidth]{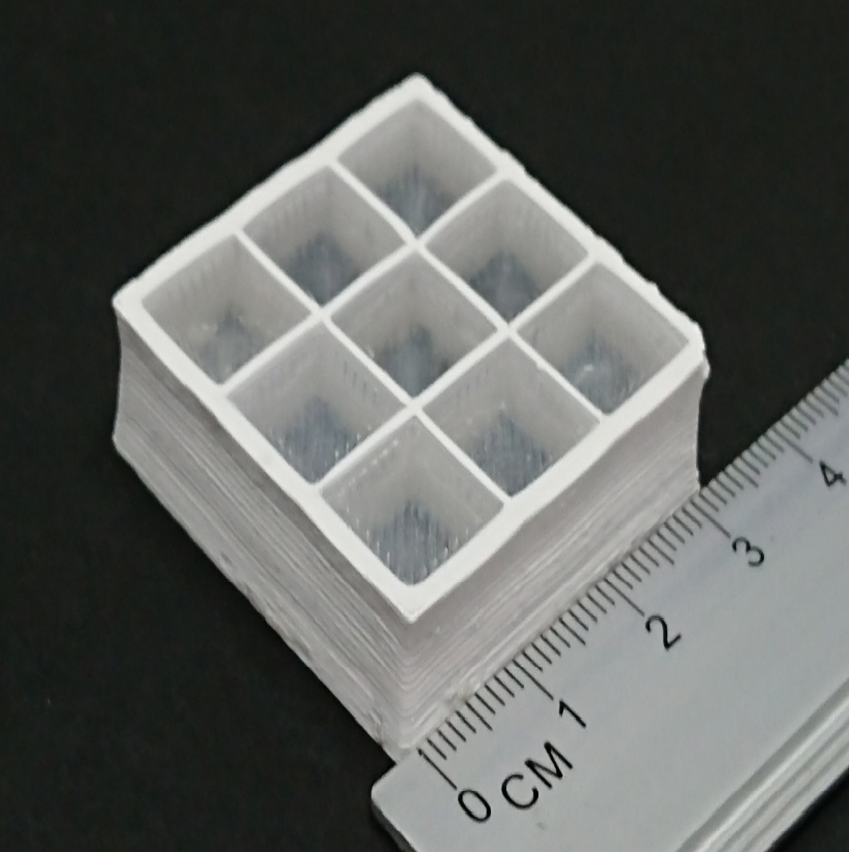}
         \caption{An array of scintillating pixels produced by the 3DET collaboration .}
         \label{fig:3DET_sample}
     \end{subfigure}
     \hfill
        \caption{Samples of the additively manufactured scintillator.}
        \label{fig:AM_Scintillators}
\end{figure}

\section{Advanced Photon Sensors and Collectors}

\subsection{LAPPDs}
\label{sec:LAPPD}

For decades, the high energy physics (HEP) community has relied on photomultiplier tubes (PMTs) to provide low cost, large-area coverage for a wide variety of detector systems. Increasingly, the demands of HEP experiments are pushing for new imaging capabilities, combined with temporal resolutions far better than PMTs can typically offer. 

Large Area Picosecond Photodetectors (LAPPDs) are commercially available 20 cm x 20 cm microchannel plate photomultiplier tubes (MCP-PMTs)~\cite{Wiza:1979iia} now in use by the neutrino community and capable of millimeter-scale spatial resolutions, tens of picosecond sPE time resolutions, and gains exceeding $10^6$~\cite{LAPPDtiming}. They bring much needed timing and imaging capabilities to a wide range of applications in fields such as particle physics, nuclear physics, X-ray science, and medical imaging. They also hold great potential in light-based neutrino detection, particularly in combination with imaging optics and dual scintillation-Cherenkov detectors~\cite{Aberle:2013jba}. 

\begin{figure}[ht!]
\centering
\includegraphics[width=0.4\textwidth]{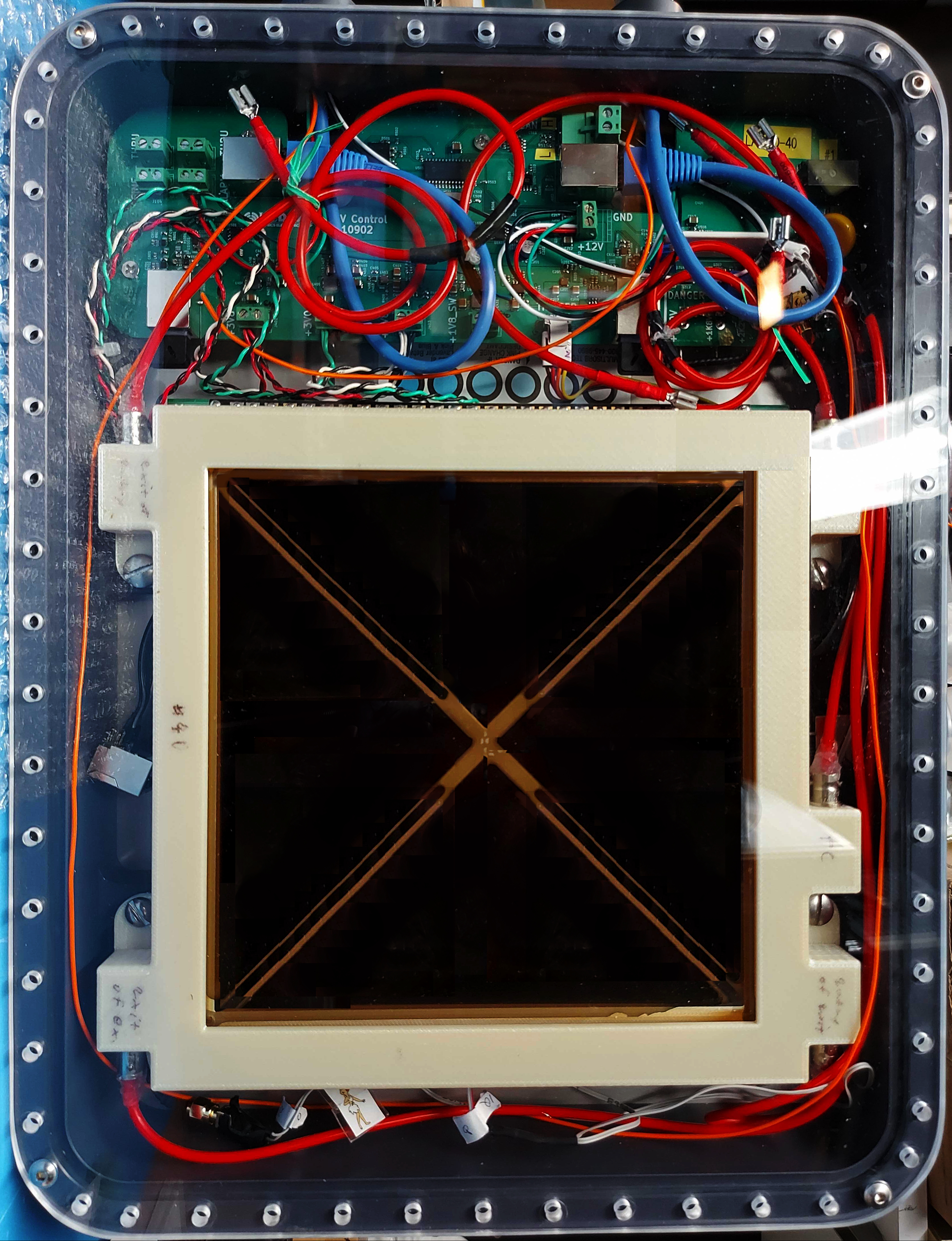}
\includegraphics[width=0.4\textwidth]{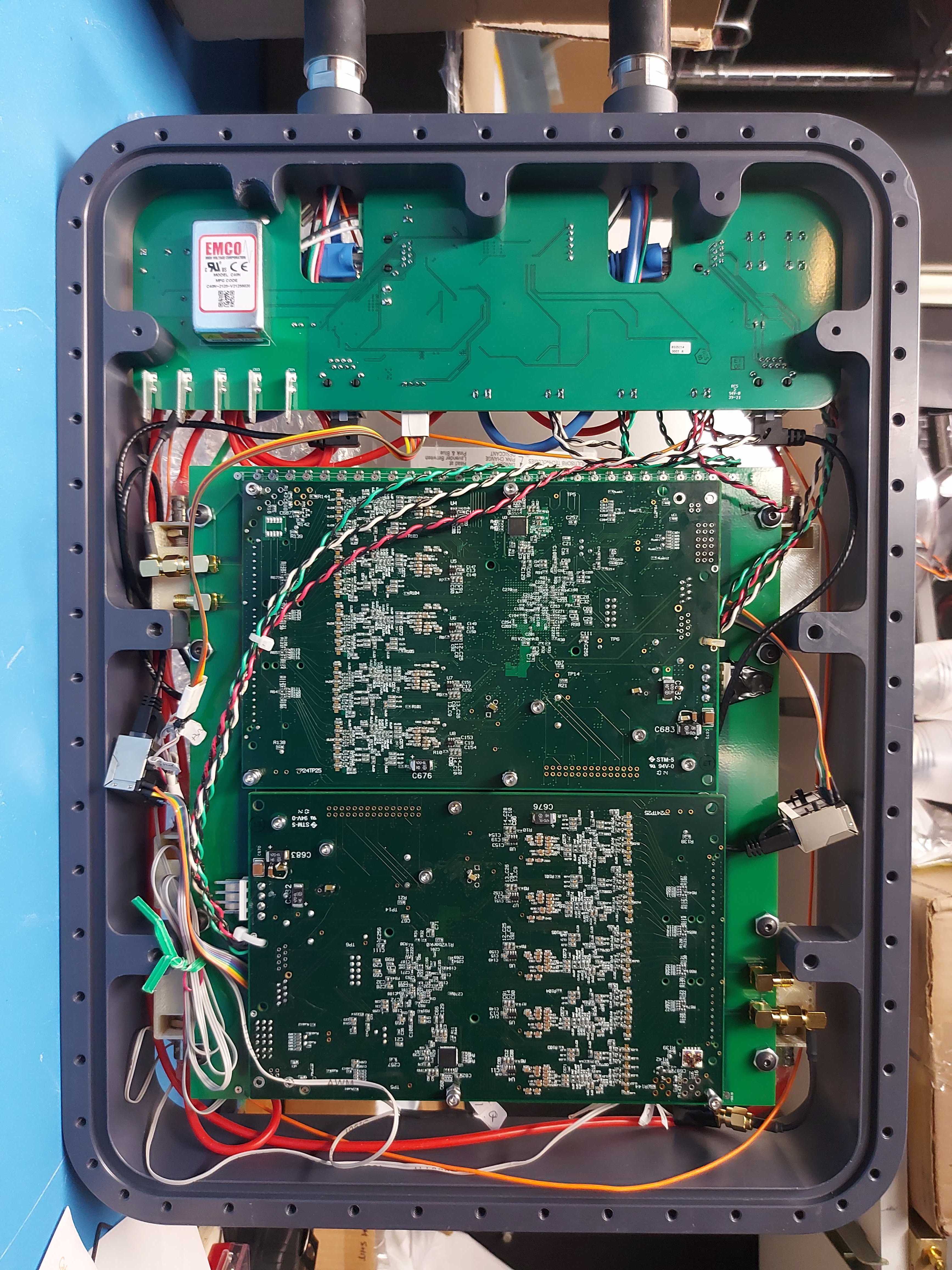}
\caption{Left: Front view of an ANNIE LAPPD Module. Right: The back of an LAPPD module with the back plate off.}
\label{fig:LAPPDmodule}
\end{figure}

As imaging photodetectors, capable of resolving correlations between position and timing in a single module, LAPPDs are fundamentally different from conventional photomultipliers. LAPPDs can be used in combination with an wide variety of imaging optics such as plenoptic lens arrays and novel mirror optics to expand area coverage, sort photons by color and polarization, and even recover directional information. The combination of LAPPDs with novel imaging optics has the potential to free future water-based neutrino detectors from typical design constraints, and break away from the idea of the neutrino detector as a simple cylindrical (or spherical) bounding volume with phototubes on the surface. 

\begin{figure}[ht!]
\centering
\includegraphics[width=0.5\textwidth]{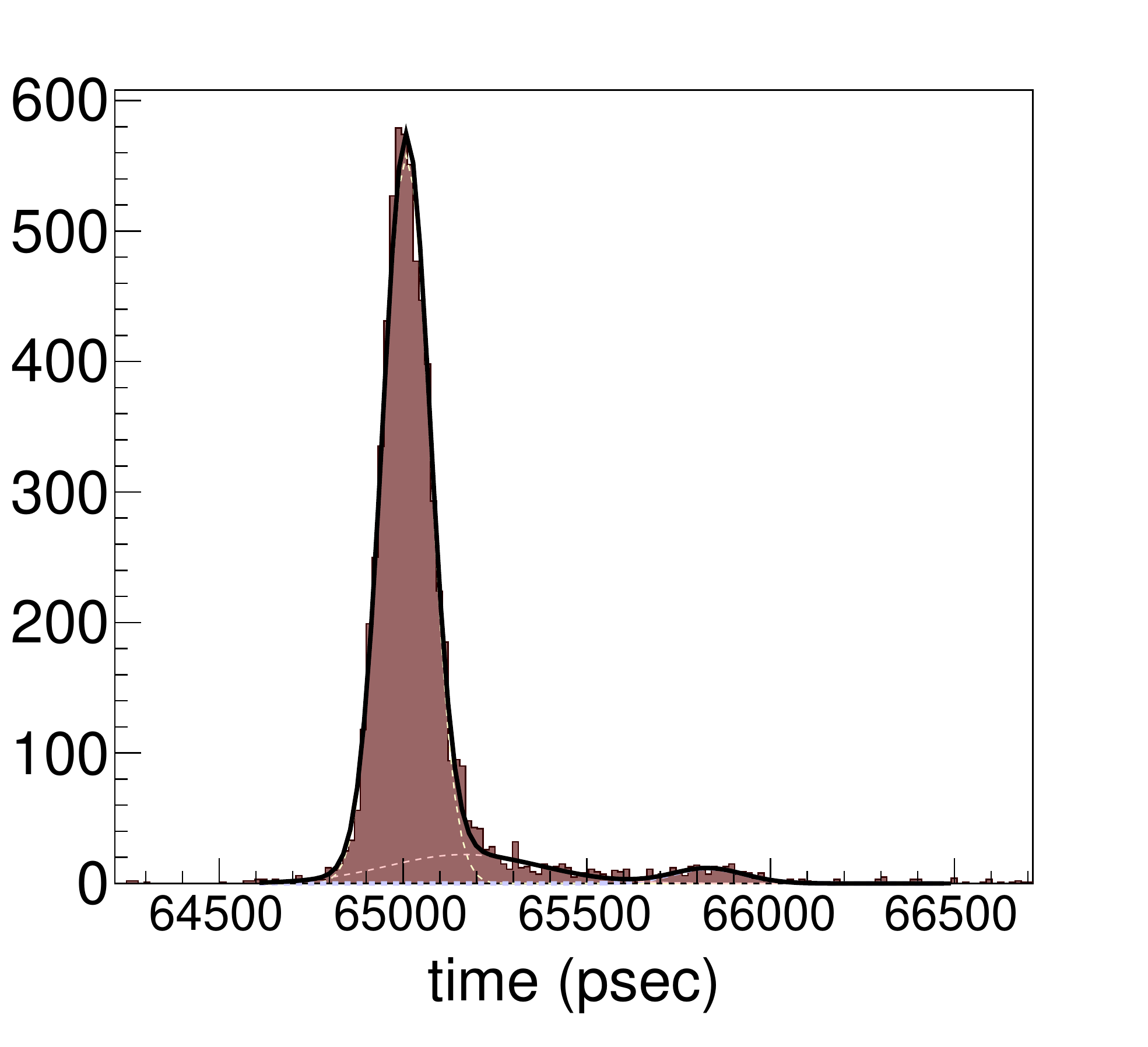}
\caption{The Transit Time Spread (TTS distribution) for ANNIE LAPPD-25. The time resolution for the prompt pulses is below 70ps, where roughly 30ps of smearing is due to the pulse duration of the diode laser used.}
\label{fig:LAPPDTTS}
\end{figure}

The combination of spatial and temporal information make LAPPDs ideal for Scintillation-Cherenkov separation. The $<$100 psec resolution of LAPPDs makes it possible to separate between the two components on the basis of timing alone. However, it is also possible to sort different colors and different polarization of light to different areas of a single LAPPD surface.

The capabilities of LAPPDs not only change how they can be used in designing future neutrino detectors but also places new requirements on Monte Carlo and reconstruction tools. The major existing Monte Carlo packages are hard-coded to assume single-pixel optical readouts. Many available reconstruction tools factorize the charge and timing likelihoods from the pattern-of-light when fitting for track hypotheses. Future development in reconstruction of light patterns should be broadened to enable simultaneous fitting of the positions and timing of the light patterns.

The precision timing capabilities of LAPPDs make possible new stroboscopic techniques that where different energy components of a wide-band neutrino beam can be selected based on the arrival of the neutrinos relative to the beam RF timing. Future neutrino expeirments, including possible upgrades to LBNF could enable sufficiently short proton bunches to make this technique viable. At the detector end, this technique requires time resolutions on the order of no more than a few hundred picoseconds. Hybrid Cherenkov-scintillation detectors with LAPPDs are naturally well suited for this level of vertex precision. 

In order to fully exploit the characteristics of LAPPDs, present and future experiments require waterproof modules containing low-cost, high-channel-count readout electronics to digitize signals close to the sensor. The leap from an individual LAPPD to a system of highly synchronized and deploy-able LAPPD modules is a massive undertaking in its own right. The ANNIE collaboration has successfully developed a first realization of such a system (see Fig~\ref{fig:LAPPDmodule}). In the coming decade, work to continue to streamline and improve on LAPPD modules is critical to their readiness for larger scale deployments. Next steps include fiberoptic communications with the surface electronics, improved self-triggering capabilities, and better optical coupling between the LAPPD and the window of the waterproof module.  

While LAPPD technology has not yet achieved an economy of scale in terms of price and yields, the HEP community can help play a significant role in continuing to bring down the cost of this technology. R\&D efforts to streamline the fabrication process would make significant headway in making larger scale production lines possible. One possible technique involves sealing top window of the photosensor before making the photocathode and introducing the multi-alkalai vapors through tubes that can later be sealed off. Above all, continued use and testing of LAPPDs in the small- and medium-scale detectors will help support this technology through the final stages of tech transfer process. In the longer run, possible developments in medical imaging and homeland security could provide large and consistent markets for this technology outside of HEP.

\subsection{New PMT Technologies}

    While photomultiplier technologies themselves are quite old, there have been many advances in recent years that will significantly improve the performance of future detectors.  High-quantum efficiencies photocathodes---with peak QEs in the range of 35\% or more---have been successfully built into large PMTs (10'' and more) by both ETL~\cite{etl} and Hamamatsu~\cite{hamm}.  Precision timing (transit time-spreads in the regime of $\sigma<1$~ns) in large-scale (8'') PMTs has also been developed~\cite{tanner}, which can help with reconstruction and with Cherenkov/scintillation separation~\cite{dichroicon1}.  In addition, extending the quantum efficiencies into the long wavelength ``red'' region, up to 700~nm, has also been successfully developed, which can be particularly useful as the long-wavelength sensor in a dichroicon (see below).
    

\subsection{Dichroicons}

Typically photon-based neutrino detectors record no more than the number
of detected photons and their arrival times. Photons may also carry
information about physics events by the direction they travel, in 
the orientation of their polarization, and their wavelength.

Next-generation neutrino detectors such as Theia plan to detect neutrino interactions via photons produced in a scintillating target medium. To ensure high energy resolution, scintillators with high light yields are preferred, as the energy resolution is typically limited by photon statistics. Consequently, the number Cherenkov photons produced in these target media is typically several orders of magnitude fewer than the scintillation. This disparity in intensity makes Cherenkov photons 
difficult to identify, however they carry information not found in the scintillation signal, including the direction and type of charged particle that produced the light.

Direction is encoded in the angular distribution of the Cherenkov photons, which are created at a constant angle with respect to the momentum of the producing particle. Reconstructing direction would allow for directional rejection of solar neutrinos or other directional sources in future neutrino programs, as well as an enhanced ability to study the directional solar neutrinos. Particle type can be inferred with the relative amount of Cherenkov and scintillation light, as alpha particles, for instance, typically produce no Cherenkov photons while still producing a bright scintillation signal. Tagging these alphas with their lack of Cherenkov photons would be beneficial in reducing radioactive backgrounds, particularly in the energy region of interest for neutrinoless double beta decay searches. One approach to separating Cherenkov and scintillation light is by discriminating photons by wavelength, as scintillation is typically within a narrow emission band, while Cherenkov is a broad spectrum of light, falling as roughly $1/\lambda^2$. 



%
The dichroicon is a device which performs such spectral sorting, by separating the photons into long wavelength PMTs (mostly Cherenkov) and short wavelength PMTs (mostly scintillation) using dichroic filters.
As shown in Figure~\ref{fig:dichroicon} below, the dichroicon follows the off-axis parabolic
\begin{figure}[ht!]
\centering
\includegraphics[width=0.35\textwidth]{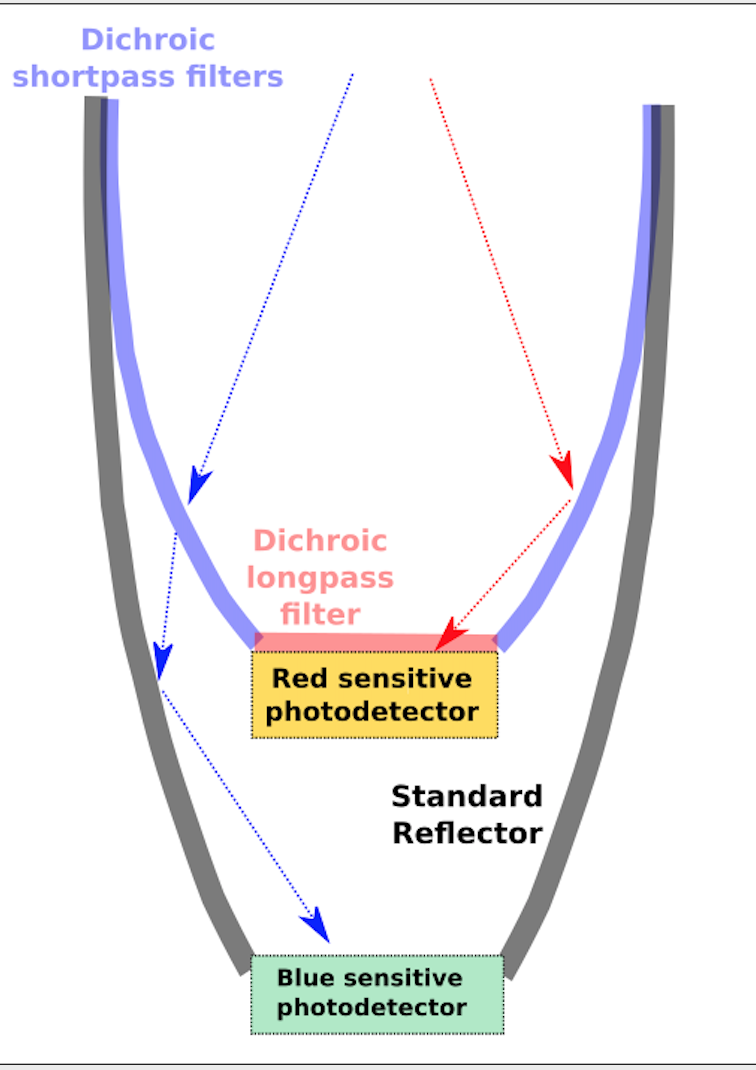}
\includegraphics[width=0.45\textwidth]{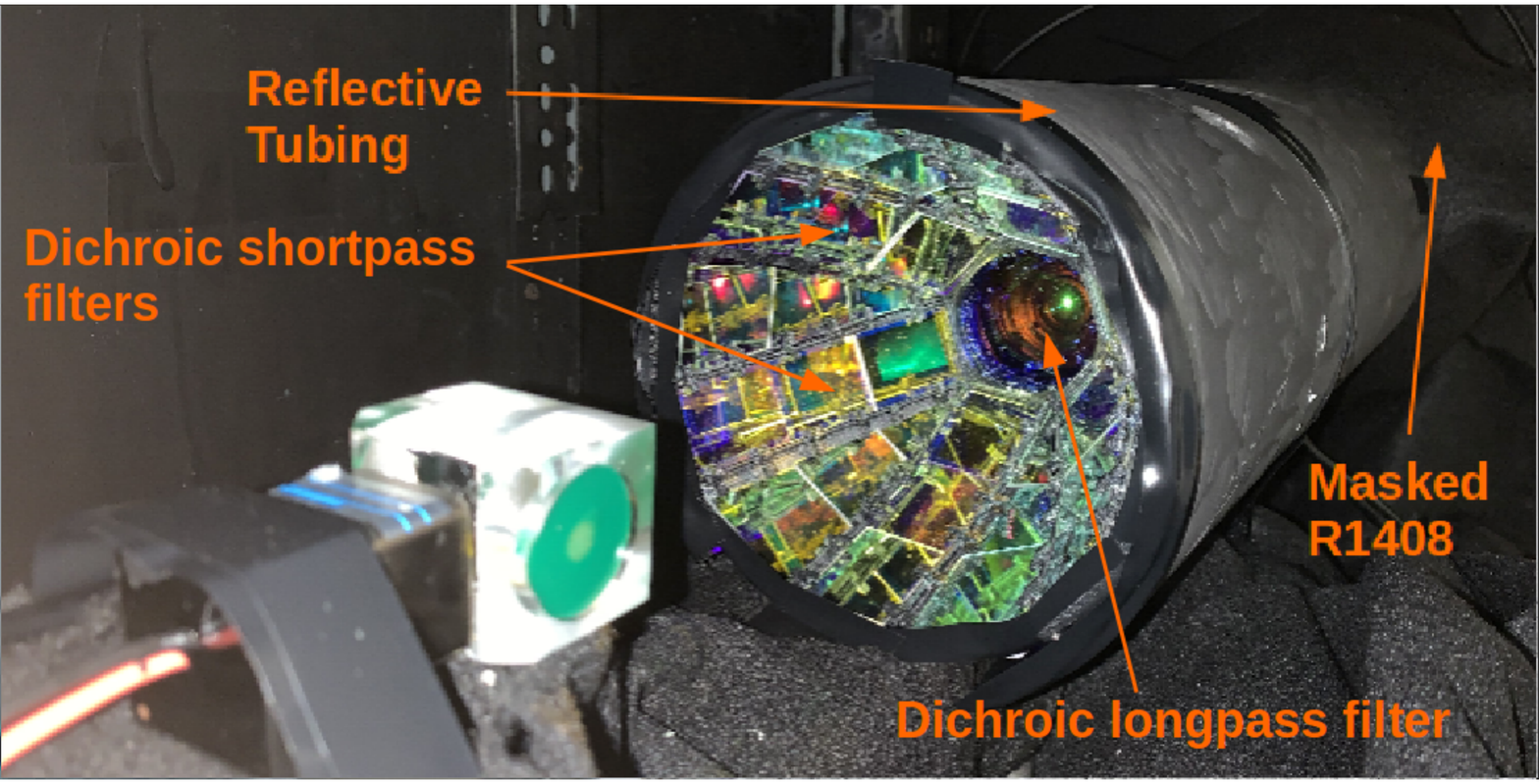}
\caption{Left: Schematic of a “nested” dichroicon configuration, with shortpass filters on the barrel and a longpass filter at the aperture. Right: Dichroicon in benchtop test setup, viewing $^{90}$Sr source embedded in an acrylic block containing LAB-PPO.}
\label{fig:dichroicon}
\end{figure}
design of an ideal Winston light concentrator but is built as a tiled set of
dichroic filters. The filters are used to direct long-wavelength light towards
a central red-sensitive PMT, while transmitting the shorter wavelength light
through the ``barrel'' of the Winston cone to secondary photodetectors. This is
possible because of the remarkable property of the dichroic reflectors, which
reflect one passband of light (below or above a ‘cut-on’ wavelength) while
transmitting its complement, with very little absorption. As shown
schematically in one possible design in Figure~\ref{fig:dichroicon}, the
barrel of the dichroicon is built from shortpass dichroic filters (cutoff
wavelength near 480~nm) and a longpass dichroic filter is placed at the
aperture of the dichroicon. The shortpass filter passes short-wavelength light
while reflecting long-wavelength light; the longpass has the complementary
response. In the ``nested'' design shown, the back PMT detects the short
wavelength light.

Tests using a low-energy ($^{90}$Sr) radioactive source in high light-yield
LAB-PPO scintillator show
extremely good ``purity'' (over 90\%) for the dichroicon’s ability to identify
Cherenkov light as distinct from scintillation light. The left panel of
Fig.~\ref{fig:cherscint}
\begin{figure}[ht!]
\centering
\includegraphics[width=0.33\textwidth]{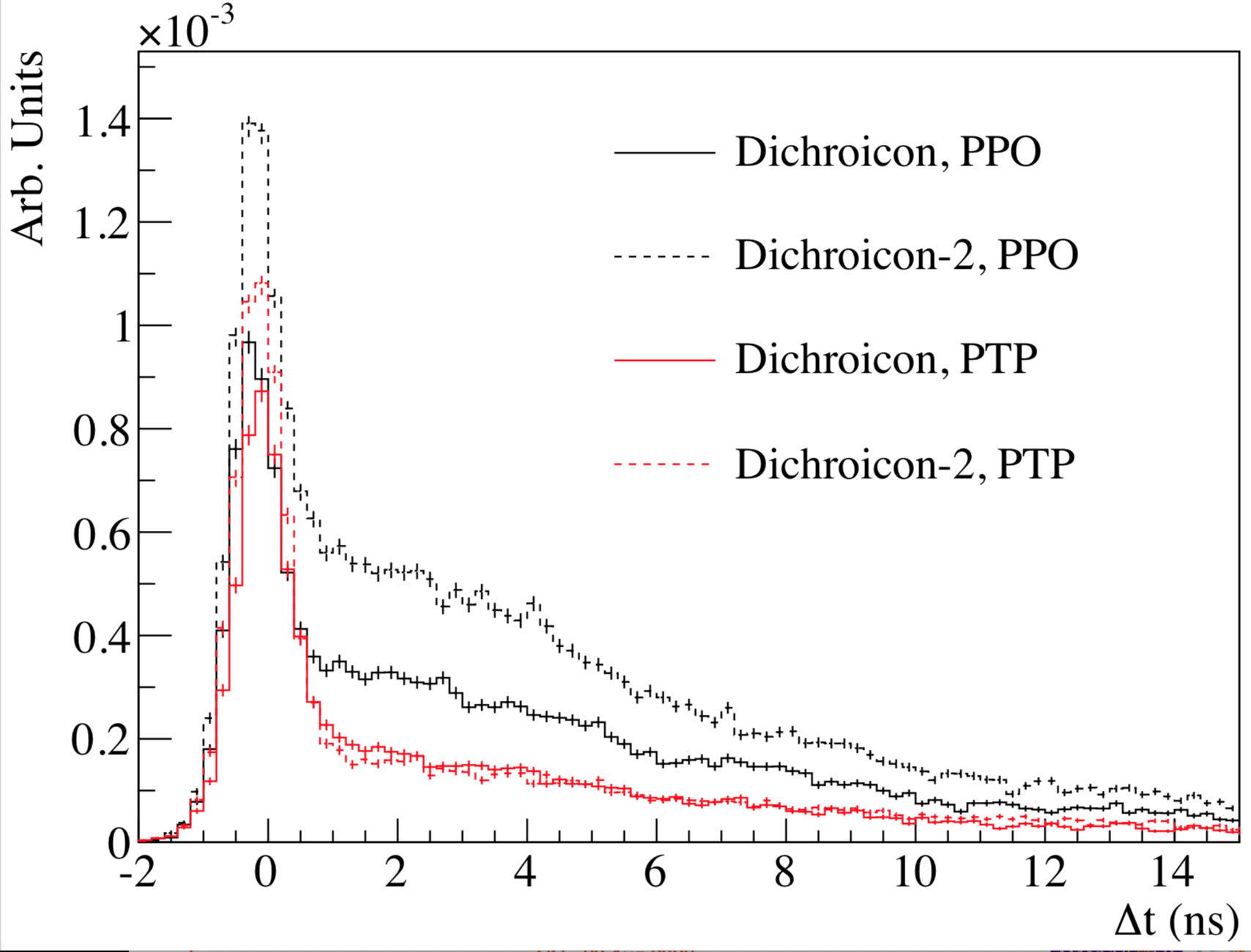}
\includegraphics[width=0.33\textwidth]{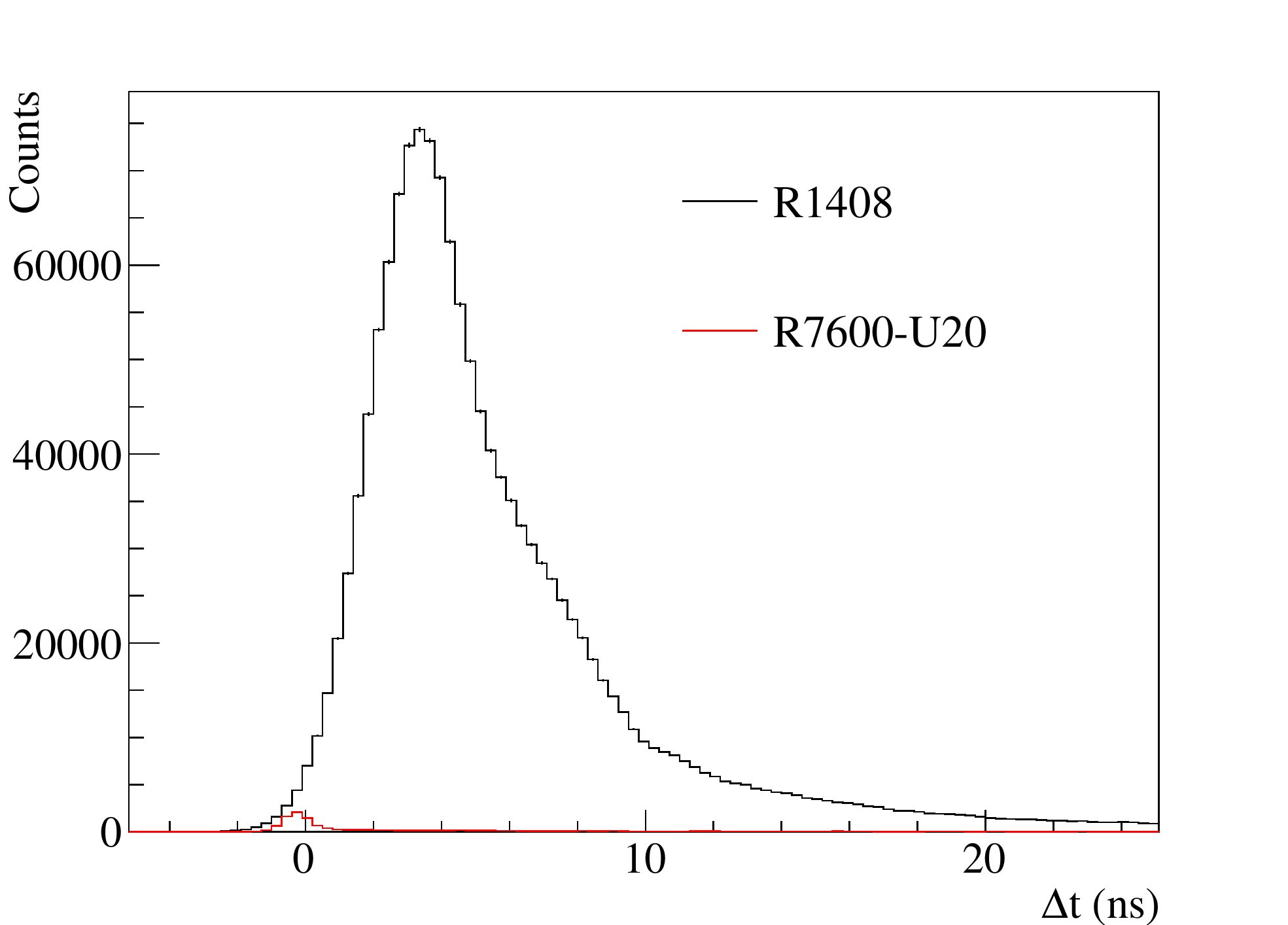}
\includegraphics[width=0.33\textwidth]{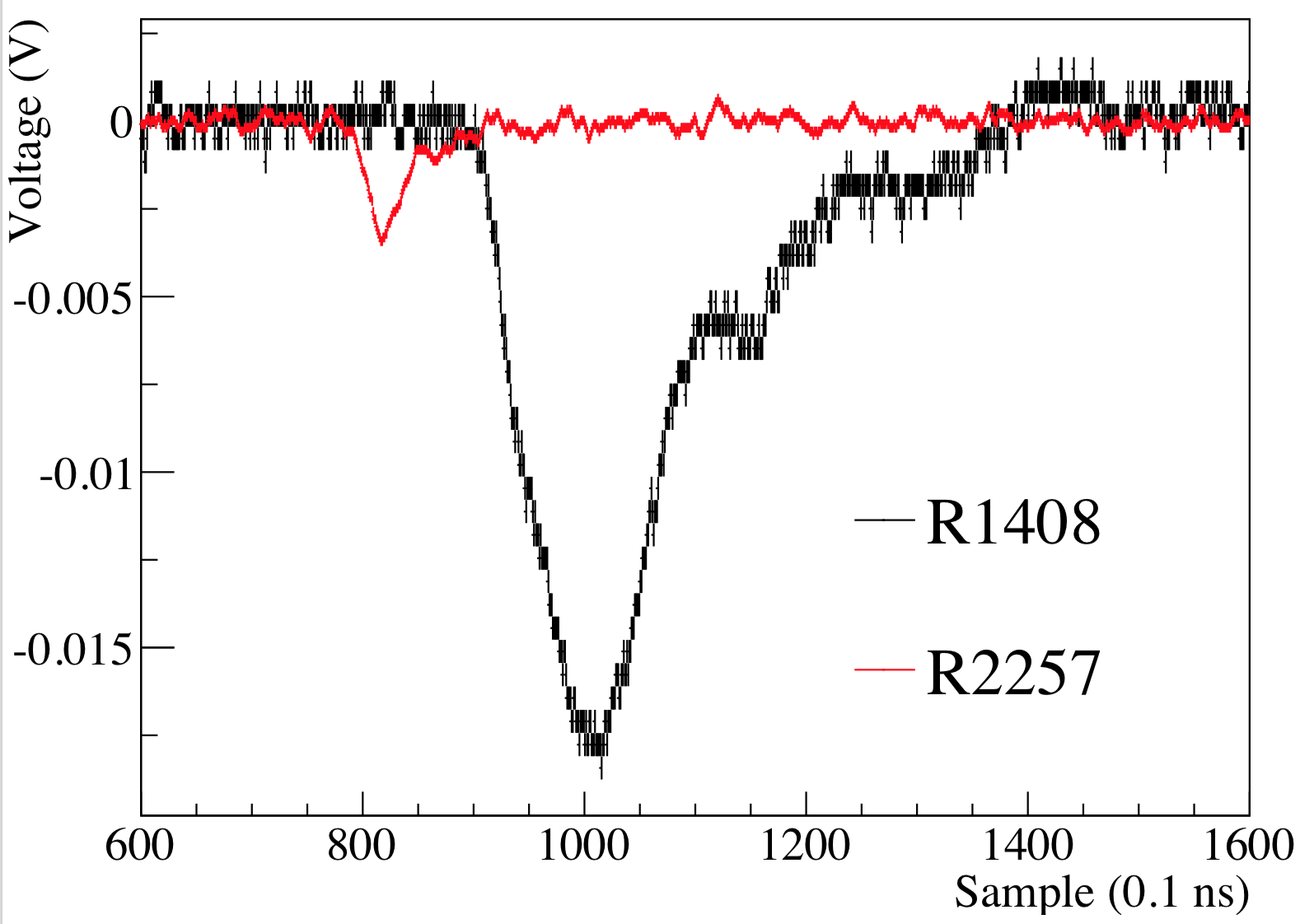}
\caption{Left: Time profile of light that is reflected by dichroicon through
aperture, dominated by Cherenkov light. The plot shows two different dichroicon
models, for two different secondary fluors (PPO and PTP) in LAB. The prompt
Cherenkov peak can be seen, with some leakage of slower scintillation light.
Middle: Timing profile for scintillation light (black curve) compared to
Cherenkov light (red curve), normalized by the total number of events. Right:
Coincidence between a Cherenkov photon (red trace) measured by aperture PMT
(R2257), with multiple scintillation photons detected by back (shortpass) PMT
(R1408).}
\label{fig:cherscint}
\end{figure}
shows the prompt Cherenkov peak seen in the long-wavelength (aperture)
dichroicon peak, and the middle panel compares the timing of the photons
observed in the back (``scintillation'') PMT (R1408) to the timing of photons in
the aperture (``Cherenkov'') PMT, normalized by the number of events and thus
representing the observed ratio of short-wavelength scintillation light to
long-wavelength Cherenkov light. The right panel shows a single coincidence
between a Cherenkov photon in the long-wavelength PMT and the multi-photons in
the scintillation PMT.

There are many possible configurations of the dichroicon; the ones built to
date are not necessarily optimal, and different detectors may have different
needs. The nested photon sensor configuration of the design above requires
more than one photon sensor and is thus most useful when the available
detection area is limited (for example, when the desired coverage is $>50$\%, or
in a segmented detector where each segment is viewed by a single sensor).
Simpler designs could simply offset the dichroicon and its aperture PMT,
collecting the low-yield Cherenkov photons while allowing the scintillation
light to be detected by the rest of the PMT array. Using a pixelated photon
sensor, such as an LAPPD or an array of SiPMs, would also work, with the pixels
then mapping to different wavelength bands. A complementary design—--with
Cherenkov light passing through the barrel and scintillation light reflected
toward the aperture—--might be most useful when ring imaging is a high
priority.  To achieve more than two passbands, the ``nested'' design could be
extended using mutliple dichroicons.

It is clear that with the
dichroicon, a truly hybrid Cherenkov/scintillation detector can be built
without the kind of compromises on scintillation light yield that come with
``lean'' scintillators, or the loss of timing that are associated with slow
fluors. Nevertheless, using dichroicons in combination with these techniques is also possible, and presents a wide variety of potential applications.

A simulation model of the dichroicon has been developed in {\it Chroma} (See Section~\ref{sec:chroma}) and calibrated against the measurements of a benchtop prototype, for use in simulating dichroicon performance large-scale detectors. This model was simplified and scaled up to use a 20" diameter large area PMT to collect photons passed by the short-pass dichroic filters, and a cylindrical 5” PMT to collect the long wavelength photons reflected by the dichroic filters shown at the top of Fig.~\ref{fig:chromadich}. This dichroicon unit was then tiled around a cylindrical volume to simulate a large-scale neutrino detector, seen from the inside at the bottom of Fig.~\ref{fig:chromadich}. In this model a 50 kt volume of the scintillator LAB with 2g/L of PPO [10] is surrounded by 90\% 
\begin{figure}[ht!]
\centering
\includegraphics[width=0.35\textwidth]{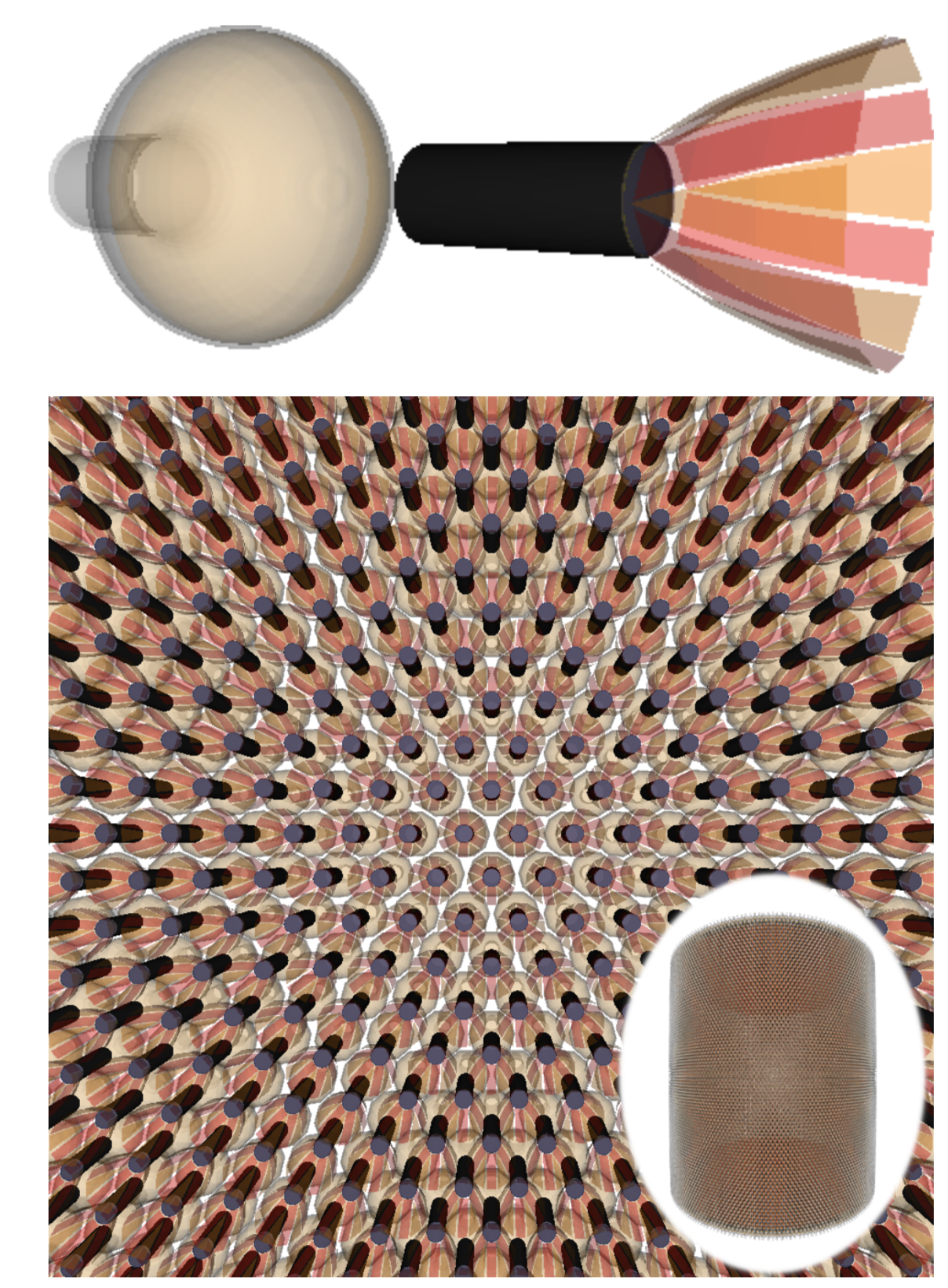}
\caption{A rendering of the {\it Chroma} geometry for (top)  a dichroicon unit with (left to right) 20” large area PMT, cylindrical 5” long-wavelength PMT, and dirchroic filter con- centrator, and (bottom) the dichroicon units tiled to create a 50 kt neutrino detector, viewed from inside the detector, with the full cylindrical geometry shown in the inset.}
\label{fig:chromadich}
\end{figure}
coverage of simplified dichroicons, which gives effectively 90\% coverage of both long and short wavelength photons, as the dichroic filters allow the long and short PMTs share the same solid angle. Using this model, we have begun to evaluate the impact of spectral photon sorting on future neutrino experiments.

A reconstruction algorithm developed in~\cite{wblsberk} has been modified such that it uses hit time information from all (short and long wavelength) PMTs to perform a position and time fit, and then uses the angular distribution of photons detected on long wavelength PMTs to perform a direction fit. An angular distribution of long-wavelength PMTs that detected photons from 2.6 MeV electrons generated at the center is shown in Fig~\ref{fig:dichcher}, which clearly shows the Cherenkov topology. 
\begin{figure}[ht!]
\centering
\includegraphics[width=0.35\textwidth]{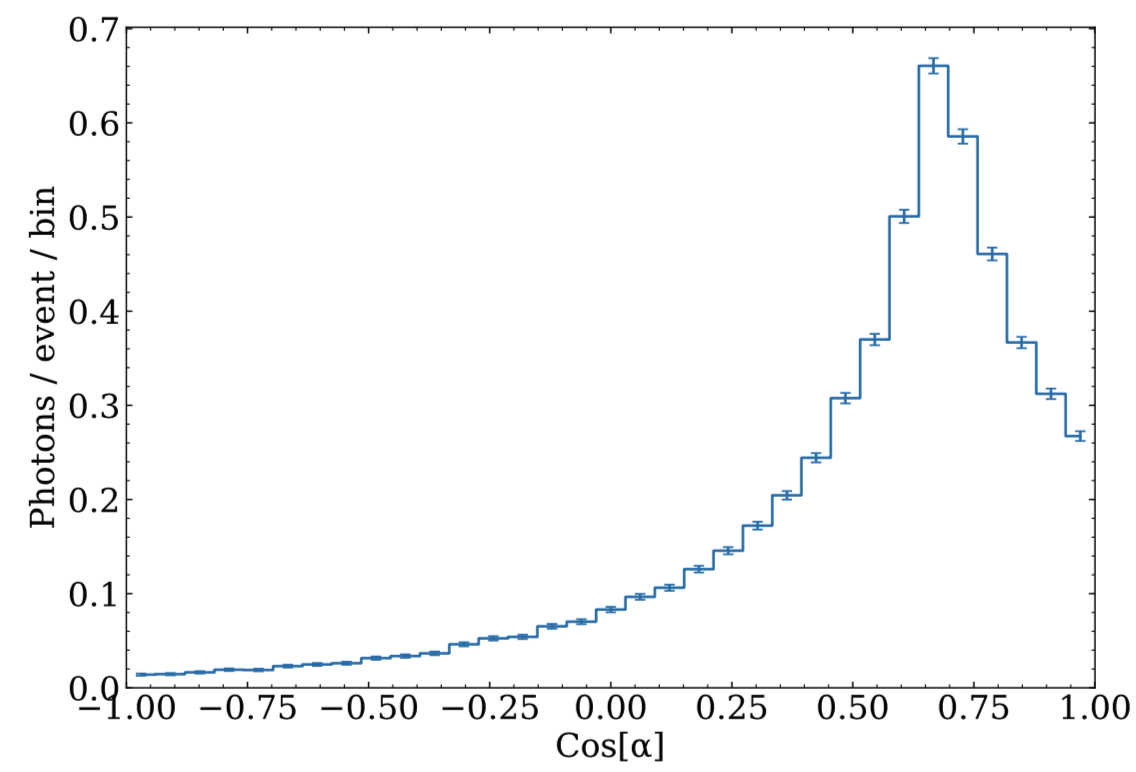}
\caption{Distribution of the cosine of the angle between photons detected by long-wavelength PMTs and the initial direction of a 5 MeV electron simulated at the center of the detector geometry. The Cherenkov topology is clearly visible.}
\label{fig:dichcher}
\end{figure}

Alpha particle identification has been explored by simulating alpha and beta particles with the same quenched energy (number of scintillation photons) and inspecting the signal on the long-wavelength PMTs, which are pri- marily sensitive to Cherenkov photons. A quenched energy comparable to neutrinoless double beta decay in 130Te is chosen to highlight potential background rejection capabilities. The mean number of long-wavelength hits is indeed higher for betas than alphas as shown in Fig.~\ref{fig:dichpid} due to Cherenkov production with betas. The detected long wavelength photons from alphas are all scintillation, and indicate that the filters could be optimized for better rejection of scintillation. Fig.~\ref{fig:dichtime} shows a clear Cherenkov signal in the hit time residuals of betas at early times, which would allow for discrimination of alphas in a neutrinoless double beta decay region of interest. A similar method provides some discrimination between single and double beta events, as can be seen in the $^{130}$Te 0$\nu\beta\beta$ plots in Figs.~\ref{fig:dichpid} and~\ref{fig:dichtime}, which could further constrain backgrounds.
\begin{figure}[ht!]
\centering
\includegraphics[width=0.35\textwidth]{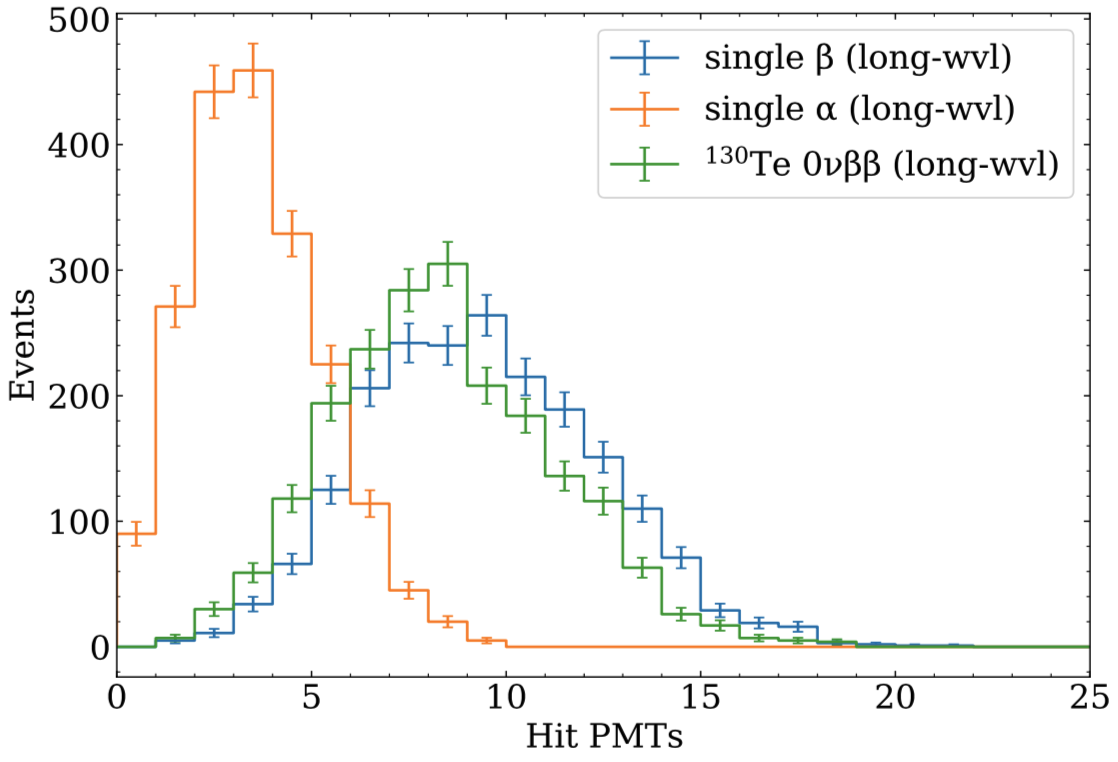}
\caption{Number of long-wavelength PMTs that detected photons for alpha and beta events with the same quenched energy as $^{130}$Te neutrinoless double beta decay, showing alpha/beta discrimination based on Cherenkov photon identification in a 50 kt liquid scintillator detector.}
\label{fig:dichpid}
\end{figure}
\begin{figure}[ht!]
\centering
\includegraphics[width=0.35\textwidth]{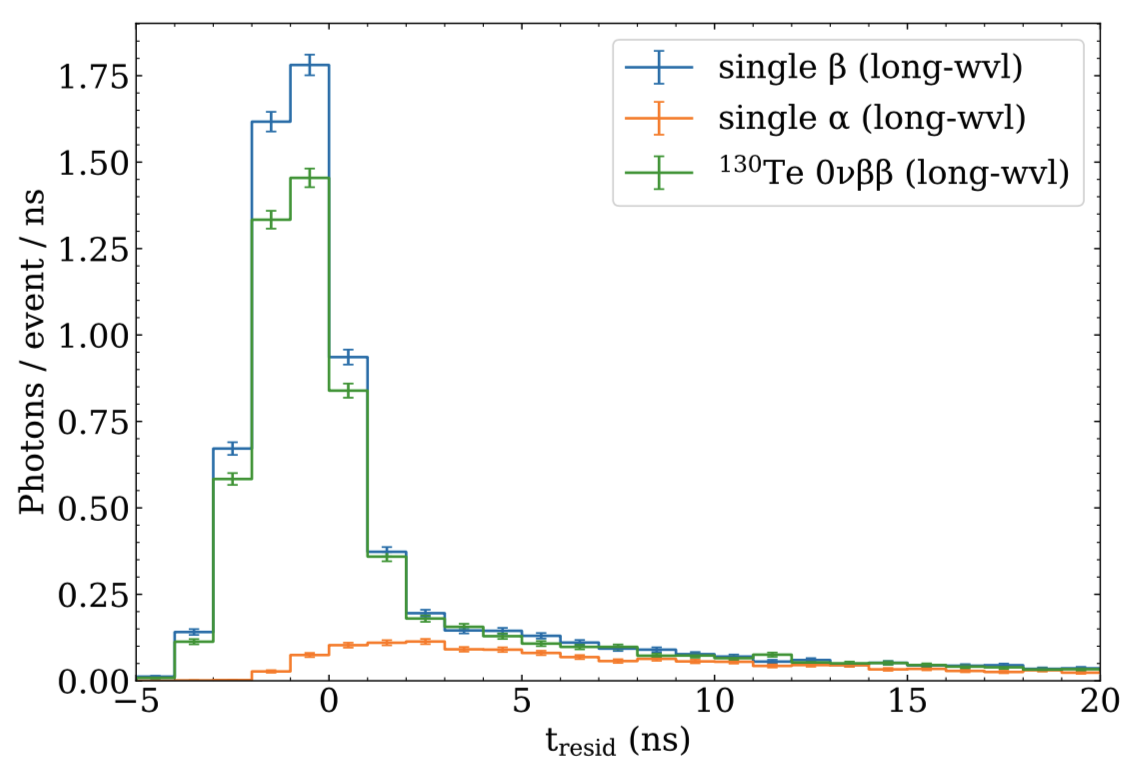}
\caption{The hit time residuals of long-wavelength PMTs for the events shown in Fig.~\ref{fig:dichpid}. The Cherenkov is clearly visible at early times for the beta events.}
\label{fig:dichtime}
\end{figure}
The developed simulation already indicates that dichroicons and spectral sorting will offer powerful background rejection in future large-scale neutrino detectors, and there are still many potential improvements to these techniques to explore. Chroma offers a flexible path forward for testing different configurations of the dichroicon concept, including other readout methods and geometries. These paths are being explored and incorporated into a robust simulation that can be used to evaluate the sensitivity of such a detector to future neutrino programs.

\subsection{ARAPUCAs}
    
        ARAPUCAs are light traps that use a combination of wavelength shifters and dichroic mirrors to trap photons from LAr scintillation.  These scintillation photons sit well into the ultraviolet, at 128~nm, and are wavelength-shifted by TPB deposited on a dichroic mirrors, which passes photons in the TPB emission spectrum. On the back side of the dichroic mirrors is PTB wavelength shifter which shifts the photons again but now into a regime in which they are reflected by the dichroic mirror.  Thus the photons are trapped, and after many bounces they can  be detected by even a small-aperture SiPM.  The performance of ARAPUCAs can be found in Ref.~\cite{dunearapucas}.

\subsection{Distributed Imaging}

    While dichroicons and ARAPUCAs leverage new information in the photons---wavelength---to expand the capabilities of photon-based detectors,  distributed imaging leverages the {\it direction} of photons to reconstruct events and reject backgrounds---for example, rejecting $\gamma$ rays while keeping electrons.  Distributed imaging requires the use of lenses, which must be carefully design but if done right can produce dramatic results.  Further details can be found in Ref.~\cite{gratta_image}.



\section{Future Approaches to Readout, Instrumentation, and DAQ}

The readout system for large photon-based neutrino experiments is largely a function of the photomultiplier tube that forms its basic detection unit.  Modern PMTs have transit time spreads (TTSs) of order 1~ns.  To take full advantage of such small TTSs, readout systems are designed to have timing resolutions that are a fraction of a nanosecond, in order to contribute a negligible amount to the detector-wide timing resolution. These readout systems often must be self-triggering and have sufficiently short deadtimes to accommodate background rates from PMT dark noise and ambient radioactivity, and to be maximally sensitive to ``bursty'' signals such as those from inverse beta decay interactions or galactic supernova explosions.

Custom electronics often form the backbone of neutrino detector readout systems, and range from sophisticated custom ASICs to high performance commercial integrated circuits, using complex custom printed circuit boards (PCBs).  Nowadays, turnkey commercial systems are only financially viable below $\mathcal{O}(100)$ channels; systems using commercial off-the-shelf (COTS) parts on custom PCBs do not require the specialized engineering needed for custom ASICs and are cost-effective below $\mathcal{O}(5\mathrm{K})$ channels; beyond $\mathcal{O}(5\mathrm{K})$ channels, custom ASICs are often the best choice.  Over time, these transition points usually move to higher channel counts because the cost of COTS parts generally decreases with time, while the cost of specialized engineering generally increases.

Systems using COTS parts on custom PCBs typically perform full waveform digitization at 250--500~MHz.  These systems are intrinsically deadtimeless, and event triggers are formed downstream of the digitizers, either in custom electronics using fast field-programmable gate arrays (FPGAs), crate-resident single-board computers, or in  online CPU/GPU farms.  Using modest PMT signal shaping and template waveform fitting, lab-based measurements show that 250~MHz ADC systems can attain $\mathcal{O}(100)$~ps single-photoelectron timing, a negligible fraction of the native PMT timing resolution.  Systems with full waveform digitization permit a wide array of signal processing options that can be updated as needed after deployment.  With COTS ADCs generally getting faster and cheaper with time, the future performance of these systems will likewise improve.

Custom ASICs with switched-capactor arrays (SCAs) can run with sampling rates an order of magnitude higher than today's cost-effective COTS ADCs.  The timing and photon-pair resolutions of such systems are superior to those using slower COTS ADCs, but SCA systems generally suffer from long readout times that can create significant per-channel and detector-wide deadtimes.  Custom ASICs that instead extract pertinent PMT pulse quantities like leading-edge time, integrated charge, and time-over-threshold, without digitization, can provide excellent performance but the intrinsically lower information content can make it more challenging to debug unanticipated signal distortions and/or noise, especially for remote systems where direct access to the signal chain is difficult or impossible.  The per-channel footprint of custom ASICs is smaller than COTS ADC systems, allowing for smaller readout electronics PCBs, although typically the high voltage cables drive the final form factor.

The FPGA is now ubiquitous in these readout systems, with its firmware handling data flow, communications and extracting relevant time and charge information in real time from PMT pulses, often in conjunction with small on-board CPUs or systems-on-a-chip (SoCs).  As these components become cheaper and more powerful, more of the readout system functionality can be moved logically and physically closer to the PMT itself, bringing attendant cost and signal quality benefits.  In the future, fully functional single-channel readout systems could be placed in or very near the PMT base, with low-voltage and data connections to the downstream readout system carried on thin cables, avoiding the signal dispersion, expense, large mass and general unwieldiness of traditional high voltage cables and connectors. 

Although the advances in FPGAs and ADCs has been impressive over the past decade or more, large detectors that use photons to detect neutrino interactions typically see only a small number of photons on each channel.  The pulses from each of the photoelectrons---typically either from a PMT or a SiPM---are well understood and measured, and need only two parameters to describe them: amplitude, and time-of-arrival.  And for any particular waveform from an individual photon sensor, the only parameters of interest are the total number of photons that were detected, and their times-of-arrival.  Therefore analysis of signals from photon-based detectors can be done precisely with analog techniques: measurements of peak times, time-over-threshold, integral, or even inflection points.  An {\it analog processor} can then provide all the useful information from a photon sensor without the need for expensive ADCs or FPGAs and their associated firmware, and can result in much smaller data sets that are easier to analyze, without the loss of any important information.  Analog processing can also result in more precise measurements, as the bandwidth of the signals can be larger without worrying about Nyquist limitations. The cost of an analog processor for photon detectors (an ``Analog Photon Processor'' or APP) might be as small as \$5/channel, making the use of these in very large scale detectors very reasonable.

\section{New Isotopic Loading Techniques}
The challenge in the development of metal loaded liquid scintillators is the addition of the inorganic metallic compounds, typically in the form of salt, to the organic scintillator solvents. It is not trivial to find a chemical complex of the metal, which dissolves in the nonpolar organic liquid scintillator without deteriorating the performance of the liquid. A simpler approach is to add the metal into aqueous solutions which can be mixed into a LS directly. This is one of the reasons making Water-based Liquid Scintillators attractive.

There are several different approaches dissolving the metal in an organic liquid. One possibility is to search for a solvent with high solubility for inorganic salt compounds as alcohols. In the scintillator of the Chooz experiment the Gd(NO)3 salt was first dissolved in hexanol. The mixture was then diluted in a mix of organic solvents. The case of the Chooz scintillator demonstrated the general difficulty in metal loaded LS production for large-scale particle physics experiments. The degradation of the attenuation length in the final liquid limited the lifetime of the Chooz detector to about 1 year. 

The most promising procedure for metal loading is probably the preparation of an organometallic complex, which is soluble in the LS. The most common choice is the use of metal carboxylates for this purpose. Among the carboxylic acids there are many different candidates and several of them were tested in neutrino experiments. Reines and Cowan were using cadmium octoate (2-ethylhexanoic acid) in their Savannah River neutrino experiment in the 1950s. The same formula of metal 2-ethylhexanoate was used again after several decades, but this time with gadolinium instead of cadmium, in the context of the Palo Verde neutrino experiment. A modest stability, degradation in less extent, of Gd-doped LS was found from Palo Verde operation. Modern successful Gd-carboxylate LS, as used in the Daya Bay and RENO experiments, are using a 9C acid, 3,5,5-trimethylhexanoic acid (TMHA). The metal salts of these acids are insoluble in water, but soluble in hydrocarbons. There were also tests with phosphor-organic compounds, which were used for stabilizing purposes or to achieve solubility. An alternative approach to the carboxylic acid systems is the application of beta-diketones. Such systems were first studied for indium-loaded liquid scintillators with potential use in LENS experiment. The first application in a large-scale neutrino detector was within the Double Chooz experiment.

\subsection{Metal-doped Water-based Liquid Scintillator}
\label{sec:mdwbls}

A completely different approach to dissolve metal in a LS is to utilize the principal of water-based liquid scintillator (WbLS) consuming a mix of surface-active agents with hydrophilic as well as hydrophobic chelating groups to bridge aqueous (polar) and organic solvents (nonpolar). The principal, performance, and characteristics of WbLS are described in a Section IV. This WbLS allows inorganic metal salt to be first dissolved in water, followed by directly blended into any type of scintillator solvents, such as pseudocumene (PC), linear-alkyl-benzene (LAB), di-isopropylnaphthalene (DIPN), 1-phenyl-1-xylyl-ethane (PXE), cyclohexylbenzene (PCH), regardless the chemical property of each scintillator. WbLS has 100\%  metallic extraction efficiency and is particularly operative in extracting the hydrophilic elements (e.g. Boron, Lithium) into organic solvents, which, due to the high electron affinity of molecular hydrophilicity, often present great challenges using conventional organometallic technology, such as carboxylate, phosphate, or diketones ligands. The metal-doped principal for oil-based WbLS ($>$80\% scintillator) has been successfully demonstrated by PROSPECT experiment, a $\sim$0.1\% 6Lithium-doped DIPN-based WbLS, with good light yield ($>$10,000 ph/MeV) and superior pulse shape discrimination s/B~3).

Metal-doped WbLS devises in several applications for nuclear and particle physics experiments and can be fabricated and deployed at the ton scale---an example is shown for development of the Prospect experiment in Fig.~\ref{fig:wblsprospect}.  Development to further advance its performance and implication continues at BNL and other institutes (e.g., NIST, LLNL). A summary of applications and competences of metallic targets doped in WbLS, either being deployed or still under development, is presented in Table~\ref{tbl:wblsload}.
\begin{figure}[htp]
\centering 
\includegraphics[width=0.2\textwidth]{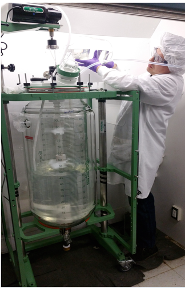}
\includegraphics[width=0.24\textwidth]{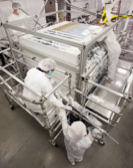}
\includegraphics[width=0.42\textwidth]{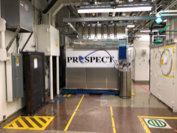}
\caption{Ton-scale fabrication and deployment of metal-doped liquid scintillator for Prospect} 
\label{fig:wblsprospect}
\end{figure}

\begin{table}[hb]
\begin{centering}
\begin{tabular}{|l|l|l|}
\hline\hline
Target	&  Loading (mass) & Potential Applications\\
\hline
 Indium		 & $>$8\% In  &  Solar $\nu$ \\ 
\hline
 Tellurium 		 & $>$ 6\% Te  &  $0\nu\beta\beta$\\ 
\hline
 Lithium 		 & $0.1$\% $^6$Li  & Reactor $\bar{\nu}$; excellent PSD  \\ 
 &  $>$0.2\% $^6$Li  & Reactor $\bar{\nu}$; super PSD with improved optics \\ 
\hline
 Boron		 & $>$0.5\%   & Dark Matter veto, reator $\bar{\nu}$ \\ 
\hline
 Potassium 		 & $>$1\%  &  Calibration for LS detectors\\ 
\hline
 Iron, Strontium & ppm to 1\%  & Nuclear waste management, \\ 
 &   & enviromental tracers\\ 
\hline
 Gadolinium & 0.1\% Gd  & Dark matter veto \\ 
 &   & Reactor monitoring\\ 
 &   & Reactor $\bar{\nu}$ oscillations\\ 
\hline
\hline
 High-Z elements & 10-15\%Pb  & Solar $\nu$ \\ 
 &   & Calorimeters \\ 
 &   & Medical QA/AC \\ 
\hline\hline
\end{tabular}
\caption{Examples of metallic targets loaded in Liquid Scintillator and Water-based Liquid Scintillator with projected applications\label{tbl:wblsload}}
\end{centering}
\end{table}

\subsection{Tellurium Loading}
A method for loading tellurium into organic liquid scintillator has been developed based on the formation of soluble organic compounds derived from telluric acid (Te(OH)6, hereafter TeA) and 1,2-butanediol (BD) in conjunction with N,N-dimethyldodecylamine (DDA), which acts as a stabilisation/solubilisation agent. The chemicals involved can all be purified to high levels, have high flash points and are relatively safe to work with in underground environments. The loading process results in acceptable optical absorbance and light output in larger detectors for loading levels up to several percent Te by weight, as shown in Figure~\ref{fig:TeLoading4}. Stability of the loading has been demonstrated to be at least on the timescale of years. 

\begin{figure}[htp]
\centering 
\includegraphics[width=0.6\textwidth]{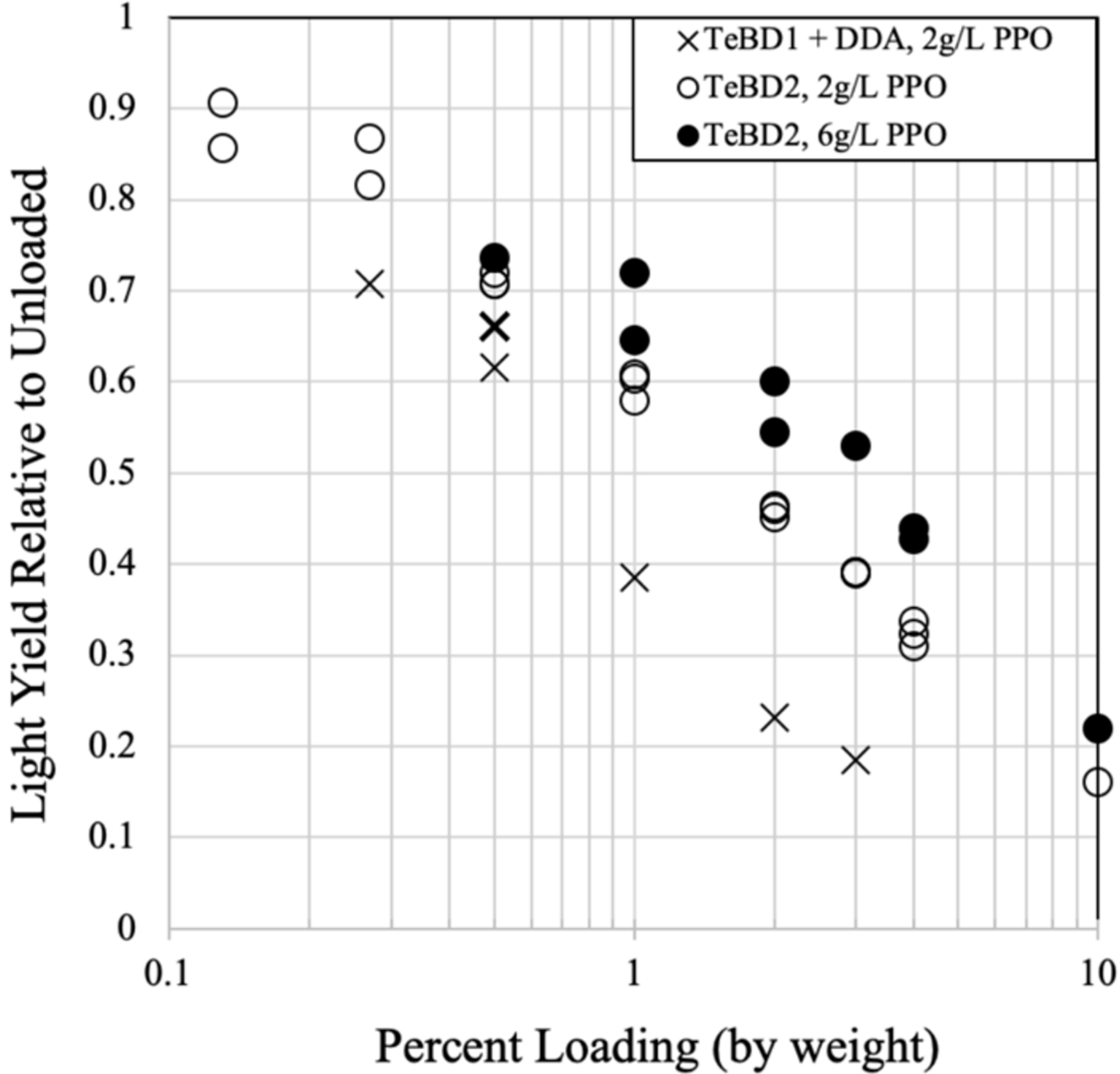}
\caption{Relative light levels as a function of percent Te loading for two variants of the technique: TeBD1 (heated solubilization) and TeBD2 (DDA-induced solubilization.} 
\label{fig:TeLoading4}
\end{figure}

This is an important advance that opens the door to a highly scalable and economical approach for neutrinoless double beta decay. Further advances in Te loading and associated purification techniques could provide a practical path to realising sensitivity to the non-degenerate normal mass ordering.

\subsection{Quantum Dots}
In recent years new methods for the production of metal loaded LS in neutrino physics were studied involving semiconducting nanocrystals, which are known as “quantum dots” [69]. The optical and electrical properties of the quantum dots are directly proportional to their size, which is typically in the order of few nanometers. The emission band consists of a narrow resonance around the characteristic wavelength of the dot.  Since the dot size can be controlled to high precision in the synthesis, the absorption and re-emission spectrum of the dots can be tuned and optimized for a respective application. In some synthesis methods the quantum dots are already delivered in colloidal suspension with the aromatic solvent toluene at concentrations of several grams per liter.

The most commonly used quantum dot cores are binary alloys such as CdS, CdSe, CdTe, and ZnS. Alternatively, there are also phosphor-based rare-earth dots. Therefore, quantum dots provide a method to dope scintillator with various metals and rare-earth elements. For a cadmium (Cd) based LS there are two different applications in the field of neutrino physics: neutron-enhanced isotopes (113Cd) and $\beta\beta$-decay candidates (106,116Cd). Also Se, Te, and Zn, which are present in common quantum dot cores, have $\beta\beta$-decay candidates. There is also a possibility to tune the scintillator emission spectrum in a way that allows to separate scintillation from Cerenkov light by narrowing scintillation emission from the quantum dots separated from the Cerenkov contribution.

The basic limitations of quantum-dot-doped liquid scintillator in use for particle physics detectors are probably cost and availability in large quantity ($\sim$ ton). The stability tests on typical solutions of quantum dots in concentration at few g/l indicate that larger particles are formed by aggregation in the concentrated solutions over long time scales. This could explain the fact that filtering improves the attenuation length as well as the observation of transparency degradation after several weeks. However, there is room for optimization on the stability and optical performance by improvements in the loading process of quantum dots in scintillator solution. Instead of suspension, incorporating chelating agents or WbLS surface active agents into the mixing procedures could load the quantum dots in organic solvents homogeneously with lengthy stability.

\section{Improvements in Simulation and Analysis Approaches}

\subsection{Generative Neural Networks}
Photons, emitted from a vertex where a particle deposits energy, illuminate specific photon detectors and form a hit pattern on the photon detection system. The number of total detected photons can be used as a calorimetric energy measurement and improve the detector energy resolution while the distribution of the illuminated photon detectors can be used to locate the 2D position of the interaction point. Traditionally, the amount and hit pattern of the detected photons, or in another word the photon detection probabilities, are explored by simulating the transport of photons in detectors using \textsc{Geant4}. However, such simulation is extremely challenging for experiments using huge detectors that record GeV-level energy depositions due to limited computing resources. Modern machine learning techniques have enabled new ways to emulate the results from full \textsc{Geant4} simulation and a generative model based on generative neural networks is one of the most promising approaches to predict the photon detection probabilities.

The generative model can be trained ahead of time using full Geant4 optical photon simulation with photons, emitted from random vertex in the detector, and then be deployed to the production environment. In this case, GPUs can be used for the network training, but the evaluation will be performed, in general, on CPUs as a part of the experiment data reconstruction chain. Traditional generative models based on deep neural networks (\textsc{DNN}) have shown great promise in generating accurate predictions, but can be too slow when deployed without GPUs. As the neural network inference speed strongly correlates to the network complexity, a balance between the inference speed and precision must be found when choosing the depth of the network.

A recently developed 1D generative neural network [], which is much shallower alternative to a traditional \textsc{DNN}, can efficiently predict photon detection probabilities at lower computational cost. The much simpler 1D generative model successfully addresses the balance between the inference speed and precision. Unlike common generative models, which use the upsample (UpSampling2D) layer or the transpose convolutional (Conv2DTranspose) layer for the input dimension expansion, the 1D generative model uses a $\mathrm{OuterProduct}$ layer to produce features and expand the dimension of inputs, which significantly reduces the inference time on CPUs while keeping similar prediction accuracy. The general network structure can be illustrated by FIG.~\ref{fig:genn}
\begin{figure}[htp]
\centering 
\includegraphics[width=0.7\textwidth]{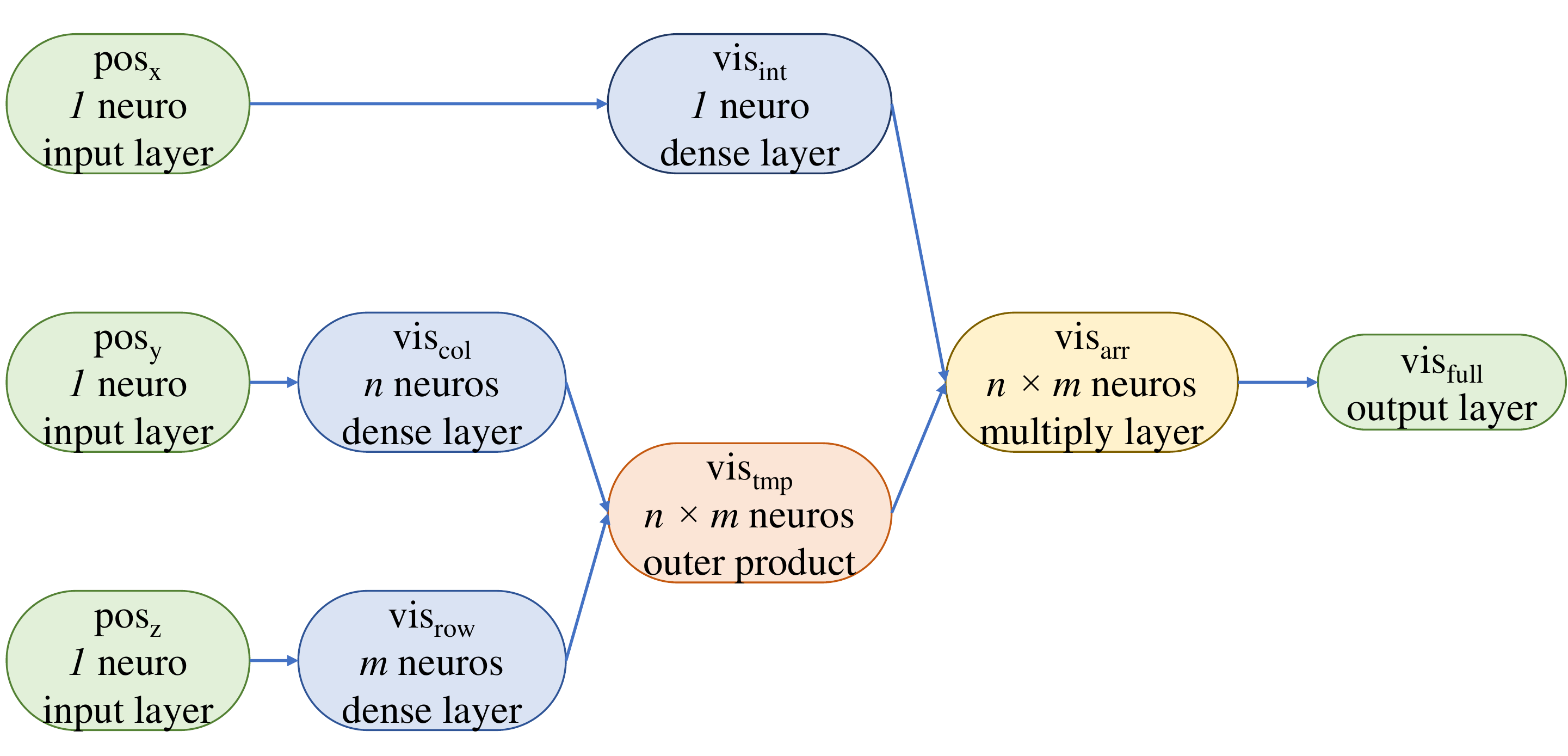}
\caption{General \textsc{Genn} architecture. ${pos_y}$-layer and ${pos_z}$-layer are used to predict the photon distribution on the photon detection system, while ${pos_x}$-layer is a normalization factor for the number of total detected photons. $\mathrm{OuterProduct}$ layer expands the dimension of the inputs.} 
\label{fig:genn}
\end{figure}

This general generative model has been applied to photon detection systems of ProtoDUNE-like [] and DUNE-like geometries []. In ProtoDUNE-like geometry, two different types of photons sensors, in total 90, are deployed asymmetrically on two photon detection array, while in DUNE-like geometry, more identical photons sensors (480) are deployed symmetrically on one photon detection array, as shown in FIG.~\ref{fig:pddist}.
\begin{figure*}[htp]
\centering
\includegraphics[width=0.7\textwidth]{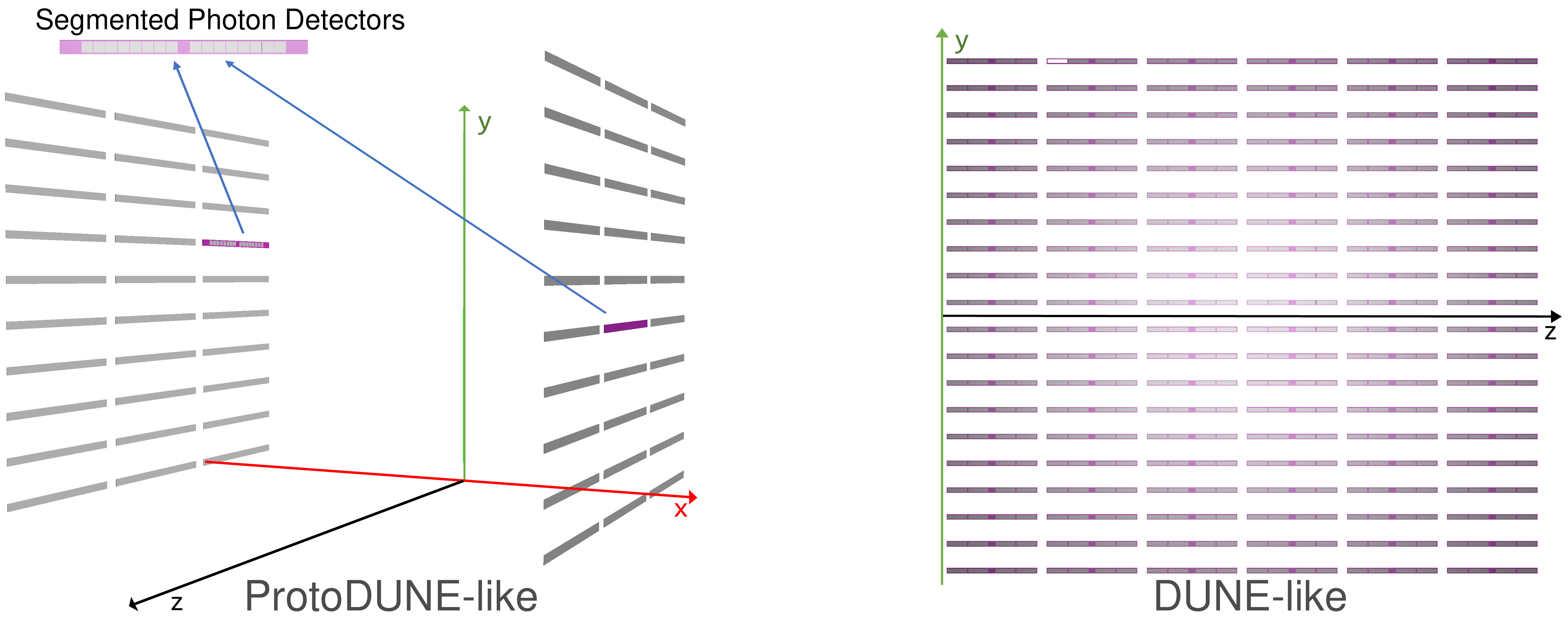}
\caption{Photon detection systems in ProtoDUNE-like (left) and DUNE-like geometries (right).}
\label{fig:pddist}
\end{figure*}

The two sample models show the prediction speed is 20 to 50 times faster than \textsc{Geant4} simulation while keeping the same level of detail on particle tracks, such as number of energy depositions, and precision. The model built for ProtoDUNE-like photon detection system demonstrates that shallow neural networks are able to learn features from training samples represented by 1D data structures, even for complex photon detection systems. The sample for DUNE-like photon detection system indicates the generative model gives fast and precise prediction of photon detection probability using limited memory, showing it can be a powerful new tool to bypass the full \textsc{Geant4} simulation in a production environment, especially for detectors with huge volumes, such as the DUNE far detector. The successful application to the ProtoDUNE-like and DUNE-like geometries shows this general \textsc{Genn} architecture is stable and easy to generalize.

The model inference also requires a relatively small amount of memory. The samples for ProtoDUNE-like and DUNE-like geometries show the required memory for the model inference is around 15\% of the \textsc{Geant4} simulation. Further, this memory use is not directly correlated to the volume of the detectors, unlike using lookup libraries where the available memory on the machine limits the potential granularity and hence precision.

\subsection{{\it Chroma}}
\label{sec:chroma}
As large-scale neutrino detectors become ever larger, photon coverage
becomes higher, and photon sensor technology becomes more complex with devices 
like the ARAPUCA~\cite{arapucas}, the dichroicon~\cite{dichroicons}, or 
distributed imaging~\cite{gratta_image}, a major bottleneck in both 
simulation and reconstruction of physics events is the propagation of photons 
through the detector geometry. 
Originally created for the water Cherenkov option for the LBNE experiment, a
fast photon ray-tracer was developed by Stan Seibert and Anthony
LaTorre~\cite{chroma} that improved photon simulation speeds by a factor of 200 over
what GEANT4 itself could do.  In order to achieve such high performance, {\it
Chroma} combines techniques from 3D rendering algorithms with the massively
parallel calculation hardware inside GPUs.  Existing 3D rendering libraries,
while quite sophisticated, cannot be used directly for physics simulation
purposes, as those libraries tend to rely on physically-unrealistic
approximations and shortcuts to improve the appearance of the produced images.
Nevertheless, {\it Chroma} takes similar ideas and applies them within the
context of a physics simulation to improve performance without sacrificing
precision physics.  These techniques map well onto the massively parallel GPUs
found in today's workstations.  A high-end GPU costs approximately \$5000
(twice the cost of a fast consumer-grade CPU), yet provides forty times the raw
floating point performance and ten times the memory bandwidth.  {\it Chroma}
uses the CUDA toolkit, provided by NVIDIA, to directly access the GPU resources
and perform all major calculations.  CUDA-compatible GPUs are being used more
and more in the construction of large supercomputing clusters, which will make
it easier in the future for the work on next-generation neutrino detectors such
as \textsc{Theia}~\cite{theiawp} to use {\it Chroma}.

Nearly all fast 3D rendering systems represent the world geometry using a mesh
of triangles.  Triangle meshes are very simple to represent, and can be used to
approximate any surface shape, limited by how much memory can be devoted to
triangle storage.  With only one surface primitive, there is only code path to
optimize.  In particular, we have adopted the Bounding Volume Hierarchy (BVH)
technique from the graphics world to speed up ray intersection tests with
triangle meshes.  A BVH is a data structure that organizes a spatial
arrangement of shapes (triangles in our case) into a tree of nested boxes.
Rather than test for ray intersection with every triangle in a geometry, the
photon propagator tests for intersection with boxes in the BVH.  If the ray
does not intersect the box, then all of the children of that box can be
skipped, leading to a large reduction in the number of intersection tests
required.  For example, a model of a large, 200 kton water Cherenkov detector
consists of 62 million triangles, but the BVH reduces a typical propagation
step for a photon to 130 box intersection tests and only 2 triangle
intersection tests.

{\it Chroma} implements the most important physics processes for optical photons, which include:
\begin{itemize}
\itemsep0em 
\item Wavelength-dependent index of refraction
(chromatic dispersion)
\item Absorption in the bulk
\item Reemission from a wavelength shifter
\item Rayleigh scattering
\item Diffuse reflection at surfaces
\item Specular reflection at surfaces
\item Arbitrary angle, wavelength dependent transmission and reflection (dichroic)
\item Standard Fresnel reflection and refraction
\item Detection at a photosensitive surface (e.g. a PMT photocathode)
\end{itemize}

One collateral bonus of {\it Chroma's} speed is that it provides
remarkably beatiful, {\it realtime} detector displays. It is quite easy to
``fly through'' a detector and see it rendered in all of its detailed geometry,
exactly the way the photons themselves will see the detector.
\ref{fig:snoplus} shows a rendering of the SNO+~\cite{snoplus} detector which,
in a static pdf document like this proposal cannot be examined dynamically in
realtime, but was created from that realtime fly-through and simply captured
by screen-shot. The rendering is in fact a 3D image; with red-blue glasses one
can see the relief in the image.

\begin{figure}[htb!]
\centering
\includegraphics[width=0.5\textwidth]{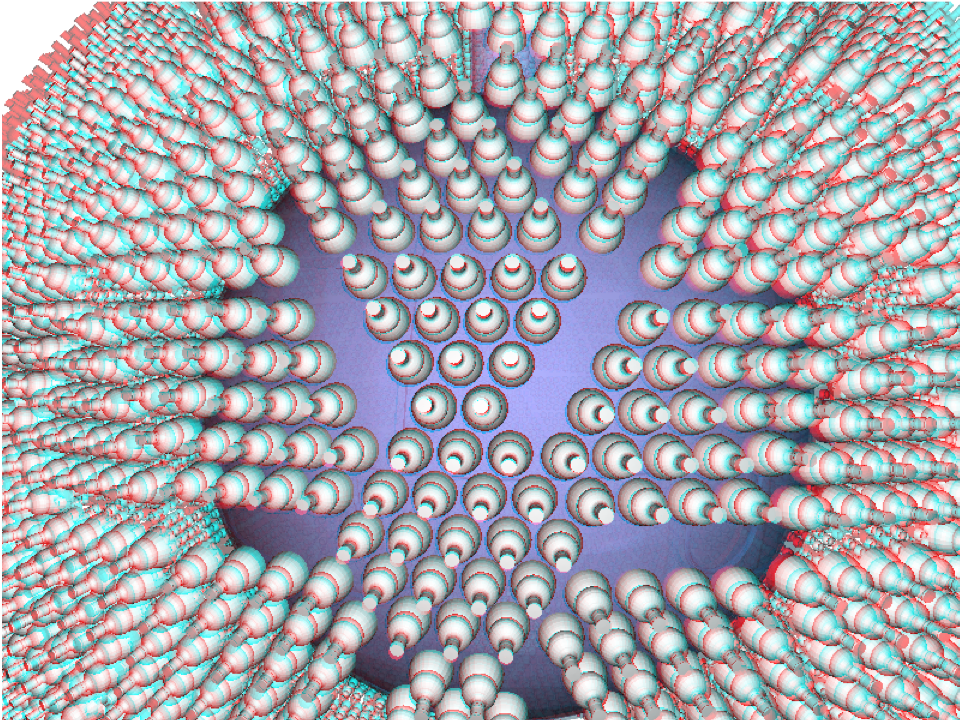}
\caption{A screen capture from a realtime, three-dimensional ``movie'' of the
SNO+~\cite{snoplus} detector created with {\it Chroma}, which is best viewed with red-blue 3D glasses. The model shown is exactly the model
used by {\it Chroma}'s physics simulation, albeit with false-color optics for
the display rather than the complete physics the simulation uses.  The PMTs are
fully rendered, including their Winston-cone light
concentrators.\label{fig:snoplus}}
\end{figure}

\begin{figure}[htb!]
\centering
\includegraphics[width=0.3\columnwidth]{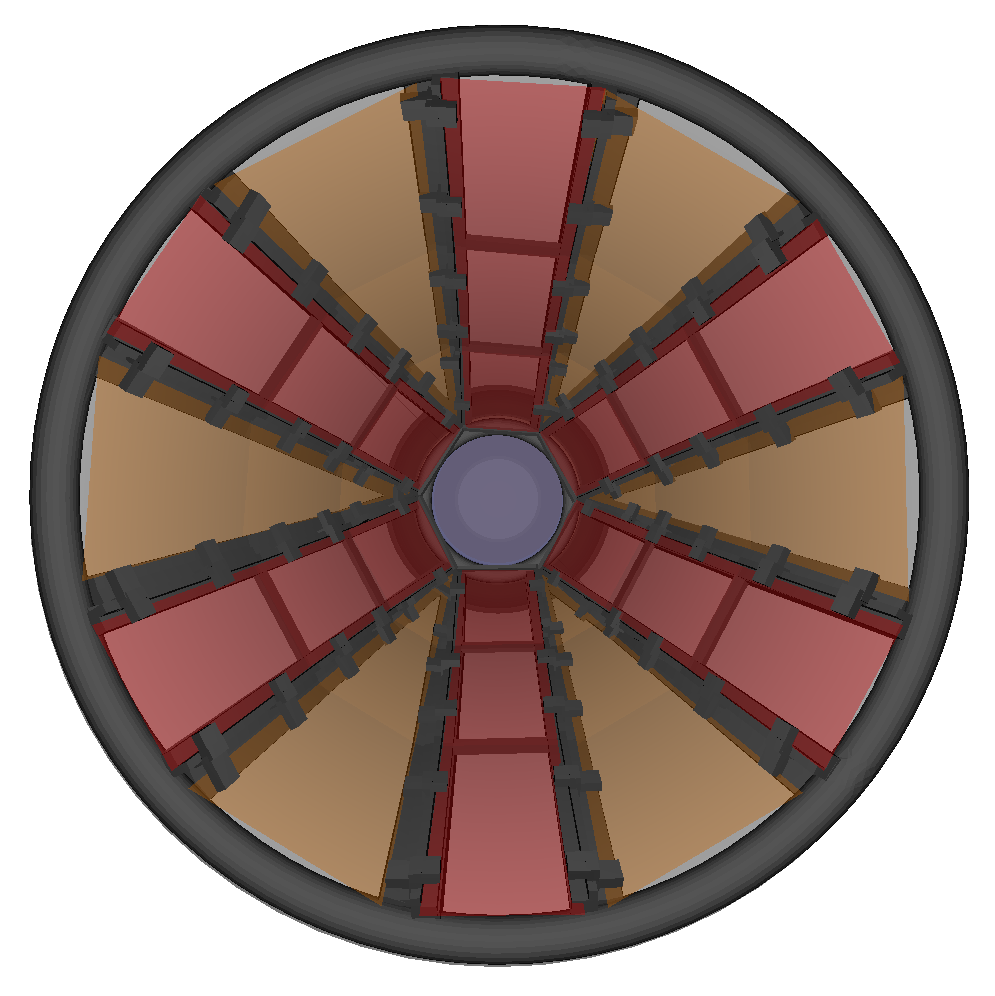}
\caption{\label{fig:holder}
The CAD drawing for the 3D-printed dichroic filter holder for the dichroicon~\cite{dichroicons} was directly imported into {\it Chroma}, and used to accurately simulate the orientations of the dichroic filters (shown in red and orange) with respect to the benchtop experiment.}
\end{figure}

\begin{figure}[htb!]
\centering
\includegraphics[width=0.3\textwidth]{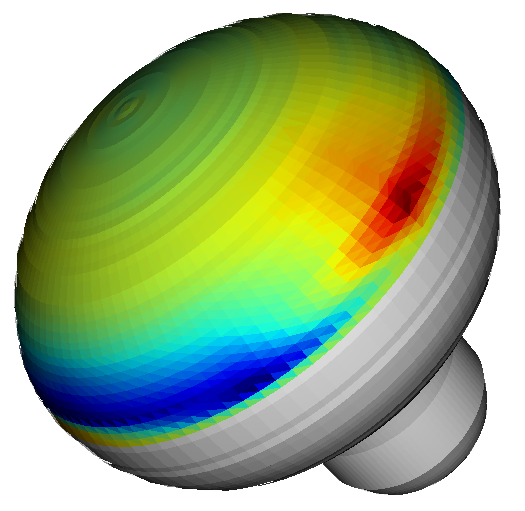}
\caption{\label{fig:pmt_prop}
From~\cite{chroma}, a visualization of per-triangle control of the PMT transit time property, which represents the delay in detection of a photon due to the time it takes the photoelectron to travel from the photocathode to the first dynode. 
Red areas are up to 3~ns late, and blue areas 3~ns early, relative to the mean detection time.
All {\it Chroma} properties can be defined at this level. }
\end{figure}

\begin{figure}[htb!]
\centering
\includegraphics[width=0.40\textwidth]{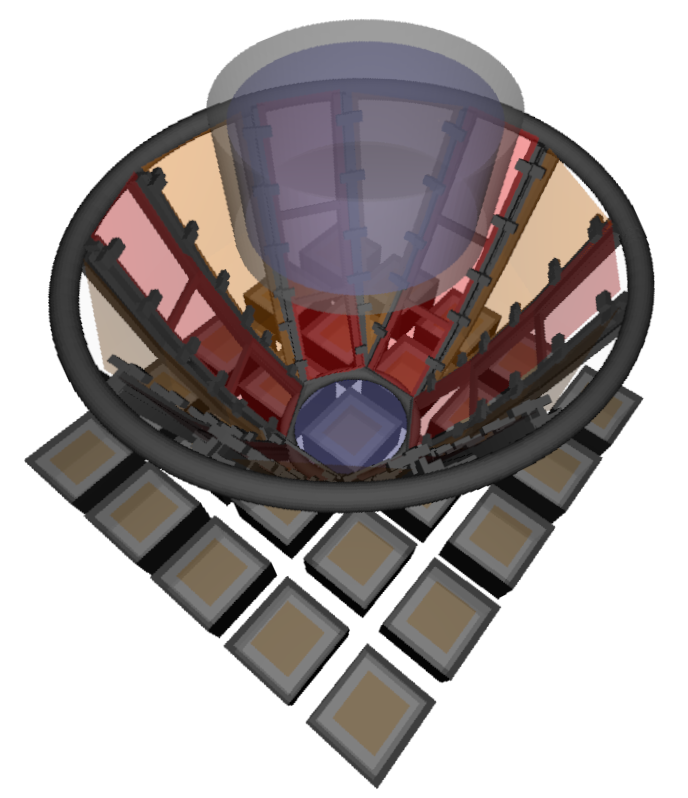}
\caption{\label{fig:chess_dichroicon}
{\it Chroma} is also applicable to benchtop-scale experiments like CHESS~\cite{chess} where application of the Cherenkov/scintillation separation with the dichroicon~\cite{dichroicons} is being explored.
A preliminary geometry is shown here with a liquid scintillator target, a dichroicon, and an array of fast PMTs from CHESS.
GEANT4 is used to simulate cosmic muons traversing this geometry, while {\it Chroma} simulates the optical and PMT response, all within the framework of {\it Chroma}.}
\end{figure}

The choice of representing geometries in {\it Chroma} as triangle mesh makes it straightforward to import CAD drawings into the optical simulation, as shown in \ref{fig:holder}.
All that is required is that the triangles on the mesh have optical properties assigned to them, which is a trivial operation in the case of a CAD model with uniform properties over its entire surface, but allows for fine-grain position dependent control of material properties as well, as shown in \ref{fig:pmt_prop}.
This allows anyone with CAD experience to quickly create an arbitrarily complicated simulation without having to learn a new way to represent geometries. 
The ability to rapidly prototype designs makes {\it Chroma} well suited for benchtop studies as well, as can be seen in \ref{fig:chess_dichroicon}, which shows preliminary application of the dichroicon to the CHESS~\cite{chess} experiment.
Combining these features allows easy modeling of new photon-detector technologies, as shown in \ref{fig:theia25}, which includes 3D Large Area Picosecond Photodetectors (LAPPD)~\cite{LAPPDtiming} mixed with standard PMTs in a {\it Chroma} model of the \textsc{Theia}25~\cite{theiawp} detector.
\begin{figure}[htb!]
\centering
\includegraphics[width=0.4\textwidth]{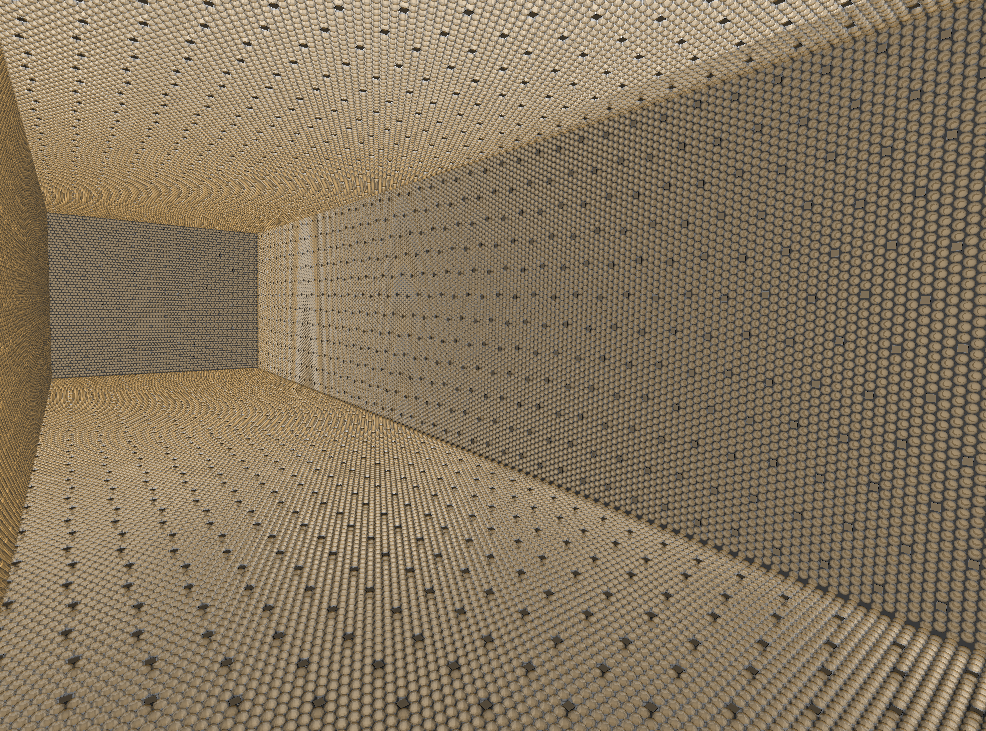}
\caption{\label{fig:theia25}
Next-generation photon detectors like LAPPDs~\cite{LAPPDtiming} are straightforward to simulate and include in larger geometries, as shown in this internal view of a \textsc{Theia}25~\cite{theiawp} model made with {\it Chroma}, which has mixed PMTs and LAPPDs for detecting photons.
The {\it Chroma} framework makes it quick and easy to explore new detector geometries and photon detecting devices.}
\end{figure} 

{\it Chroma} was designed to be easy to use, is now maintained in a public Github repository~\cite{chroma} complete with Docker containers for easy deployment.
Primarily written in Python, {\it Chroma} interfaces well with popular datascience packages in that ecosystem (Matplotlib, Scipy, Tensorflow, etc.), allowing for rapid development of ideas with minimal overhead.
{\it Chroma} is also well connected to GEANT4, which it uses to simulate physical interactions that produce photons, and ROOT, which serializes {\it Chroma} events allowing for traditional ROOT-style analyses of the simulations.

\subsection{RAT-PAC}

RAT-PAC is an open-source GEANT4-based toolkit that offers both micro-physical simulation capabilities and analysis tools for high-precision event modeling, evaluation, and characterization, from benchtop test stands to large-scale detectors.

The RAT-PAC Monte Carlo simulation and analysis suite~\cite{ratpac} is a free and open-source version of the RAT toolkit.  RAT was first written for the Braidwood reactor experiment~\cite{braidwood},
and is now the official simulation and analysis package for  SNO+~\cite{snoplus}, DEAP, and MiniCLEAN~\cite{miniclean} experiments, thus benefiting from shared efforts in development and verification. A GEANT4-based package~\cite{geant4}, RAT-PAC (standing for ``RAT Plus Additional Code'') was branched off from the core RAT development some years ago, to form an open-source version of the code, available for public use.  RAT-PAC forms the basis of the official software for the \theia collaboration~\cite{theiawp}, the proposed third phase of ANNIE~\cite{annie-results}, and for the WATCHMAN collaboration, who are developing a design for the NEO detector to be located at the AIT facility in the UK~\cite{watchman}.

One of the great advantages of the RAT-PAC approach is that its procedural geometry description allows the same code to be used to simulate or analyze data from a large-scale experiment and a small benchtop test-stand.  Figure~\ref{fig:SNOp} shows the detailed geometry of the full ktonne-scale SNO+ detector, and the even larger \theia detector, and Fig.~\ref{fig:CHESS} shows the much smaller CHESS detector at UC Berkeley/LBNL~\cite{chess}. In addition to the flexible geometry descriptions, RAT-PAC takes a micro-physical approach, relying on physical, rather than phenomenological models. For example, individual photons are simulated hitting photon sensors and the resulting timing and charge are evaluated photon-by-photon, rather than by application of a phenomenological risetime correction.  Therefore, simulating both benchtop test stands and large-scale detectors with the same micro-physical detail and the same code means that parameter measurements made by the benchtop are more easily translated into the larger-scale detector.  A measurement of, for example, the light yield of a scintillator cocktail performed in a small-scale setup can be straightforwardly propagated to predict performance in large detectors, complete with systematics associated with optical models or even data acquisition approaches. Comparisons between simulations of Cherenkov and scintillation light generated using RAT-PAC and data from test stands, such as at Penn, CHESS at LBNL, and FlatDot at MIT, show good agreement.  An example from FlatDot is shown in Fig.~\ref{fig:CHESS}~\cite{flatdot}. 

RAT(-PAC) is based on GEANT4.10~\cite{geant4} and the GLG4Sim package written by Glenn Horton-Smith, with custom code for scintillation and neutron absorption processes as well as a complete model of  PMT optics.
RAT(-PAC) handles all stages of event simulation: from the propagation of primary particles; production of optical photons via Cherenkov and scintillation processes; individual photon propagation, including a full optical model of all detector materials;  photon detection at the single PE level, including individual photon detector charge and timing response; and data acquisition including full
customizable simulation of  front end electronics, trigger systems, and event builders.  It also allows root-formatted data to be used as input, and provides simple analysis tools and ways to include many more, as well as a macro command structure for control. Lastly, RAT-PAC also includes the
ability to 
dynamically generate detector configurations via an external database.
Thus, RAT-PAC is a complete package that can be used with small modifications for entire experiments.

As experiments grow in scale and use increasing numbers of photodetectors, RAT-PAC will need to progress to reflect these needs. Planned improvements include:
\begin{itemize}
    \item Updating dependencies to reflect currently-used versions (Python3, ROOT6, and Geant4.10.6)
    
    \item Adding generators for rare physics processes, such as the addition of a neutrinoless $^{124}$Xe positron-emission/electron-capture ($0\nu\beta+/EC$) decay generator \cite{graham_thesis}, and double-beta decays to excited states. 
    
    \item Incorporating new photodetector types in the public code, such as the Large-Area Picosecond Photodetectors (LAPPDs) implemented in the ANNIE and CHESS branches.
    
    \item Improvements to simulation efficiency, which will be needed to speed up simulations of experiments with $\mathcal{O}(10^5)$ channels, including the ability to parallelize aspects of the simulation. 
\end{itemize}
These improvements will ensure that RAT-PAC continues to meet the needs of the liquid scintillator  and water Cherenkov community.

\begin{figure}[ht!]
    \centering
    \includegraphics[width=0.4\textwidth]{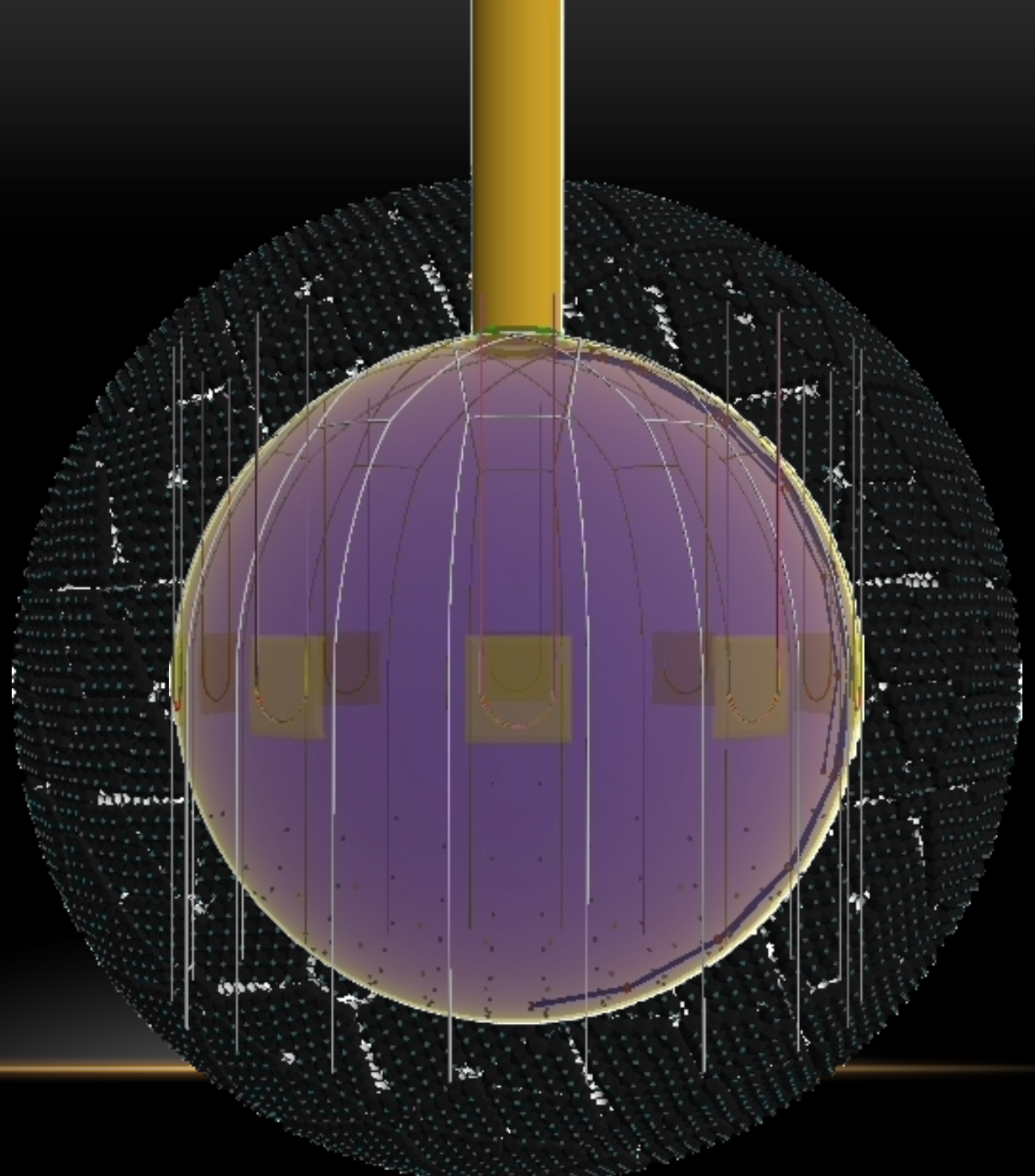} 
    \includegraphics[width=0.39\textwidth]{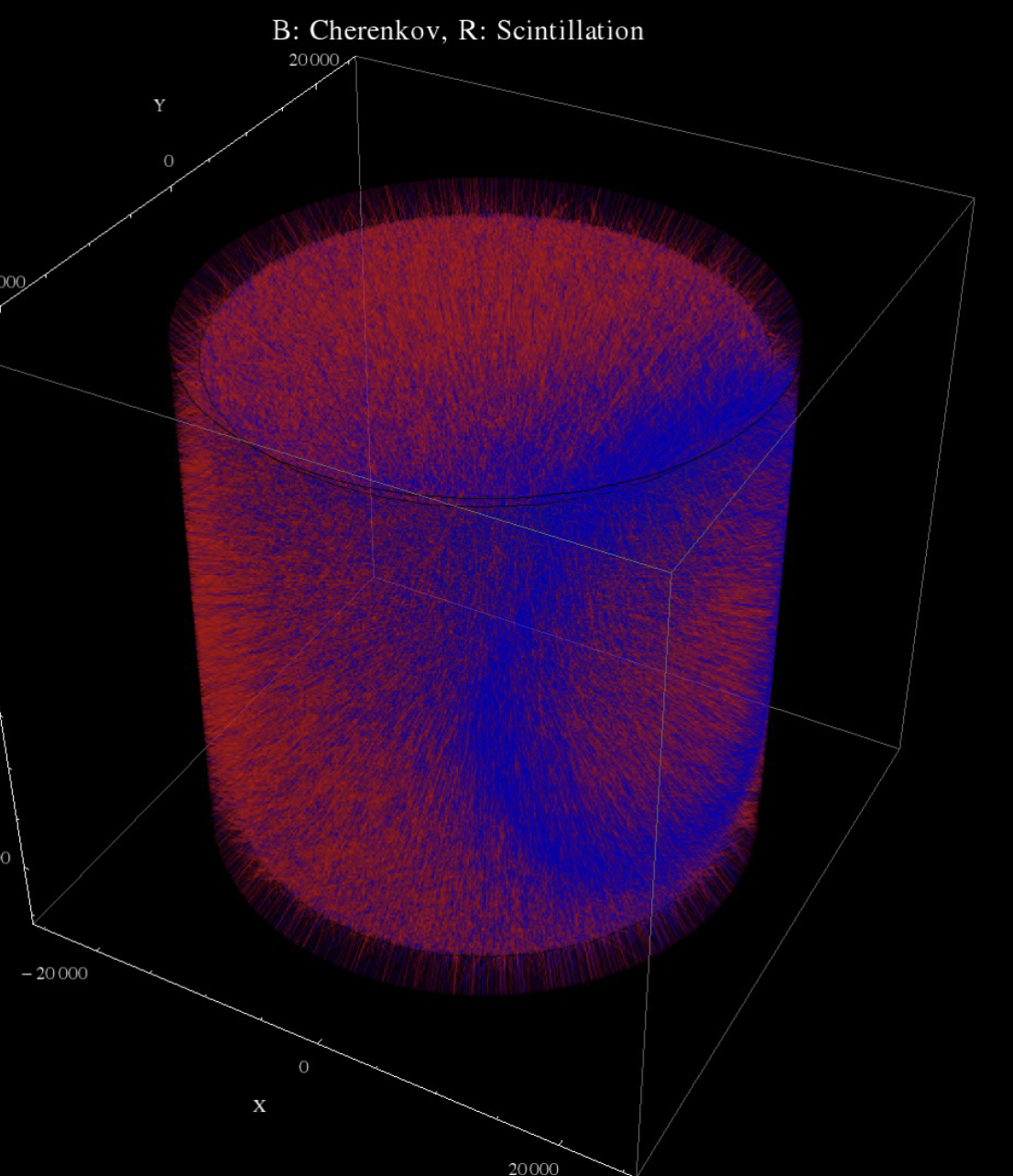} 
    \caption{(Left)  RAT-PAC generated image of the ktonne-scale SNO+ detector. (Right) RAT-PAC simulation of a high-energy (GeV) electron in the 50-ktonne \theia detector, including full photon tracking.  Blue shows Cherenkov photon track and red shows scintillation.}
    \label{fig:SNOp}
\end{figure}

\begin{figure}[ht!]
    \centering
    \includegraphics[width=0.5\textwidth]{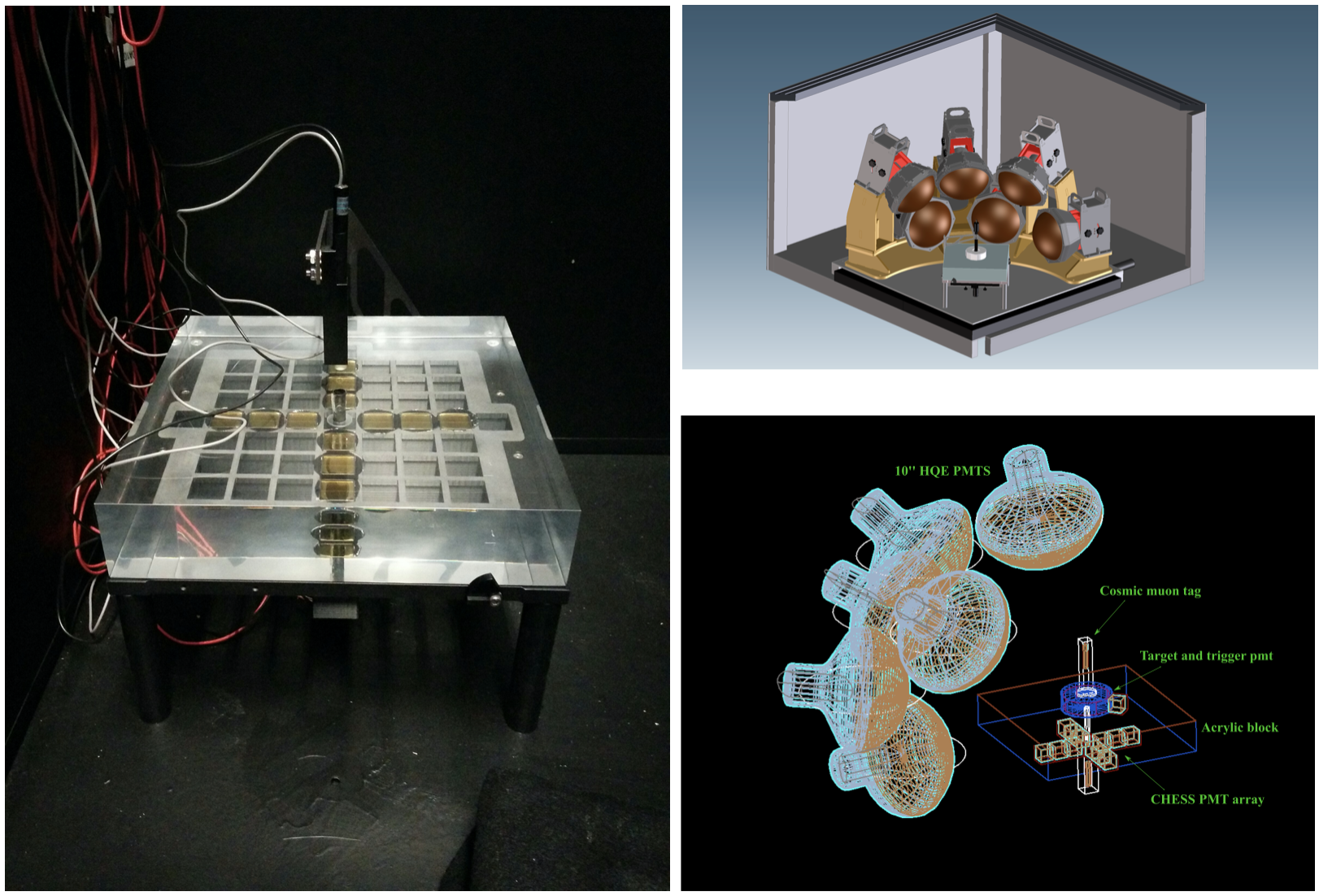} 
    \includegraphics[width=0.48\textwidth]{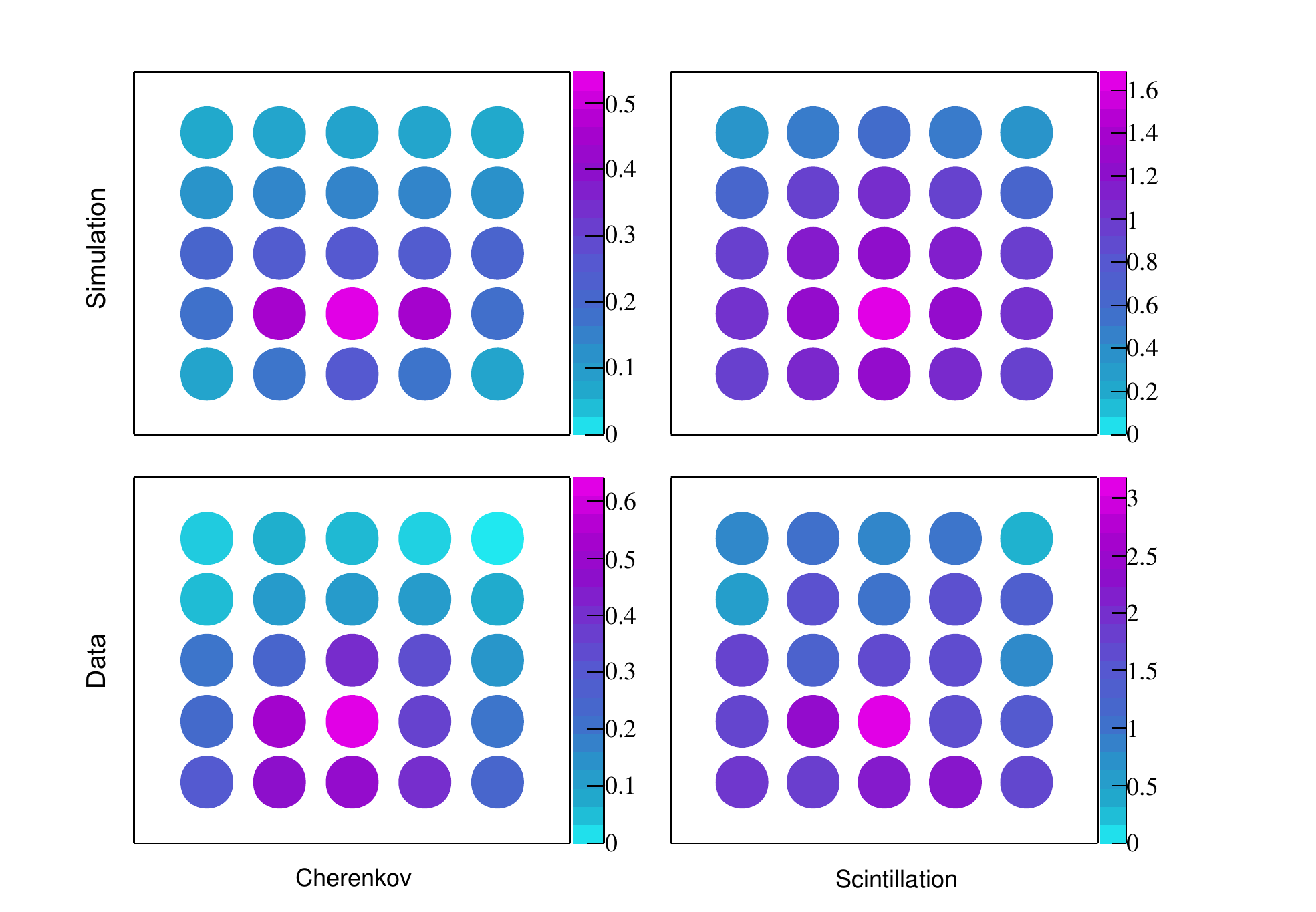}
    \caption{(Left) Photograph of the CHESS PMT array.  (Top centre) CAD image of the full CHESS detector.  (Bottom centre) RAT-PAC generated image of the full CHESS detector.  (Right) RAT-PAC simulations of both Cherenkov \textit{(left)} and scintillation \textit{(right)} signals show good agreement with data from the FlatDot experiment \textit{(bottom)}, up to a normalization factor reflecting the absolute light yield of the liquid scintillator.}
    \label{fig:CHESS}
\end{figure}

\subsection{KamNet}
\label{sec:KamNet}
The challenge of photon detection in monolithic LS detectors is that they function primarily as calorimeters and lack the sophisticated tracking and topological information provided by other technologies like Time Projection Chambers, Cherenkov ring imaging detectors, or silicon strip trackers. Enhancing monolithic LS detectors with the capability to discriminate between different event types based on tracking and topology would be a revolutionary advancement. In our previous work\cite{conventionalCNN}, we applied conventional convolutional neural network~(CNN) to reject muon spallation backgrounds. However, the conventional CNN lacks the ability to harness certain symmetries embedded in KamLAND-Zen data which leads to a learning bias during training that suppresses the performance of neural network. Therefore, we invented a novel spatiotemporal neural network model KamNet~\cite{KamNet} which harnesses breakthroughs in geometric deep learning and spatiotemporal data analysis to maximize the information extraction from LS detectors.

\begin{figure}[htp!]
    \centering
    \includegraphics[width=0.54\textwidth,trim={3pc 0 4pc 0pc},clip]{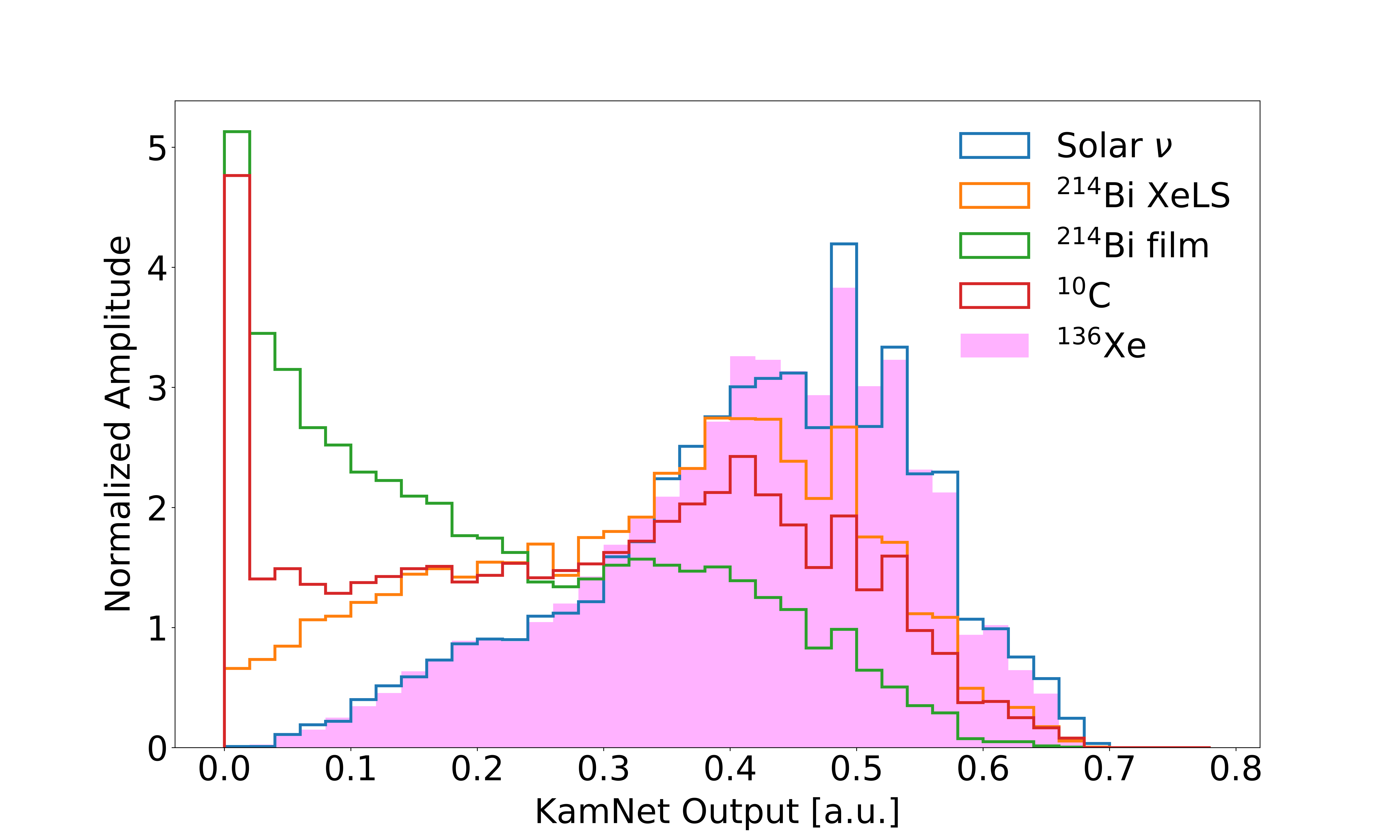} 
    \includegraphics[width=0.4\textwidth,trim={3pc 0pc 1pc 0pc},clip]{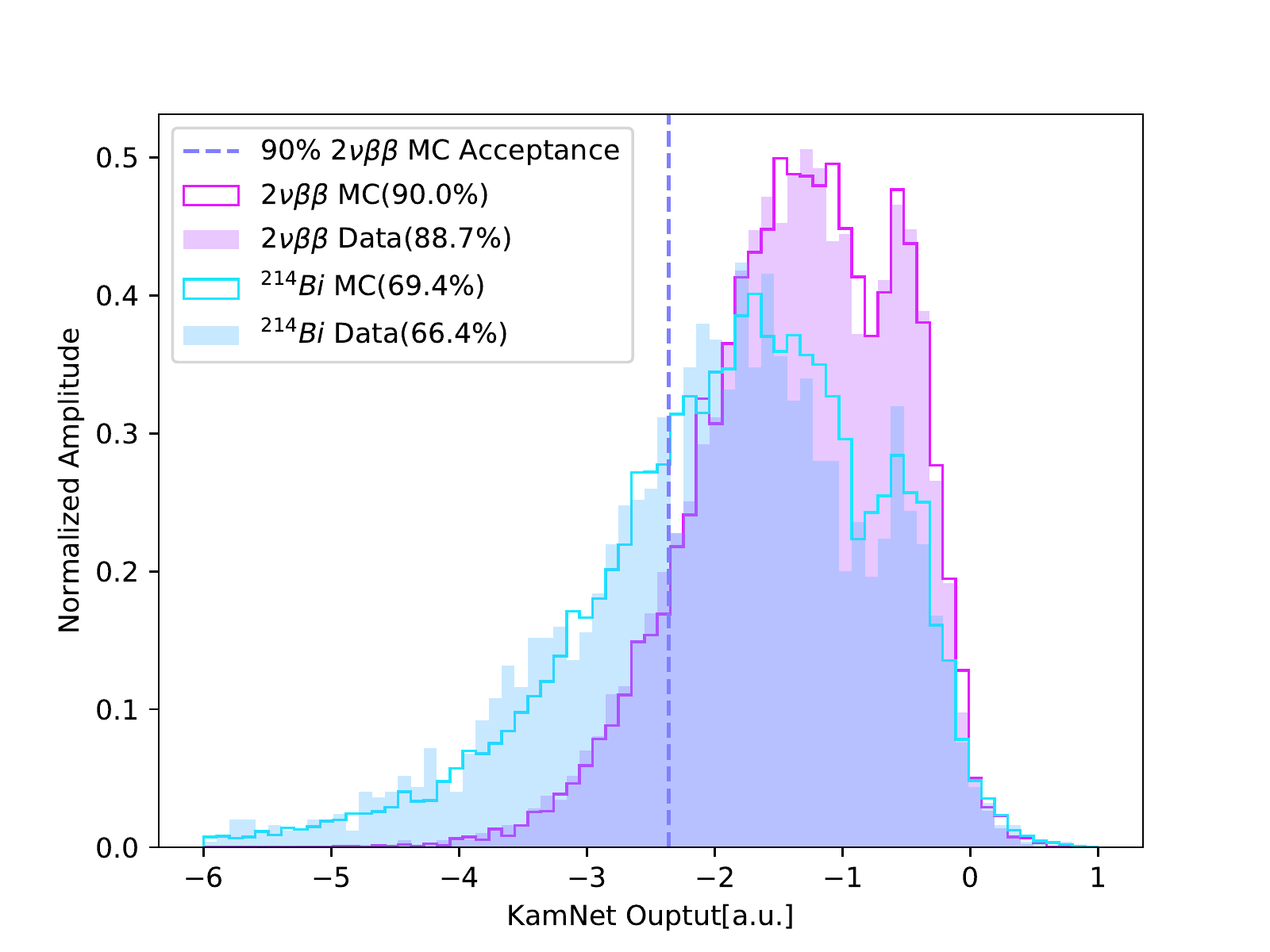}
    \caption{(Left) KamNet output spectrum on common types of backgrounds in KamLAND-Zen detectors. (Right) KamNet output spectrum showing the comparison between detector data and MC simulations.}
    \label{fig:KamNet}
\end{figure}
With the power to discriminate between different event topologies, KamNet rejects 27\% percent of XeLS backgrounds and 60\% of film backgrounds in KamLAND-Zen detector as shown in Figure~\ref{fig:KamNet} left. Our network interpretability study shows that the rejection power of KamNet comes from distinguishing strictly single-vertex events from closely spaced multi-vertex events such as $\beta^{\pm}$ decay with $\gamma$ cascade. There has not been any algorithm powerful enough to capture those information, including $\gamma$ casacade and compton scattering, in LS detector. Therefore, KamNet provides an unprecedented way to reject background events without any coincidence tagging. A conservative study shows that KamNet can increase the $0\nu\beta\beta$ sensitivity of KamLAND-like detectors by at least 10.8\% without any hardware upgrade. After data-MC agreement study shown in Figure~\ref{fig:KamNet} right, KamNet is merged into the new KamLAND-Zen 800 analysis to tag critical backgrounds, allowing the first search for the Majorana nature of neutrinos in the inverted mass ordering region~\cite{KamLAND-newresult}.

\subsection{Generative Models} \label{sec:generative_models}
Simulation of physical events is always the cornerstone of any neutrino experiments, from detector design to research and development analysis to comparison with theoretical models. The traditional simulation based on first-principle starts with prior knowledge and uses theoretical calculations to make predictions. This includes the widely used Monte Carlo simulation. However, the complexity of the interactions in nuclear physics, coupled with a high degree of randomness, makes these traditional methods time-consuming and computationally expensive. So a data-driven generative method based on machine learning is designed to solve these problems.
These new developed generative models have the ability to extract major information of given data samples and construct very similar simulation data based on that. 

Our generative models based on both Variational Autoencoders (VAEs) and Generative Adversarial Networks (GANs). The power of these generative models to reconstruct signal events is illustrated in Figure \ref{fig:gen_model}, where the distributions of hit times, hit charges, numbers of triggered PMTs of multiple events generated from our models are shown, together with the original KamLAND-Zen 800 MC simulation. The generative models achieve a decent agreement between  the original simulation and the forged dataset from machine leaning. These generative models can improve the simulation that has lack of precise knowledge with minimal assumptions. 
\begin{figure}[htb!]
    \centering
    \includegraphics[width=1.0\textwidth,trim={0pc 0pc 0pc 0pc},clip]{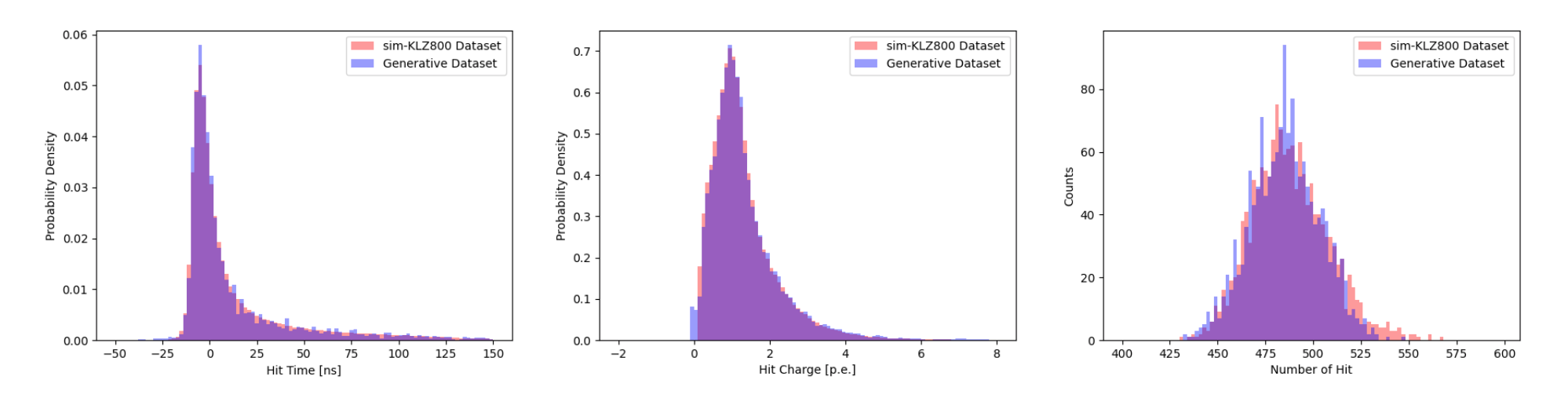}
    \caption{(Left) Distributions of hit time. (Middle) Distributions of hit charge. (Right) Distributions of number of triggered PMTs.}
    \label{fig:gen_model}
\end{figure}

%
%

\section{Prototypes and Large-scale R\&D Platforms}
	The effort on new, enabling technologies has progress to the point
where there are several large-scale platforms to perform integrative tests.
Among these are NuDot, \eos, and ANNIE~\cite{annieLOI}. The
goals of these platforms spans a broad range including demonstrations of
Cherenkov/scintillation separation at both high and low energies, exploration
of new target materials such as water-based liquid scintillator and new fluors,
and verification of large-scale detector models and new reconstruction
techniques. 	

\subsection{ANNIE}
\label{sec:ANNIE}

The ANNIE (Accelerator Neutrino Neutron Interaction Experiment)~\cite{annieLOI,annie-results,pershingdiss} is located in the Booster Neutrino Beam at Fermilab. This beam is about 93\% pure $\nu_{\mu}$ in neutrino mode and has a spectrum that peaks at about 700 MeV, an energy scale of great interest to current and future neutrino oscillation experiments. While ANNIE's current Phase 2 investigates the combination of a Gd-loaded water target read out by ultrafast LAPPDs, a first insertion of a small WbLS-filled vessel (SANDI) is planned for the upcoming summer break, adding scintillation to the Cherenkov signal at the neutrino interaction vertex. In an eventual Phase 3, the full detection volume is to be filled with WbLS, making ANNIE the first experiment exploring the benefits of the new hybrid detection technique in the reconstruction of GeV-scale beam neutrinos.

\begin{figure}[t]
    \centering
    \includegraphics[width=0.225\textwidth]{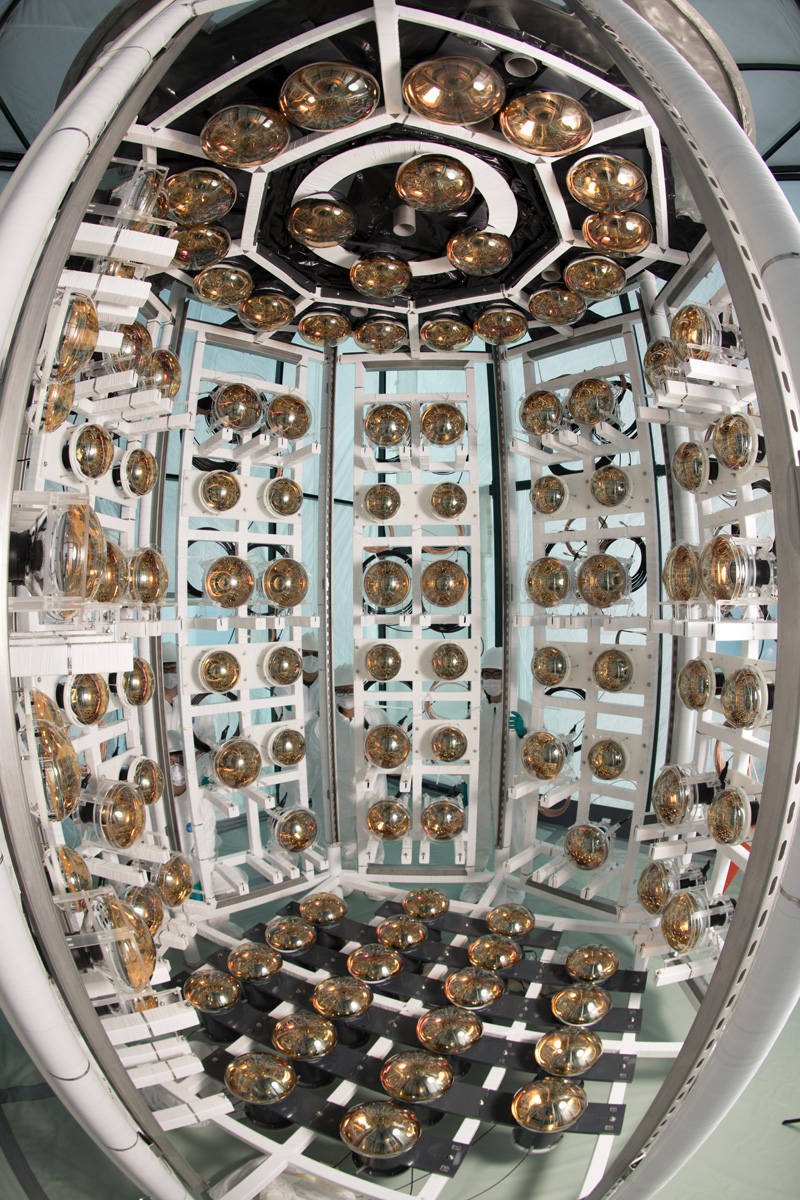}
    \hfill
    \includegraphics[width=0.43\textwidth]{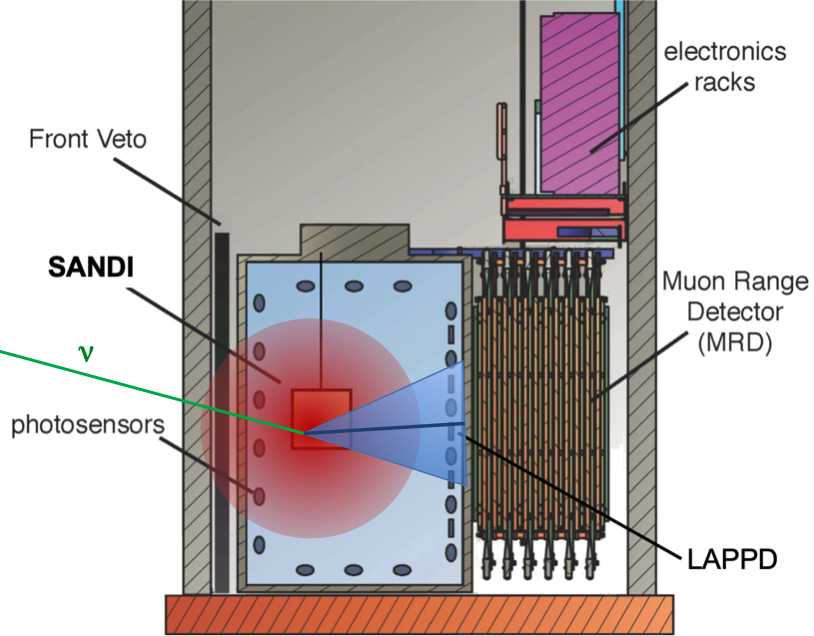}
    \hfill
    \includegraphics[width=0.255\textwidth]{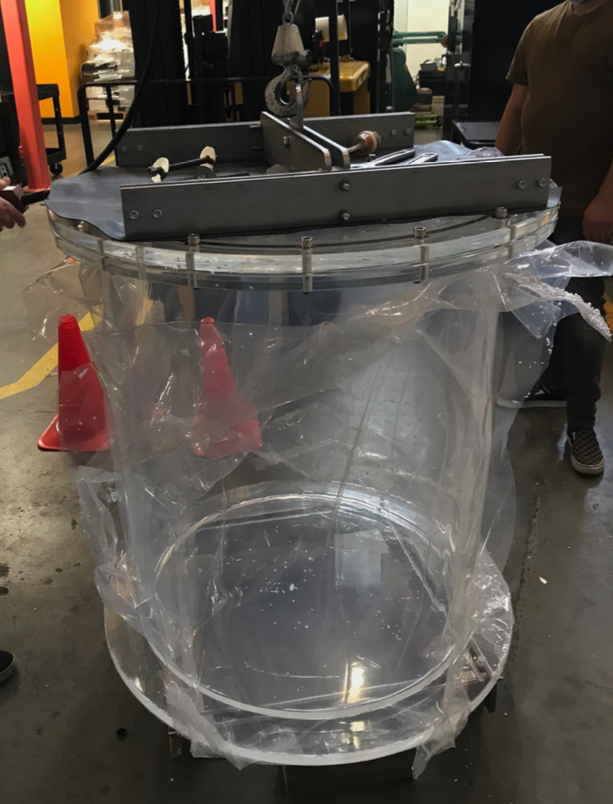}

    \caption{(left) NTT before installation. The rails allow LAPPDs to be inserted between the PMTs. (center) ANNIE detector setup with envisaged SANDI vessel containing 350\,kg of GdWbLS. (right) Test of SANDI acrylic vessel wrapped in protective plastics at UC Davis.} 
    \label{fig:ANNIE}
\end{figure}
\medskip
\noindent{\bf ANNIE program with Gd-water target:} The two main goals of the current ANNIE Phase 2 are the precision measurements of the neutron yield from neutrino interactions at the 1\,GeV scale, and a test of the effectiveness of Large Area Picosecond PhotoDetectors (LAPPDs) in a realistic physics environment (Sec.~\ref{sec:LAPPD}). As illustrated by Figure~\ref{fig:ANNIE}, the current setup of the ANNIE detector comprises a 26-ton Neutrino Target Tank (NTT) equipped with 128 conventional PMTs, a Front Muon Veto (FMV) to reject ``dirt'' muons, and a Muon Range Detector (MRD) to track and range out muons. The water of the NTT was loaded with a mass fraction of 0.1\% of gadolinium (GdS) in early 2019. In January 2021, ANNIE started to take physics data in this basic configuration, while the deployment of first LAPPDs is foreseen for early 2022. The aim is to install in total five LAPPDs before the end of Phase 2 to evaluate their impact on the reconstruction of CC interaction final states as well as vertex reconstruction of the low-energy signals of neutron capture on gadolinium. The conventional PMT array provides a vertex reconstruction of $\Delta r=38\;$cm for CC final-state muons. The full configuration including the LAPPDs and their sub-nanosecond timing is expected to substantially improve this vertex resolution to about $\Delta r=12\;$cm.
\medskip\\
\noindent{\bf Scintillator for ANNIE Neutron Detection Improvement (SANDI):} From simulation studies, we know that $-$ even in the presence of LAPPDs $-$ the vertex resolution suffers from the need to image the ring edge with the conventional PMT array. There is a timing degeneracy in single track particles that makes the vertex location parallel to the track entirely dependent on ring edge imaging. One way to improve on this is to use a slightly scintillating material as neutrino target, resulting in a comparatively small amount of isotropic light emitted from the vertex that can be separated from Chereknov light by timing and topology. 

To explore the potential benefit for event reconstruction, ANNIE plans to insert a 3'$\times$3' cylindrical vessel (SANDI) containing about 350\,kg of gadolinium-loaded Water-based Liquid Scintillator (GdWbLS) into the center of the NTT water volume (Figure~\ref{fig:ANNIE}). In addition, one or more of the five LAPPDs may be moved to the backward hemisphere. A deployment of several weeks will allow us to confirm models for track reconstruction and to identify additional scintillation light emitted by the hadronic recoil at the vertex. First simulation studies show that the inclusion of this signal will indeed lead to an improvement in reconstruction capability. While the muon energy reconstruction is dominated by the information from the (unchanged) MRD, the additional information from the GdWbLS volume improves the neutrino energy reconstruction from ($\Delta E/E\sim15\%$ to $11\%$).  
\medskip\\
\noindent {\bf Future full GdWbLS phase:} Following a successful SANDI test, ANNIE plans to propose a Phase 3 that features a complete replacement of the Gd-water with GdWbLS. Neutron captures in GdWbLS feature a substantially increased light output (factor $\sim$3) compared to Gd-water, greatly improving neutron detection efficiency ($\to\,\geq90\%$) and spatial resolution for the capture vertex ($\to\,\sim40\,$cm). This will enhance not only the precision of neutron multiplicity measurements but also bears the chance to reconstruct the neutron energy spectrum by their capture position relative to the production vertex. A full calorimetric measurement of the cross-section (including vertex hadronic energy) may also be possible. 

With these ambitious goals in mind, ANNIE is investigating the possibility to reconstruct the inner part of the NTT to make it compatible with WbLS, including encapsulation of PMTs and other components, in addition to the first-ever deployment of a WbLS liquid recirculation system. Systems based on nanofiltration and phase separation technology are being developed for Eos and Theia (Sects.~\ref{sec:eos} and \ref{sec:Theia}). ANNIE Phase 3 will be a significant step towards realization of hybrid optical detectors in addition to making neutron-inclusive neutrino cross-section measurements of unprecedented scope and quality. 

\subsection{NuDot}
\label{sec:NuDot}
NuDot is a mid-scale prototype designed to demonstrate timing-based Cherenkov/scintillation separation in a realistic experimental geometry, focusing on techniques applicable to searches for neutrinoless double-beta decay ($0\nu\beta\beta$). It builds on the successful demonstration of this approach in the FlatDot test stand~\cite{flatdot}. It is currently undergoing detector commissioning, with its initial physics data-taking campaign expected to begin by Summer 2022. 

\begin{figure}[htb]
    \centering
    \includegraphics[width=\textwidth]{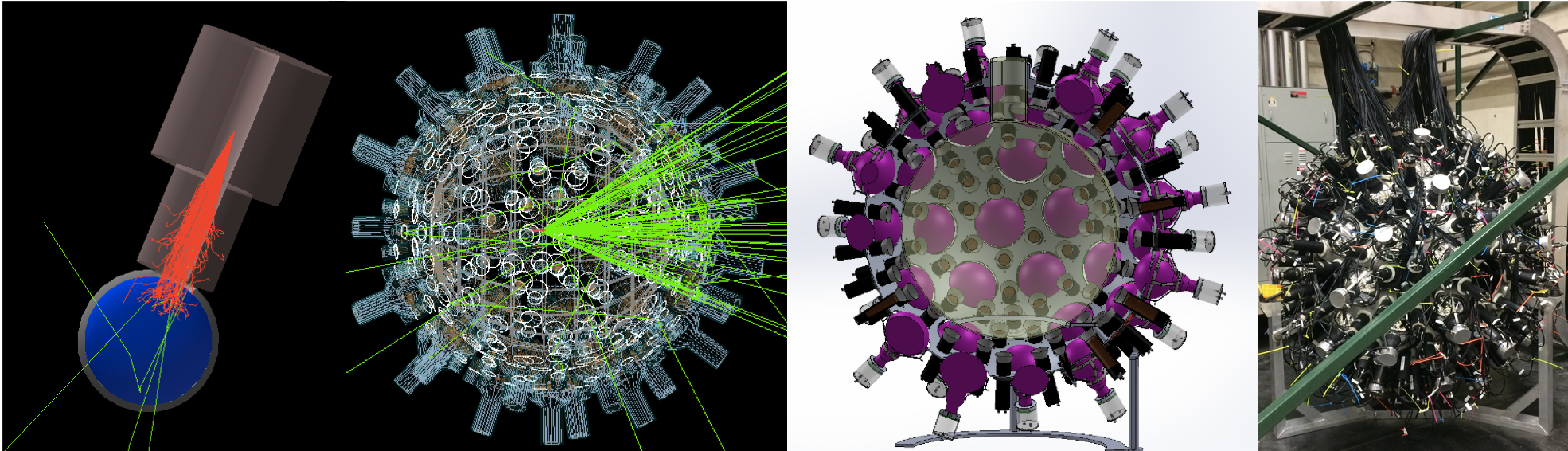} 
    \caption{\textit{From left to right:} A RAT-PAC simulation of electron tracks in the collimated $\beta$ source and coupled liquid scintillator cuvette used for ``dry" surface data-taking; a RAT-PAC simulation of Cherenkov light tracks for a 1\,MeV $\beta$ in NuDot; a rendering of a section view of the NuDot detector; a photo of NuDot, before the calibration system was deployed and the sphere was made light-tight for commissioning runs.}
    \label{fig:NuDot}
\end{figure}

The NuDot design, shown in Fig.~\ref{fig:NuDot}, relies on a combination of 59 8"-diameter PMTs (primarily a combination of EMI Model D642 and Hamamatsu Model R1408 PMTs previously used by the MACRO experiment) and 151 2"-diameter PMTs (Hamamatsu Module R13089, with an average transit time spread of 200\,ps) to achieve high light collection efficiency and fast-timing capability. These detectors surround a 36"-diameter acrylic sphere that will be filled with a LAB-based liquid scintillator cocktail, with the entire assembly immersed in a mineral oil tank to provide shielding and optical coupling. Detector calibration and the first two phases of data-taking will be conducted using a remotely-controlled $4\pi$ calibration system with 3 independent axes of motion. In addition to an LED source used for gain and pulse-shape calibrations, this system holds a collimated $\beta$-emitting $^{90}$Sr needle source that can be adjusted to shine on any point on the NuDot sphere from any point along the sphere diameter, allowing tests of the NuDot Cherenkov/scintillation separation capability as a function of position and direction. 

NuDot has 3 planned phases of operation, the first two of which are conducted prior to underground deployment:
\begin{enumerate}
    \item ``Dry" runs at Bates Research and Engineering Center: a small cuvette of liquid is coupled directly to the collimated source, allowing for calibration of the PMT timing response and pulse shape with water Cherenkov data, followed by C/s separation tests with a variety of liquid scintillator cocktails.
    \item Liquid scintillator- and mineral-oil-filled runs at Triangle Universities Nuclear Laboratories: following the completion of phase 1 data taking, NuDot will be moved to TUNL for any needed detector upgrades and liquid-filling. The goal of this phase is to demonstrate NuDot's C/s separation and energy reconstruction capability in the final operational configuration using the calibration system. 
    \item Underground proof-of-concept measurement: in phase 3, NuDot will be re-deployed underground for a proof-of-concept measurement. The current baseline is to conduct a $2\nu\beta\beta$ decay measurement, but alternatives like $\beta + $/EC searches are under consideration.
\end{enumerate}
The isotope and isotope-loading technique that will be used in phase 3 has not yet been selected. The NuDot Collaboration is currently conducting R\&D measurements of quantum-dot-loaded liquid scintillator cocktails that use perovskite wavelength shifters (see Ref.~\cite{perovskite}), but is also considering high-concentration Te loading and pressurized Xe loading options.

As described in Section~\ref{sec:KamNet}, the novel spatiotemporal neural network model KamNet has the ability to maximally extract information based on event topology in spherical LS detectors. We plan to leverage KamNet to achieve various analysis goals in NuDot including event tagging, position/directionality reconstruction and Cherenkov/Scintillation separation.



\subsection{\eos}
\label{sec:eos}

Successful detection of Cherenkov light from highly scintillating media has been achieved in CHESS, and complemented with an extensive characterisation program at LBNL, BNL, LLNL, and Davis.  These measurements inform large-scale Monte Carlo models, which are used to predict performance in kton-scale detectors. Ton-scale deployments of novel scintillators (ANNIE, BNL) are planned to test production, deployment, and recirculation.  

While bench-top measurements have been, and will continue to be used to measure microphysical properties of novel scintillators and photon detection technology, all demonstrations of event reconstruction and the resulting background rejection capabilities are purely Monte Carlo driven.  Data-driven demonstrations of the event imaging capabilities of hybrid detector technology is a critical step to realising a large detector for both fundamental physics and nonproliferation applications.

The proposed \eos prototype is a few-ton scale detector, designed to hold a range of novel scintillators, coupled with an array of photon detection options and the ability to deploy a range of low-energy calibration sources.  \eos will be constructed, calibrated and tested in Berkeley.  Assuming a successful surface deployment,  \eos could later be re-deployed underground, for example at SURF or SNOLAB, or at an alternative off-site location such as a reactor or test beam for further characterization of detector response to a range of particle interactions. \eos represents significant risk reduction for a large-scale deployment of (Wb)LS and novel detection technology.

\eos will be sufficiently large to use time-of-flight based reconstruction, and to fully contain a range of low-energy events ($\alpha$, $\beta$, $\gamma$, n) for detailed event-level characterisation.  \eos represents a balance of sufficient size for full event characterization, complemented with economy of scale, and flexibility to adapt for multiple target materials and photon detection options.  

The primary goal of \eos is to validate performance predictions for large-scale hybrid neutrino detectors by performing a data-driven demonstration of low-energy event reconstruction leveraging both Cherenkov and scintillation light simultaneously.  By comparing data to model predictions, and allowing certain detector configuration parameters to vary -- such as the fraction of LS in a WbLS target cocktail, or by using PMTs with differing TTS, or deployment of dichroicons -- the predictive power of the model can be validated.  This validated microphysical model of hybrid neutrino detectors can then be used by the neutrino community for design optimization of next-generation hybrid detectors.

 \eos will provide an important test bed to the community, for testing alternative target media, photon detectors, and readout technology and methodology, and assessing the impact of these novel developments on detector performance.  
 
After the conclusion of operations at LBNL, \eos could be redeployed at an off-site location.  This would take place after the end of the primary project period, and successful completion of the project objective. Options include:
\begin{itemize}
\item Underground deployment, for example at SURF or SNOLAB, where scintillator handling procedures are well vetted and understood, and we have relationships of long standing.  This would allow more precise measurements in a low-background environment.
\item Deployment at a reactor for low-energy neutrino reconstruction.  This provides a near-field demonstration of the remote reactor monitoring concept.
\item Deployment at a test beam for hadronic event reconstruction, or a neutron source such as the Spallation Neutron Source at ORNL for neutron studies, useful for advanced event identification and background characterization.
\end{itemize}

\begin{figure}[htb!]
\centering
\includegraphics[width=0.5\textwidth]{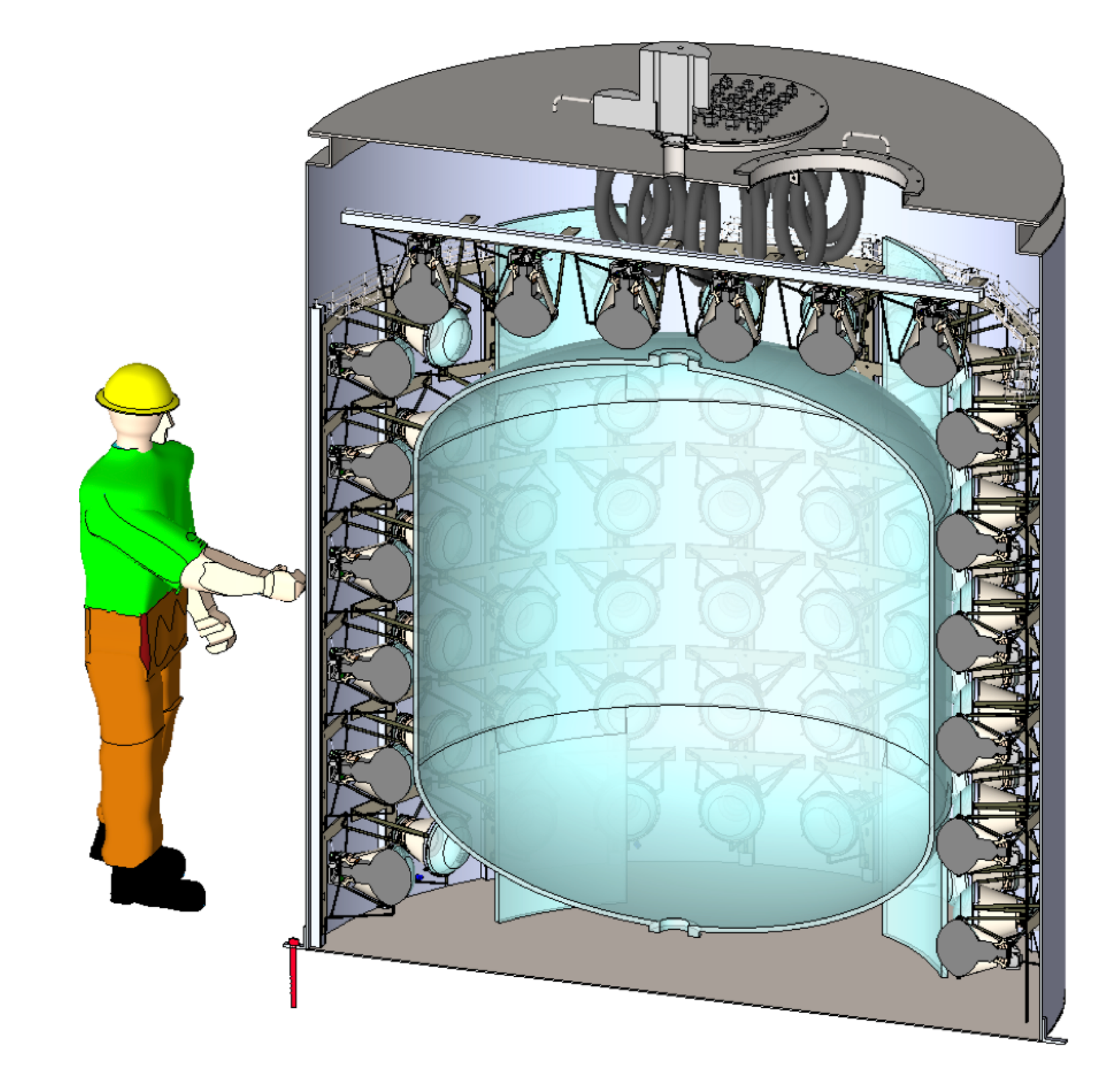}
\includegraphics[width=0.38\textwidth]{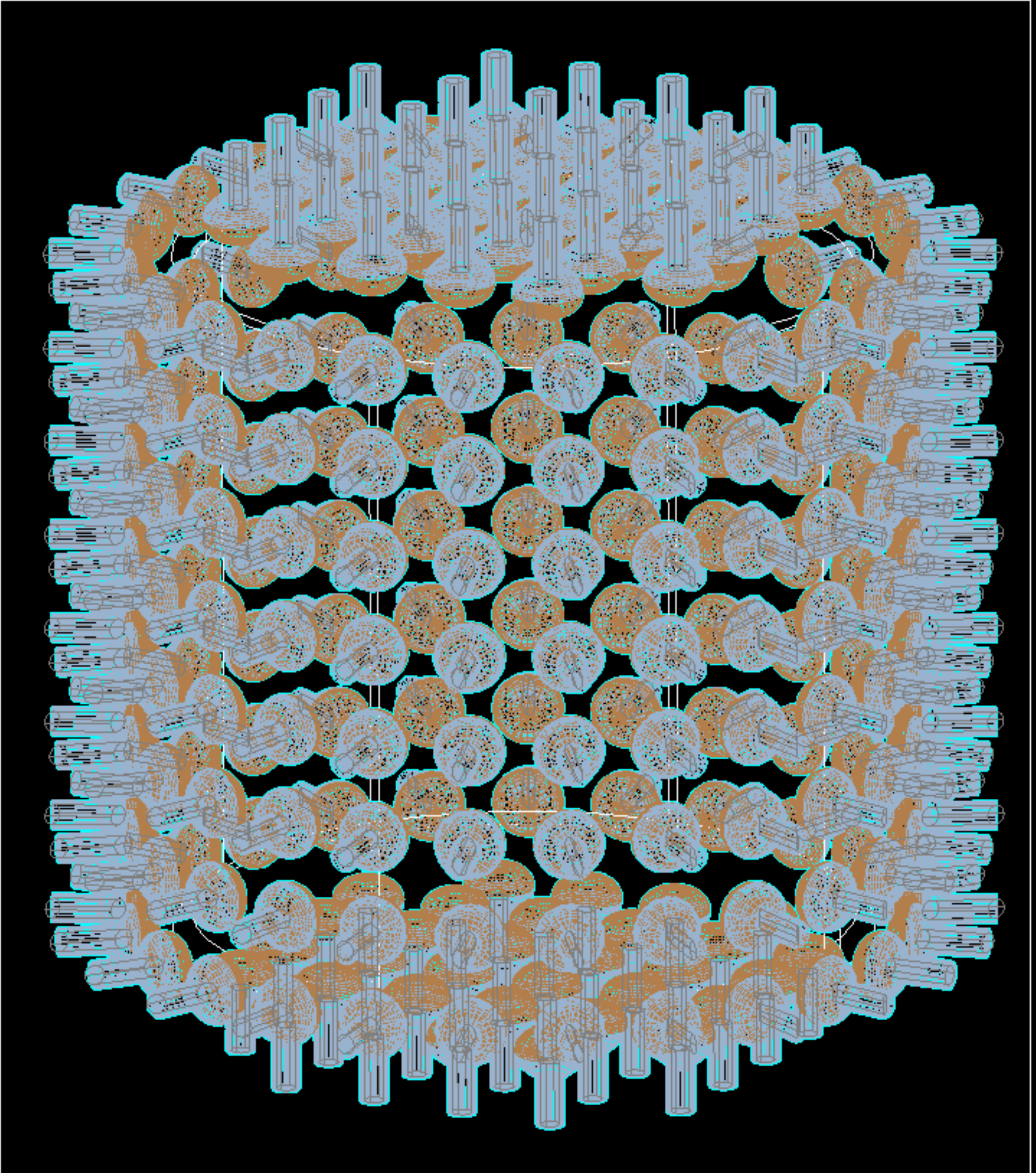}
\caption{\label{fig:eos}
(Left) a preliminary sketch of \eos, courtesy of Joe Saba.  (Right) a MC image of a possible design for \eos, from the RAT-PAC software package.}
\end{figure}

\section{Novel Detector Ideas}

\subsection{Theia}
\label{sec:Theia}

New developments in liquid scintillators, high-efficiency, fast photon detectors, and chromatic photon sorting have opened up the possibility for building a large-scale detector that can discriminate between Cherenkov and scintillation signals. Such a detector could exploit these two distinct signals to observe particle direction and species using Cherenkov light while also having the excellent energy resolution and low threshold of a scintillator detector. Situated in a deep underground laboratory, and utilizing new techniques in computing and reconstruction techniques, such a detector could achieve unprecedented levels of background rejection, thus enabling a rich physics program that would span topics in nuclear, high-energy, and astrophysics, and across a dynamic range from hundreds of keV to many GeV. The scientific program would include observations of low- and high-energy solar neutrinos, determination of neutrino mass ordering and measurement of the neutrino CP violating phase $\delta$, observations of diffuse supernova neutrinos and neutrinos from a supernova burst, sensitive searches for nucleon decay and, ultimately, a search for NeutrinoLess Double Beta Decay (NLDBD) with sensitivity reaching the normal ordering regime of neutrino mass phase space.

\theia is a detector concept that incorporates these new technologies in a practical and affordable way to accomplish the science goals described above. We consider two scenarios, one in which \theia would reside in a cavern the size and shape of the caverns intended to be excavated for the Deep Underground Neutrino Experiment (DUNE) which we call \theia-25, and a larger 100 ktonne version (\theia-100) that could achieve an even broader and more sensitive scientific program.

    \begin{figure}[htp!]
\centering
\includegraphics[width=0.95\textwidth]{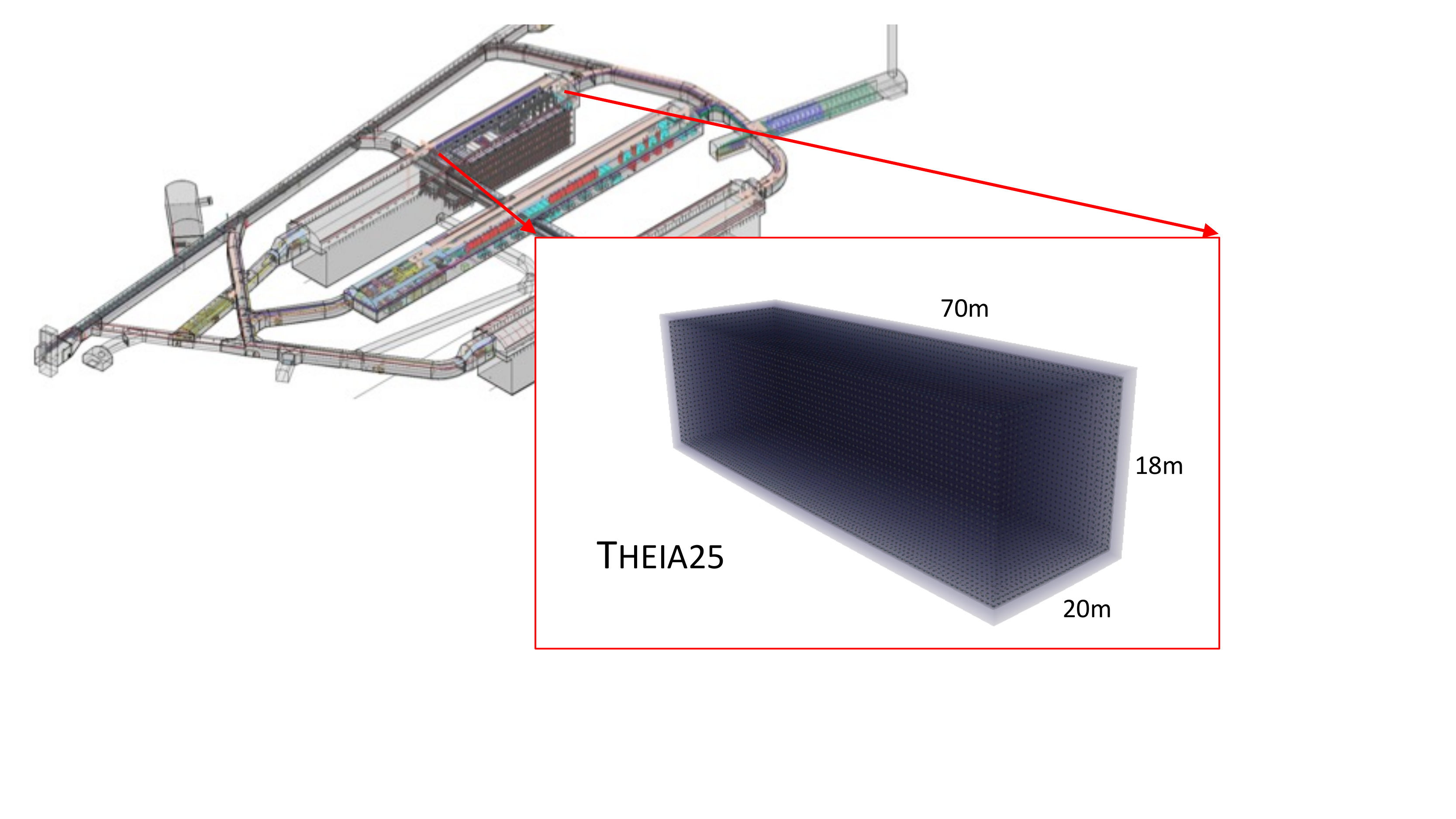}
\includegraphics[width=0.36\textwidth]{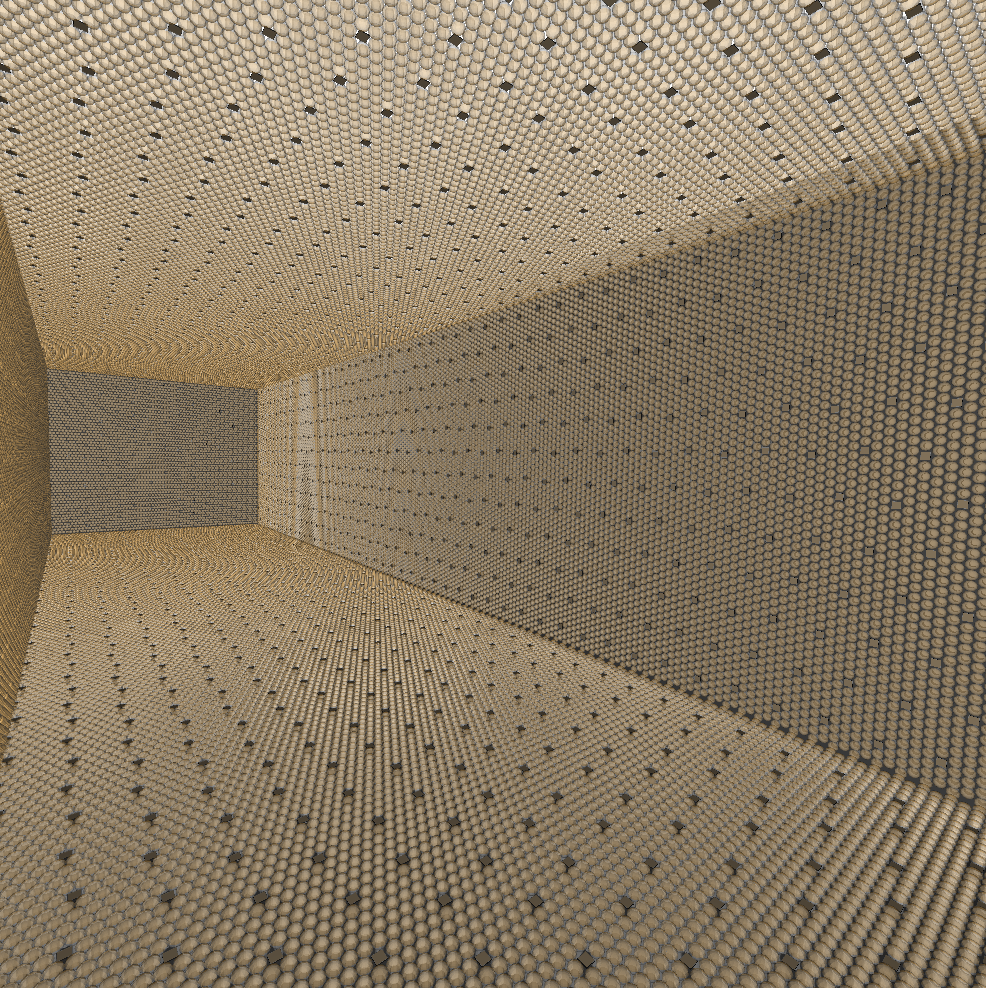}
\includegraphics[width=0.26\textwidth]{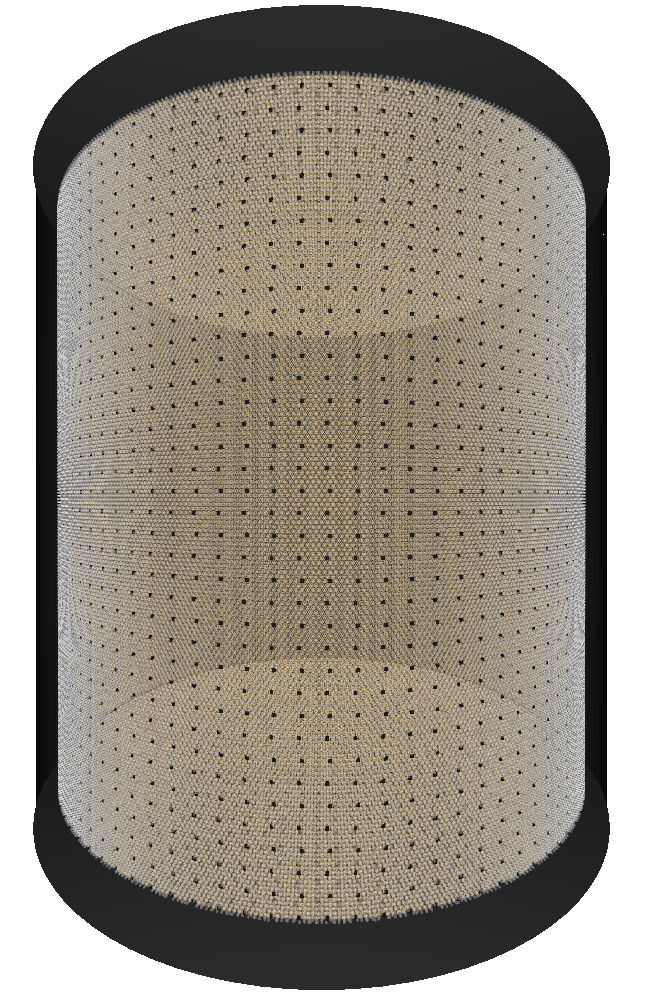}
\includegraphics[width=0.36\textwidth]{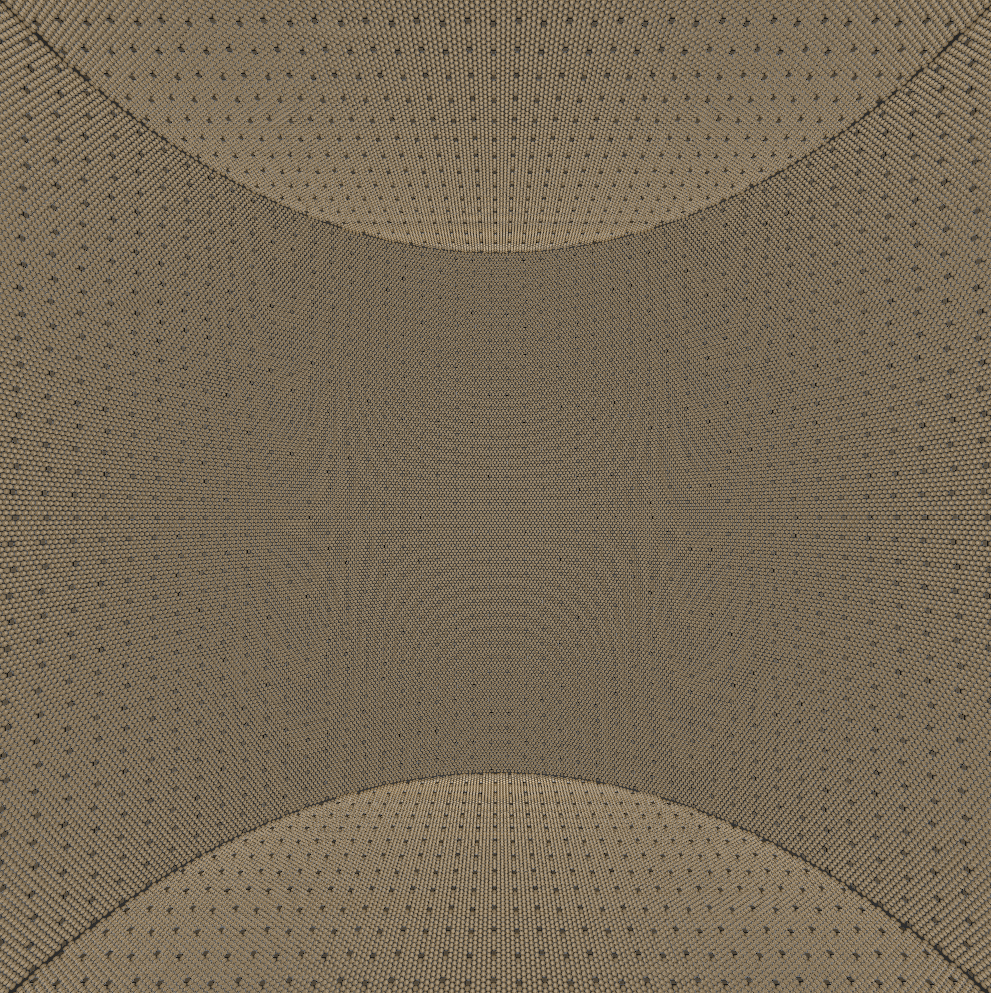}
\caption{The \theia detector.  
{\it Top  panel:} \theia-25 sited in the planned fourth DUNE cavern; 
{\it Lower left panel:} an interior view of \theia-25 modeled using the Chroma optical simulation package~\cite{chroma}; 
{\it Lower middle panel:} exterior view of \theia-100 in Chroma;
{\it Lower right panel:} an interior view of \theia-100 in Chroma.  In all cases, \theia has been modelled with 86\% coverage using standard 10-inch PMTs, and 4\% coverage with LAPPDs, uniformly distributed, for  illustrative purposes.  Taken from~\cite{theiawp}.}
\label{f:detector}
\end{figure}

    
In the \textsc{Theia} reference design, the target material would be water-based liquid scintillator (WbLS) described  in Section~\ref{sec:WbLS}, which has an advantage for a big detector in its  long attenuation lengths. For much of the low-energy program of \textsc{Theia}, the WbLS would need to be made radiopure at levels not far from what can be done in an organic scintillator.   

The broadband neutrino beam being built for the Long Baseline Neutrino Facility (LBNF)
~\cite{lbnf} and the Deep Underground Neutrino Experiment (DUNE)~\cite{dunecdr} offers an opportunity for world-leading long-baseline neutrino oscillation measurements. Due to advances in Cherenkov ring reconstruction techniques, a \theia detector in the LBNF beam would have good sensitivity to neutrino oscillation parameters, including CP violation (CPV), with a relatively modestly-sized detector. 
In addition to this long-baseline neutrino program, \theia will also contribute to atmospheric neutrino measurements and searches for nucleon decay, particularly in the difficult $p\rightarrow K^+ + \overline{\nu}$ and $N\rightarrow 3\nu$ modes.

	\theia will also make a definitive measurement of the solar CNO neutrinos,
which have recently been detected exclusively by Borexino~\cite{bxcno}, but without enough precision to distinguish between competing models of the Sun's metallicity.
\theia will also provide a high-statistics, low-threshold ($\sim$ 3~MeV)
measurement of the shape of the $^8$B solar neutrinos and thus search for new
physics in the MSW-vacuum transition region~\cite{friedland2004, minakata2012}. Antineutrinos produced in the
crust and mantle of the Earth will be measured precisely by \theia with statistical uncertainly far exceeding all detectors to date. 

Should a supernova occur during
\theia  operations, a high-statistics detection of the $\bar{\nu}_e$ flux will be
made---literally complementary to the  detection of the $\nu_e$ flux in the DUNE liquid argon
detectors. The simultaneous detection of both messengers---and detection of an optical, x-ray, or gamma ray component will enable a great wealth of neutrino physics
and supernova astrophysics. With a very deep location and with the detection of a combination of scintillation and Cherenkov light, \theia will have world-leading sensitivity to make a detection of the Diffuse Supernova Neutrino Background (DSNB) antineutrino flux. The most ambitious goal,
which would likely come in a future phase, is a search for Neutrinoless double beta decay (NLDBD), with a total isotopic mass of 10 tonnes or more, and with decay lifetime sensitivity in excess of $10^{28}$ years~\cite{theiawp, biller_normal}.

\theia is able to achieve this broad range of physics by exploiting new
technologies to act simultaneously as a (low-energy) scintillation detector and
a (high-energy) Cherenkov detector. Scintillation light provides the energy
resolution necessary to get above the majority of radioactive backgrounds and
provides the ability to see slow-moving recoils; Cherenkov light enables event
direction reconstruction which provides particle ID at high energies and
background discrimination at low-energies. Thus, the scientific program benefits in many cases on
the ability of \theia to discriminate efficiently and precisely between the ``scintons'' (scintillator photons)
and ``chertons'' (Cherenkov photons). 

    A major advantage of \theia is that the target can be modified in a phased program to address the science priorities. In addition, since a major cost of \theia is expected to be photosensors, investments in \textsc{Theia25} instrumentation can be transferred directly over to \textsc{Theia100}. Thus, \theia can be realized in phases, with an initial phase consisting of lightly-doped scintillator and very fast photosensors, followed by a second phase with enhanced photon detection to enable a very low energy solar neutrino program, followed by a third phase that could include doping with a $0\nu\beta\beta$ isotope and perhaps an internal containment vessel. 


\subsection{LiquidO}

\subsubsection{The LiquidO Concept}
LiquidO is a new approach to detecting neutrinos that, in contrast with conventional liquid scintillator detectors, relies on using an opaque scintillator medium as the primary neutrino target~\cite{Cabrera:2019kxi}. The scintillators that can be best used by LiquidO have a short scattering length and a medium to long absorption length, an example of which has already been successfully produced~\cite{Buck:2019tsa}. In such a medium, the photons produced by the opaque scintillator undergo a random walk process near their creation point and are trapped in so-called {\it light balls} around each energy deposition point. The light is collected by a dense array of wavelength-shifting fibres that traverses the volume and that is readout by photo-sensors in the periphery. Silicon photomultipliers (SiPMs) are well-suited to this purpose given their affordable price, high efficiency, and excellent time resolution. 

\begin{figure}[!htbp]
\begin{center}
\includegraphics[width=0.90\textwidth]{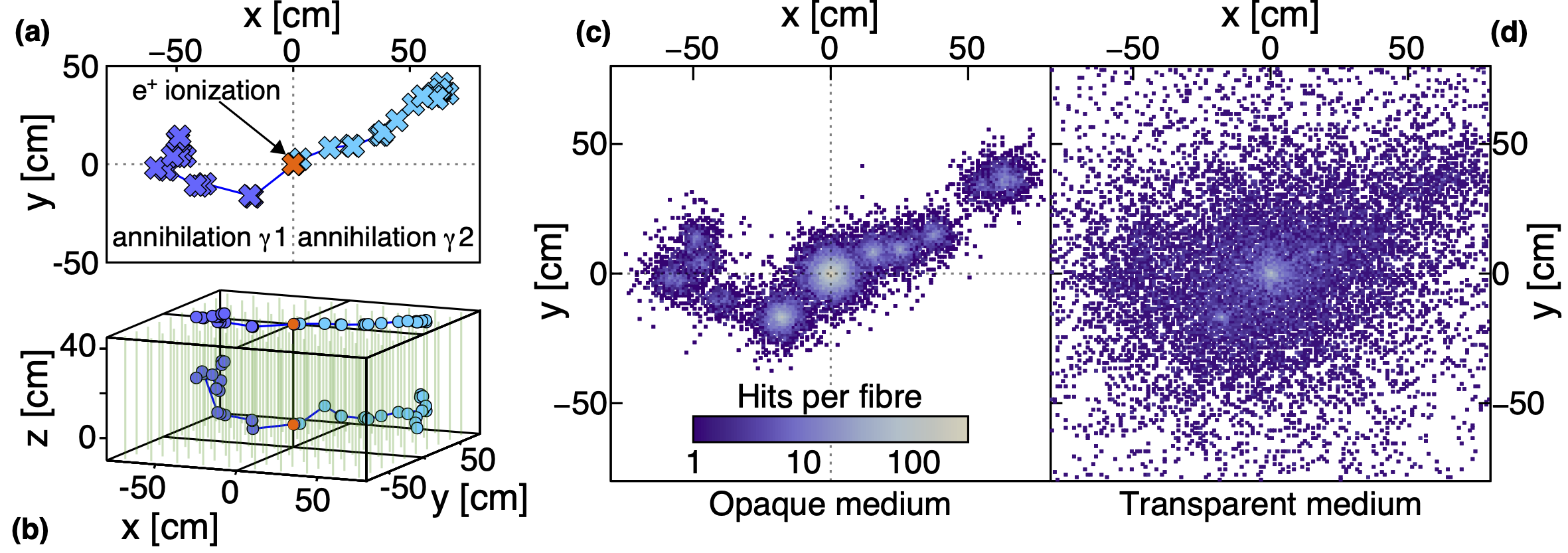}  
\caption{\label{fig:lozfig} Left: energy depositions of a simulated 1~MeV kinetic energy positron in a LiquidO detector with a regular 1~cm fibre pitch running along the $z$ direction. Panel (a) shows the $x$-$y$ projection, while panel (b) shows the full three-dimensional extent. The fibres are represented in green. Right: true number of photons hitting the fibres, each of which is represented by a pixel, in the opaque and transparent scintillator scenarios. In the former case, the scintillator is assumed to have a 5~mm scattering length and a 5~m absorption length. Figure obtained from Ref.~\cite{Cabrera:2019kxi}.}
\end{center}
\end{figure}

\begin{figure}
    \includegraphics[width=8cm]{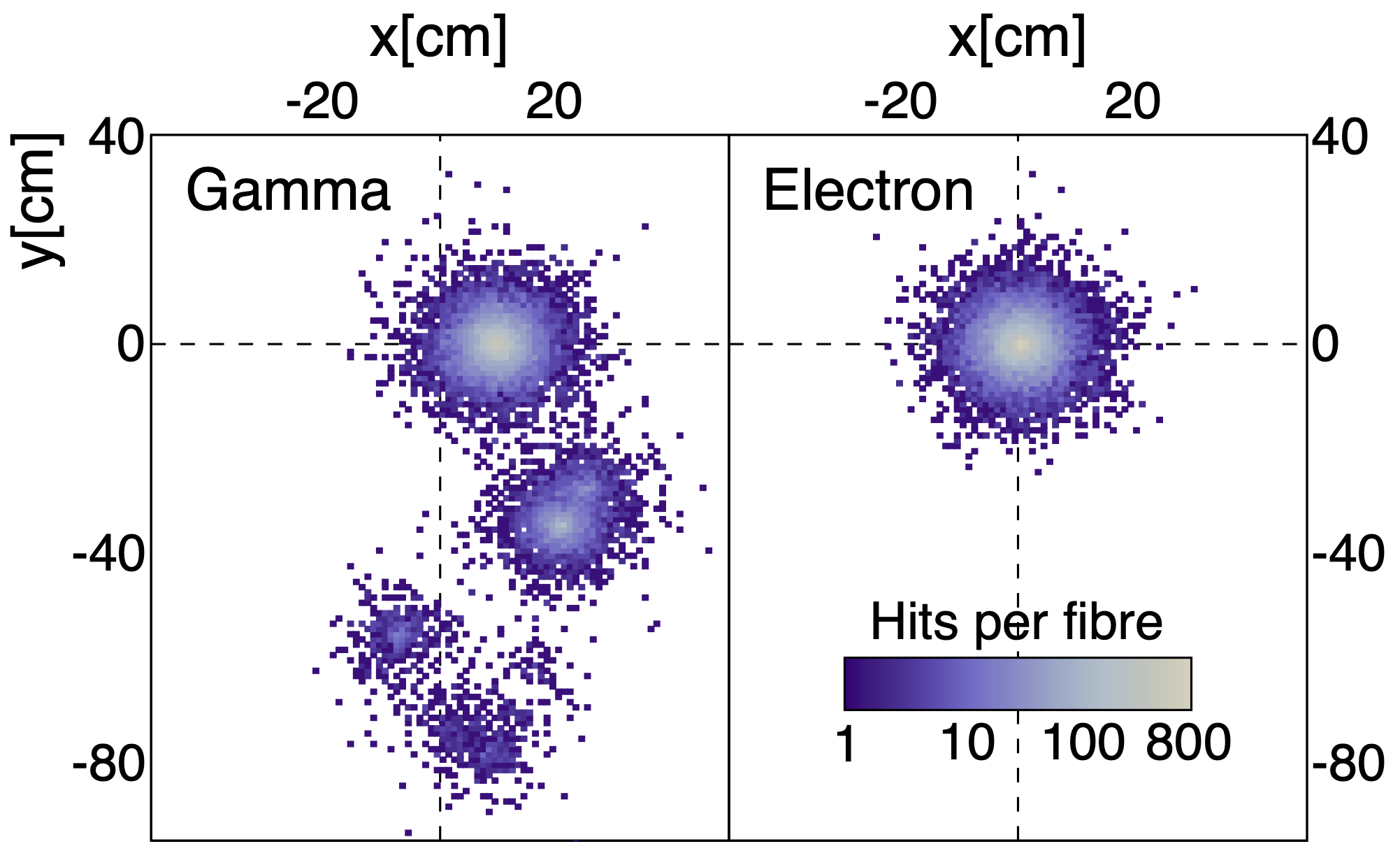}
    \caption{Simulated 2~MeV gamma (left) and electron (right) in the same detector configuration of Fig.~\ref{fig:lozfig}. Figure obtained from Ref.~\cite{Cabrera:2019kxi}.
    \label{fig:gammaelectron}}
\end{figure}

 A full description of LiquidO, its expected performance, first experimental demonstration, and potential applications, can be found in Ref.~\cite{Cabrera:2019kxi}. Fig.~\ref{fig:lozfig} illustrates LiquidO's performance using a simulated positron with 1~MeV of kinetic energy. Here, the simplest configuration with fibres running only along one direction ($z$) is assumed. The true energy depositions of the positron are shown on the left panels, with (a) showing the $x$-$y$ projection and (b) the full three-dimensional extent. Panel (c) shows the number of true photons hitting each fibre in a 1-cm-pitch lattice when a scintillator with a 5~mm scattering length is used. The positron's loss of kinetic energy produces a light-ball at the vertex of the event. The two back-to-back gamma-rays resulting from its annihilation lose energy via Compton scattering, leaving two trails of smaller light balls that detach from the central one. A comparison is made in panel (d) of the light pattern collected by the same fibre array when using a transparent scintillator. Despite the use of fibres, the event topology is almost entirely washed out, illustrating the key role played by the scintillator's opacity in self-segmenting the detector.

The clear event topology of $\sim$MeV positrons in LiquidO stands in contrast with that of other events, as illustrated in Fig.~\ref{fig:gammaelectron}. At these energies, gammas lose their energy primarily via the Compton effect and produce trails of light balls over many tens of cm, whereas electrons produce single light balls. At higher energies (more than $\sim$10~MeV for electrons), charged particles have enough kinetic energy to travel several cm or more in the detector, producing sequences of point-like energy depositions that form clear tracks. As a result, many other interactions, from cosmic ray muons to charged and neutral current neutrino interactions of various energies, could also be precisely reconstructed in LiquidO. In this way, this detector technology combines some of the advantages of conventional liquid scintillator detectors with those of tracking detectors.\\

\subsubsection{LiquidO's Advantages}

LiquidO builds on the decades-long experience with liquid scintillator detectors and inherits some of their main advantages, such as the relatively high light levels compared to other technologies. Estimating the total photon detection efficiency at 3\%, which is dominated by the $\sim$10\% probability that a photon re-emitted by the wavelength-shifting fibres is trapped and carried to the SiPMs~\cite{Ref_FibreAcceptance}, and assuming a scintillator with conventional light-yield and absorption length like the one of Ref.~\cite{Buck:2019tsa}, yields 400~photo-electrons per MeV for a small 1~cm-pitch lattice detector. When scaling to larger sizes this amount is reduced due to the several-meter attenuation length of wavelength-shifting fibres. The light collection could be optimized by considering elongated geometries, modularisation, and/or  double-ended readout, whose cost can be strongly mitigated by multiplexing. A notable aspect of LiquidO is that it makes scintillating materials that are naturally opaque ideal, opening up a whole landscape of substances to explore. Known scintillators with substantially higher light output present promising avenues of research, alongside the possibility of finding new materials that have not yet been carefully studied because of their poor transparency.

LiquidO also brings new advantages to the table, two of which in particular could unlock new opportunities in neutrino physics:
\begin{itemize}
    \item {\bf Unprecedented background rejection}: LiquidO's unique imaging capabilities enable unprecedented event-by-event identification down to the MeV scale. Studies have been carried out in a detector with the same configuration of Fig.~\ref{fig:lozfig} and with a total detection efficiency of 3\%. The latter accounts for all losses of light and is dominated by the fibre trapping efficiency ($\sim$10\%) and the Si-based photo-sensor (SiPM) quantum efficiency ($\sim$50\%). It is found that 2~MeV electrons can be separated from gammas with a contamination better than $10^{-2}$ and an efficiency $>85\%$. As expected, further improvements can be obtained by increasing the amount of light, confining the light more tightly, or running the fibres along more than one direction. It is also anticipated that incorporating the timing information of the light pulses coming from each fibre will further increase the particle identification capabilities. 
    
    \item {\bf Naturally enhanced affinity for doping}: maintaining the required transparency of the scintillator typically limits the doping concentrations that can be achieved in conventional liquid scintillator detectors. In contrast, LiquidO requires opacity to confine the light and therefore allows for more possibilities, be it to load new materials or to achieve significantly higher levels of doping. 
\end{itemize}

\subsubsection{Possible Applications and Status}

LiquidO could find a wide range of applications within neutrino physics. Thanks to its strong separation power between positrons and electrons, LiquidO would greatly reduce the cosmogenic backgrounds in experiments detecting antineutrinos via the Inverse Beta Decay reaction ($\bar{\nu}_e + p \rightarrow e^+ + n$), which typically bear the brunt of the systematic uncertainty. Similarly, LiquidO's self-segmenting effectively keeps events localized, allowing the detector to tolerate much higher rates. This means that a major reduction of the overburden, shielding, and buffer requirements would be possible. In fact, this feature makes LiquidO a promising technology to monitor reactor antineutrinos for non-proliferation purposes~\cite{Huber:2004xh}.

Moreover, the ability to dope a LiquidO detector with various elements at concentrations that would be prohibitive in conventional detectors could enable new measurements from a variety of sources that include the sun, supernovae, pion decay-at-rest beams, and radioactive elements. Using Indium in a LiquidO detector, for instance, could enable $pp$ solar neutrino spectroscopy with a threshold of 114~keV via the $\nu_e + \,^{115}\mbox{In} \to e^- + \,^{115}\mbox{Sn*}$ reaction proposed by Raghavan~\cite{Raghavan:1976yc}. The electron signal, followed by the decay of the tin nucleus, would provide a powerfully distinct signature. Similarly, the ability to control backgrounds, and to dope the scintillator with elements like Telurium well beyond current limits, could result in a state-of-the-art detector to search for neutrinoless double-beta decay. Some of these possibilities and others are discussed in more detail in Ref.~\cite{Cabrera:2019kxi}.

The basic principles behind LiquidO have already been validated with a small detector prototype filled with a scintillator whose opacity changes with temperature~\cite{Buck:2019tsa} and exposed to a mono-energetic 1~MeV $e^-$ source~\cite{Marquet:2015ima}. The results unmistakably showed that, as the scintillator's scattering length decreased, light was confined near the beam's energy deposition point, as expected. The details can be found in Ref.~\cite{Cabrera:2019kxi}. Further studies are ongoing with a larger prototype, and the results will be released in the near future. 

\subsection{SLiPS}
The construction of large-scale liquid scintillator detectors is complicated by the need to separate the scintillation region from photomultiplier tubes (PMTs) due to their intrinsic radioactivity. This is generally done using acrylic and/or nylon barriers, whose own intrinsic activity can also lead to substantial cuts to the fiducial detection volume for a number of low energy (~MeV) studies. Such barrier constructions also become increasingly difficult and expensive for larger detector volumes, with JUNO already pushing the boundaries of what might be achievable. The SLiPS concept \cite{Morton-Blake:2022snr} is to do away with such physical barriers entirely by instead mounting PMTs on the bottom of a wide cavity and covering them with a distillable, lipophobic liquid, above which a less dense scintillator is layered. Liquids such as various ethylene glycols are good candidates for the bottom layer as they provide a good refractive index match to a number of liquid scintillator solvents. Thin, opaque and highly reflective (>90\%) surfaces are used near the top and side areas of the detector to provide a buffer region against radioactivity from the walls and to reflect scintillation light back to the bottom PMT array, where the time-separated reflected signals are used to reconstruct the 3D vertex position as well as the event energy. A conceptual model is shown in Fig.~\ref{fig:SLIPS1}. }Initial simulation studies indicate that good position and energy reconstruction can be achieved with this approach. The notion is to use a shallow layer of scintillator relative to the cavity width, where the vertical depth of scintillator is chosen to be much less than the optical absorption length and can be optimised to balance fiducial volume, light level and reconstruction accuracy.

\begin{figure}[htp]
\centering 
\includegraphics[width=0.9\textwidth]{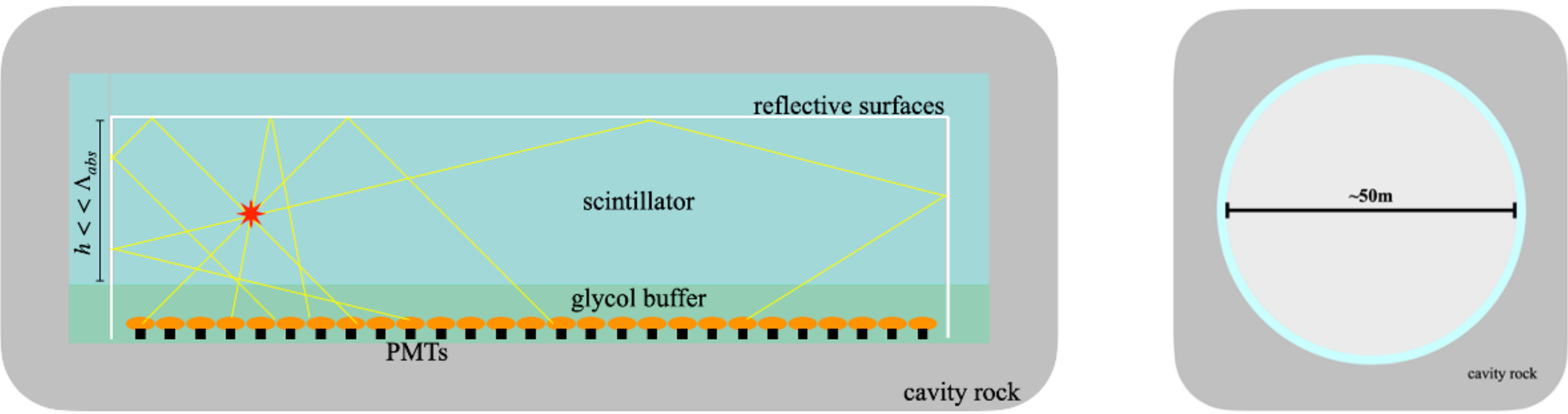} 
\caption{SLIPS design. Left - cross-sectional side view; right - top view.} 
\label{fig:SLIPS1}
\end{figure}

The dense packing of PMTs on just one surface of the cavity can potentially be made highly economical and efficient for light collection. For example, preliminary studies suggest that a similar fiducial volume and energy resolution as JUNO might be achievable with half the number of PMTs. On the other hand, the design could also be optimised to make the best use of a smaller number of PMTs, potentially using further segmentation with internal reflective partitions, which might be more useful for long baseline reactor monitoring. In principle, one could also mount PMTs on the top surface by adding a lower density buffer liquid and using a relatively high-density scintillator solvent, such as PXE, though the relative benefits of such a configuration would need further exploration. This design may allow for a much more simple and economical construction of large-scale scintillation detectors, which could have impact in a number of areas, including neutrinoless double beta decay and solar, supernova and geo neutrinos as well as long baseline monitoring of reactors, which could require detector masses of the order of $\sim$50kT.

\begin{figure}[htp]
\centering 
\includegraphics[width=0.9\textwidth]{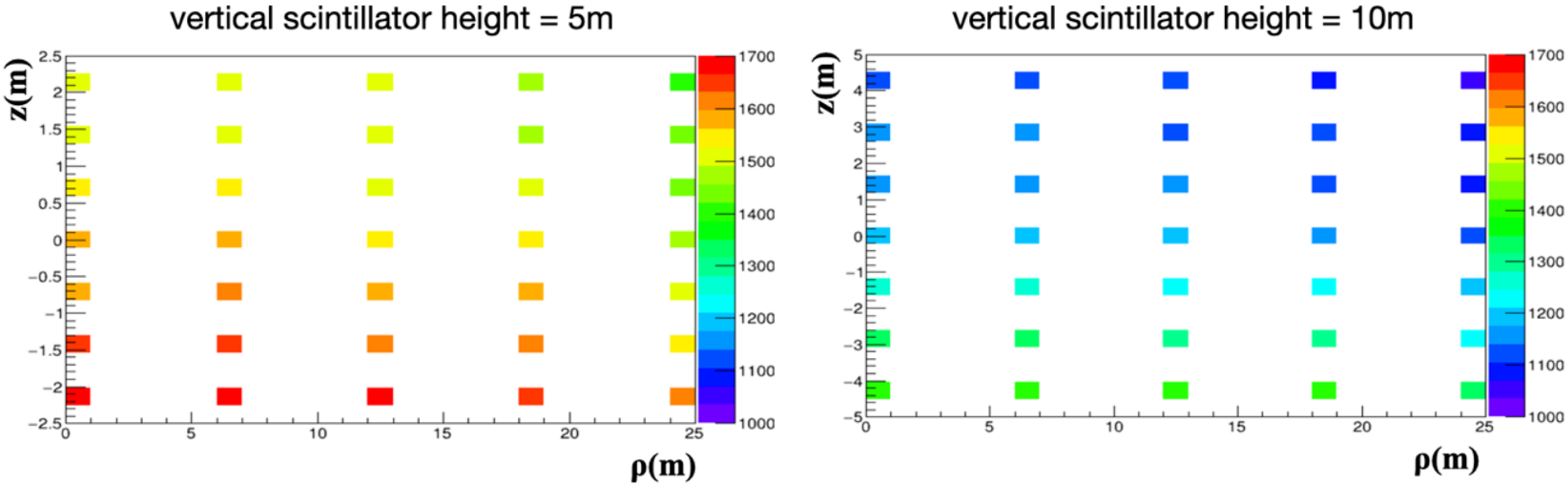} 
\caption{Detected number of photoelectrons for a 1 MeV electron for the simulated configuration as a function of $\rho$ (cylindrical radius from the centre of the detector) and $z$ (vertical height relative to the centre of the scintillator layer). The left plot is for a scintillator layer height of 5m and the right is for a height of 10m.} 
\label{fig:SLIPS3}
\end{figure}

Initial simulations have been carried out assuming a densely packed array of R5912-100 HQE PMTs in a pillbox-shaped detector with a diameter of 50m, an ethylene glycol layer extending 2m above the PMTs, a scintillator layer composed of LAB + 2g/L PPO, and 90\% specular reflective surfaces. Figure \ref{fig:SLIPS3} shows the resulting number of detected photons from a 1 MeV electron as a function of event position for vertical scintillator heights of both 5m and 10m. These correspond to fiducial volumes of $\sim$8kT and 16kT, respectively. 

\begin{figure}[htp]
\centering 
\includegraphics[width=0.9\textwidth]{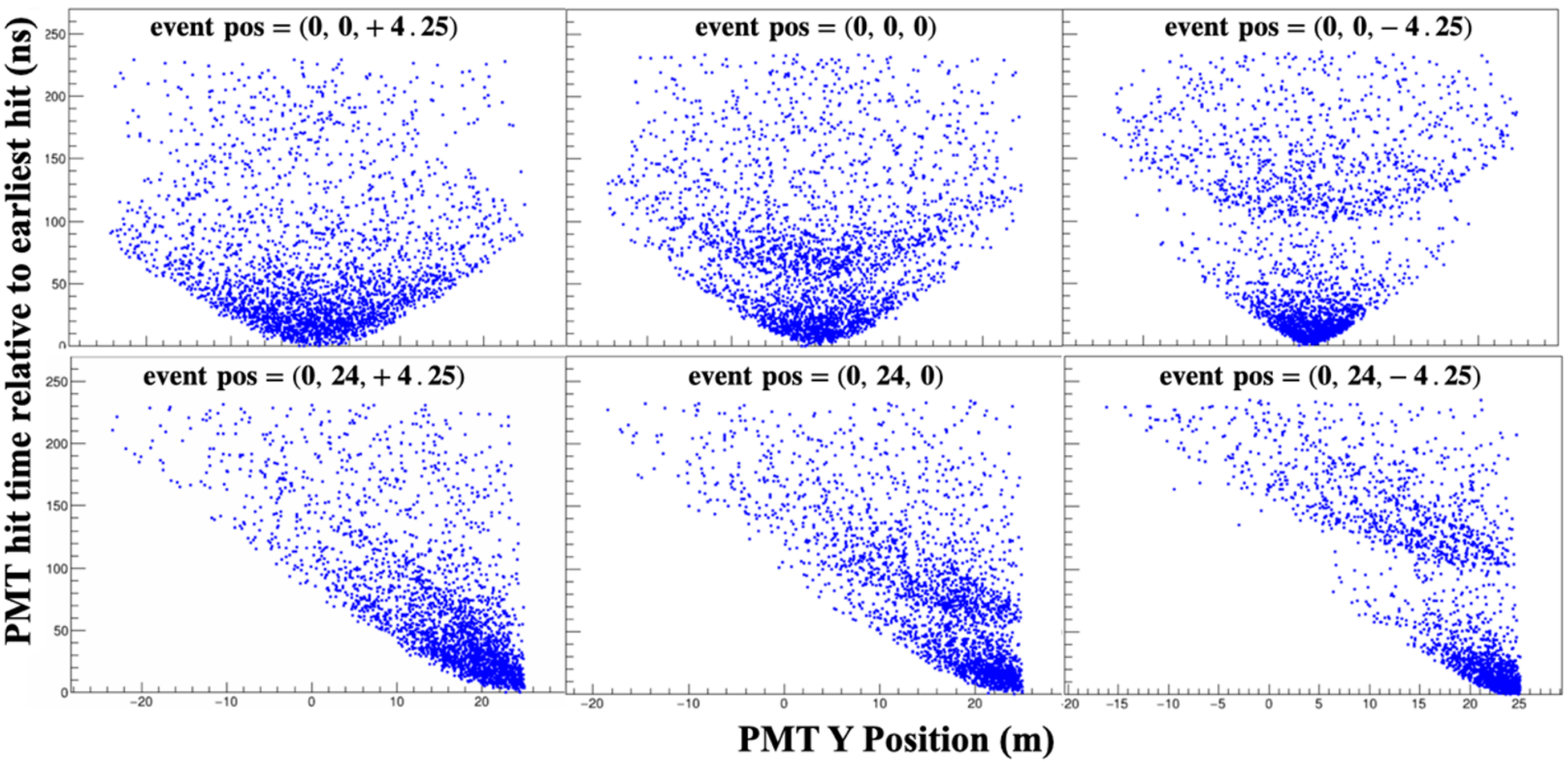} 
\caption{Relative times of PMT hits vs PMT y coordinate for a 3 MeV electron generated at different positions within a scintillator volume of height 10m.} 
\label{fig:SLIPS4}
\end{figure}

Figure \ref{fig:SLIPS4} shows timing and spatial distributions of PMT hits for an individual 3 MeV electron at different locations in the scintillator volume of 10m height. Distinctive reflection peaks and correlations with the spatial position can be clearly seen. Hit density distributions in the x-y plane are also particularly distinct for this geometric configuration and can be also used to provide good localisation. Initial studies suggest a position resolution comparable to that achieved by current conventional LS detectors.

\subsection{Slow Fluor Scintillation Detectors}
The use of slow fluors in liquid scintillation detectors \cite{Biller:2020uoi} currently offers, by far, the most efficient means of Cherenkov separation while maintaining high overall light collection (and, hence, good energy resolution). This is because it allows for excellent Cherenkov separation/detection across the entire photodetection surface and over a very wide wavelength range. This becomes particularly important at lower energies. It is also a highly economical approach that could be applied to current detectors without hardware upgrades.

As an example, the potential for using slow-fluor liquid scintillators to study low energy solar neutrinos has been explored through a series of simulations involving LAB and acenaphthene for various detector configurations \cite{Dunger:2022gif}. These studies suggest that a detector with ~50\% coverage by standard HQE PMTs could be able to make a measurement of the CNO solar neutrino flux to a precision of 10\% (enough to distinguish metallicity models) with a few kiloton-years of exposure, making use of directional Cherenkov information to distinguish the signal from other backgrounds, including 210Bi. Acenaphthene seems to be a particularly good fluor for this due to its long fluorescence decay time (~45ns) and reduced absorption at lower wavelengths compared to PPO, which is favourable for Cherenkov light.

For this application, the use of acenaphthene without bisMSB seems to offer notable advantages, as the energy and position resolution are sufficient and backgrounds can be better suppressed by the improved directional information. The use of faster PMTs appear to offer relatively modest improvement under this scenario, suggesting that currently available PMTs are more than sufficient. To explore the detection of CNO neutrinos more specifically, a background model based on Borexino Phase I levels was used and the pep flux is taken as 1.44x108 cm-2 s-1 [3]. All other normalisations, including 210Bi, were treated as free parameters in a likelihood fit. 

Examples of sensitivity curves, showing the expected measurement precision in the CNO flux as a function of exposure, are given in figures 5 and 6 below. Approximately equivalent ways of obtaining such sensitivities are indicated on the figures. The exposure at which a 10\% constraint is obtained at 1$\sigma$ is also shown. 
 
\begin{figure}[htp]
\centering 
\includegraphics[width=0.9\textwidth]{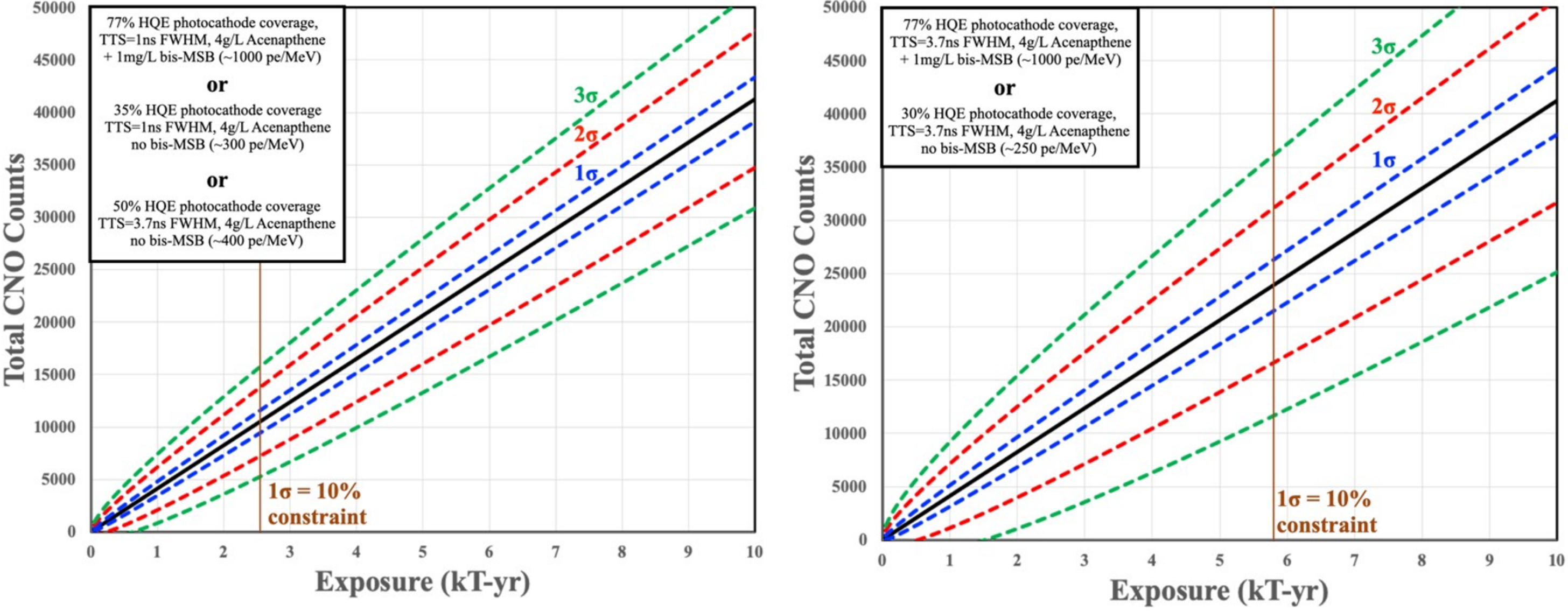} 
\caption{Expected measurement precision in the CNO flux as a function of exposure for various indicated scenarios. } 
\label{fig:SlowFluorSolar}
\end{figure}
 
Liquid scintillators using slow fluors, such as acenaphthene, therefore provide a means to improve the precision of low energy solar neutrino measurements through the use of directional Cherenkov information. This can be done using currently available phototube technology with detector volumes and photocathode coverage comparable to existing instruments. Indeed, both SNO+ and KamLAND detectors are good targets for this approach if they can achieve backgrounds levels in the vicinity of those obtained by Borexino, potentially with the use of inner containment bags. It is also an approach that is useful to consider for future ~kT-scale detectors to improved sensitivity to low energy solar neutrinos and other physics. 

\subsection{ArTEMIS\label{sec:ArTEMIS}}

Distinguishing Cherenkov and scintillation (C/S) photons is challenging in organic liquid scintillators because the short-wavelength end of the Cherenkov spectrum, where most of the photons reside, is buried by the more intense scintillation spectrum.  In contrast, liquid Ar (LAr) scintillates narrowly around 128 nm, leaving the broad Cherenkov spectrum isolated above this wavelength, which is readily detected by common devices such as PMTs.  The scintillation photons can be detected with the same devices upon coating their photosensitive surface with a wavelength shifter like TPB.  
The narrow range of scintillation photons ensures that the uncoated PMTs see purely Cherenkov light.  This pure sample of Cherenkov photons can be used to cleanly reconstruct the position, direction, and energy of events from several MeV and above, depending on the PMT photocoverage and efficiency.  The photons detected by the TPB-coated PMTs would be primarily from scintillation due to its far greater yield and could be used to independently reconstruct the position and energy of events as low as a few MeV.  Using both photon samples together would provide more robust particle ID and reconstruction. 

Relative to LAr time projection chambers like those proposed for DUNE, this type of detector would provide the same particle interaction channels while requiring simpler hardware (no charge readout) and event analysis (like a scintillator or Cherenkov detector).  Collaborators at Penn and Berkeley are studying the capabilities of a LAr detector to distinguish C/S photons.  GPU-accelerated simulations of a large liquid Ar detector with TPB-coated and -uncoated PMTs is performed using the Chroma package and pyrat wrapper.

\begin{figure}[htp]
\centering 
\includegraphics[width=0.5\textwidth]{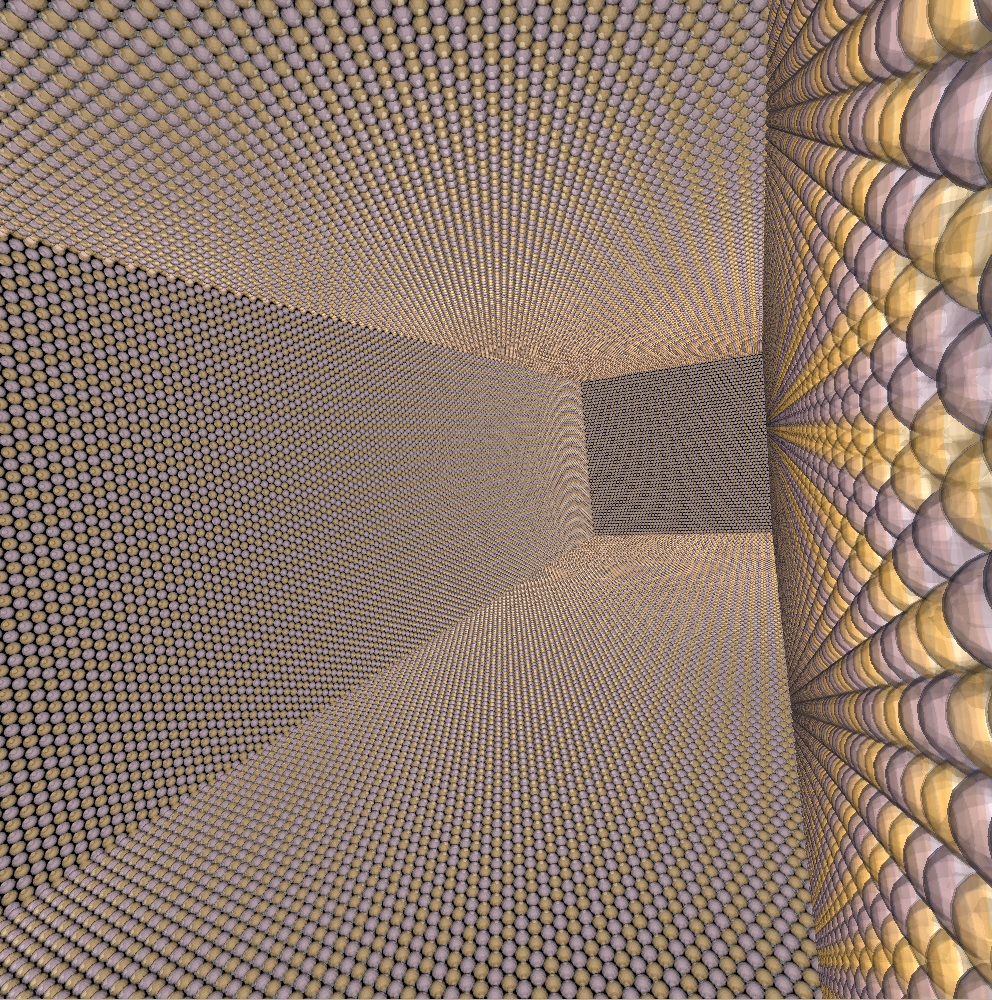} 
\caption{Rectangular detector of dimensions 13.5 x 13.0 x 60.0 m$^3$ filled with LAr and viewed by 60,000 10-inch PMTs, where half are coated with wavelength-shifter TPB (light blue). 
\label{fig:ArTEMIS}}
\end{figure}

A DUNE-sized module is simulated, filled with liquid Ar and lined with 60,000 10-inch PMTs (see Fig.~\ref{fig:ArTEMIS})
, which corresponds to 86\% photocoverage, like the proposed \theia.  The quantum efficiency of the PMTs is conservatively taken to be that of the Hamamatsu R1408 as used in SNO.  
Taking 50\% of the PMTs to be TPB-coated and the other 50\% not, the detected number of Cherenkov photons on just the uncoated PMTs is found to be as high as modern water Cherenkov detectors; namely, SNO, Super-K, and SNO+.  

In addition to distinction by wavelength, C/S photons are also distinguishable by timing since LAr scintillation rises and decays with ns-level time constants, whereas Cherenkov is instantaneous.  
In ArTEMIS, TPB coatings would enhance this distinction as TPB has a re-emission lifetime around two~ns.  Other potential coatings could enhance this further; for example, PEN, which has a re-emission lifetime that is an order of magnitude larger.

\begin{acknowledgments}

\end{acknowledgments}

\bibliography{references}

\end{document}